\documentclass{aa}
\usepackage{txfonts}
\usepackage{graphicx}
\usepackage{natbib}
\usepackage{longtable}
\begin{document}

\title{Habitable-zone super-Earth candidate in a six-planet system around the K2.5V star HD 40307}
\author{
Mikko Tuomi\thanks{\email{mikko.tuomi@utu.fi; m.tuomi@herts.ac.uk}}\inst{1,2}
\and Guillem Anglada-Escud\'e\thanks{\email{guillem.anglada@gmail.com}}\inst{3}
\and Enrico Gerlach\inst{4}
\and Hugh R. A. Jones\inst{1}
\and Ansgar Reiners \inst{3}
\and \\ Eugenio J. Rivera\inst{5}
\and Steven S. Vogt\inst{5}
\and R. Paul Butler\inst{6}
}

\institute{
University of Hertfordshire, Centre for Astrophysics Research, Science and Technology Research Institute, College Lane, AL10 9AB, Hatfield, UK 
\and University of Turku, Tuorla Observatory, Department of Physics and Astronomy, V\"ais\"al\"antie 20, FI-21500, Piikki\"o, Finland 
\and Universit\"{a}t G\"{o}ttingen, Institut f\"ur Astrophysik, Friedrich-Hund-Platz 1, 37077 G\"{o}ttingen, Germany
\and Lohrmann Observatory, Technical University Dresden, D-01062 Dresden, Germany
\and UCO/Lick Observatory, Department of Astronomy and Astrophysics, University of California at Santa Cruz, Santa Cruz, CA 95064, USA 
\and Department of Terrestrial Magnetism, Carnegie Institute of Washington, Washington, DC 20015, USA
}

\date{Received XX.XX.2012 / Accepted XX.XX.XXXX}


\abstract{The K2.5 dwarf HD 40307 has been reported to host three super-Earths. The system lacks massive planets and is therefore a potential candidate for having additional low-mass planetary companions.}
{We re-derive Doppler measurements from public HARPS spectra of HD 40307 to confirm the significance of the reported signals using independent data analysis methods. We also investigate these measurements for additional low-amplitude signals.}
{We used Bayesian analysis of our radial velocities to estimate the probability densities of different model parameters. We also estimated the relative probabilities of models with differing numbers of Keplerian signals and verified their significance using periodogram analyses. We investigated the relation of the detected signals with the chromospheric emission of the star. As previously reported for other objects, we found that radial velocity signals correlated with the S-index are strongly wavelength dependent.}
{We identify two additional clear signals with periods of 34 and 51 days, both corresponding to planet candidates with minimum masses a few times that of the Earth. An additional sixth candidate is initially found at a period of 320 days. However, this signal correlates strongly with the chromospheric emission from the star and is also strongly wavelength dependent. When analysing the red half of the spectra only, the five putative planetary signals are recovered together with a very significant periodicity at about 200 days. This signal has a similar amplitude as the other new signals reported in the current work and corresponds to a planet candidate with $M \sin i \sim$ 7 M$_{\oplus}$ (HD 40307 g).}
{We show that Doppler measurements can be filtered for activity-induced signals if enough photons and a sufficient wavelength interval are available. If the signal corresponding to HD 40307 g is a genuine Doppler signal of planetary origin, this candidate planet might be capable of supporting liquid water on its surface according to the current definition of the liquid water habitable zone around a star and is not likely to suffer from tidal locking. Also, at an angular separation of $\sim 46$ mas, HD 40307 g would be a primary target for a future space-based direct-imaging mission.}




\maketitle


\section{Introduction}

Current high-precision spectrographs, such as the High Accuracy Radial Velocity Planet Searcher \citep[HARPS;][]{mayor2003} and the High Resolution Echelle Spectrograph \citep[HIRES;][]{vogt1994}, enable detections of low-mass planets orbiting nearby stars. During recent years, radial velocity (RV) planet searches have revealed several systems of super-Earths and/or Neptune-mass planets around nearby stars \citep[e.g.][]{mayor2009a,mayor2009b,lovis2011,pepe2011,tuomi2012}.

The system of three super-Earths orbiting HD 40307 has received much attention because the planets appear in dynamically packed orbits close to mean motion resonances \citep{mayor2009a}. This has been used as an argument to suggest that low-mass planets may be found in highly compact multiple systems that are still stable in long-term, e.g. a possibility of having ten planets with masses of 17 M$_{\oplus}$ within a distance of 0.26 AU on stable orbits \citep{funk2010}. However, the physical nature of these companions as scaled-up versions of the Earths is not entirely clear \citep{barnes2009}. Their masses, between those of Earth and Neptune, suggest they are Neptune-like proto-gas giants that could not accumulate enough gas before it was blown away by the newly born star. On the other hand, recent transit observations of hot super-Earths around bright nearby stars \citep{leger2009,batalha2011,winn2011} indicate that a good fraction of these hot super-Earth mass objects can have rocky compositions.

In this article we re-analyse the 345 HARPS spectra publicly available through the ESO archive using a newly developed software tool called HARPS-TERRA \citep[template-enhanced radial velocity re-analysis application; ][]{anglada2012a}. Instead of the classic cross-correlation function method (CCF) implemented by the standard HARPS-ESO data reduction software (HARPS-DRS), we derive Doppler measurements by least-squares matching of each observed spectrum to a high signal-to-noise ratio template built from the same observations. A description of the method and the implementation details are given in \citet{anglada2012a}. In addition to an increase in precision (especially for K and M dwarfs), this method allows us to perform additional analyses and tests beyond those enabled by the CCF data products provided by the HARPS-DRS. As an example, it allows us to re-obtain the RV measurements using only a restricted wavelength range. As we show in Section \ref{sec:activity}, this capability can be instrumental in ruling out the planetary nature of prominent signals correlated with stellar activity.

We rely on the Bayesian framework when estimating the orbital parameters supported by the data, determining the significances of the signals, and the modelling of the noise in the measurements. In previous studies, radial velocities received within an interval of an hour or so have been commonly binned together in an attempt to reduce the noise caused by stellar- surface-related effects, i.e. stellar oscillations and granulation, and other factors within this timescale \citep{dumusque2010}. In principle, this would enable the detections of planets smaller than roughly 5 M$_{\oplus}$ with HARPS over a variety of orbital distances, even at or near the stellar habitable zone \citep{dumusque2010,dumusque2011}. In our approach, and instead of binning, we apply a self-consistent scheme to account for and quantify correlated noise in the Bayesian framework and use Bayesian model probabilities to show that a solution containing up to six planets is clearly favoured by the data, especially when the redmost part of the stellar spectrum is used in the RV analysis. Only the confluence of refinements in these data analysis methods (re-analysis of the spectra and Bayesian inference) allows the detection and verification of these low-amplitude signals.

We start with a brief description of the stellar properties of HD 40307 (Section 2) and describe the statistical modelling of the observations and the data analysis techniques we used (Section 3). In Section 4 we describe the properties of the RV measurements and perform a detailed Bayesian analysis that identifies up to three new candidate signals. We discuss the stellar activity indicators and their possible correlations with the RV signals in Section 5. In this same section, we find that one of the candidates is spuriously induced by stellar activity by showing that the corresponding periodic signal (P $\sim$ 320 days) is strongly wavelength dependent. When the RVs obtained on the redmost part of the spectrum are analysed (Section 6), the 320-day signal is replaced by a signal of a super-Earth-mass candidate with a 200-day period with a minimum mass of about $\sim 7 M_\oplus$ orbiting within the liquid water habitable zone of HD 40307. The analysis of the dynamical stability of the system (Section 7) shows that stable solutions compatible with the data are feasible and the potential habitability of the candidate at 200 days (HD 40307 g) is discussed in Section 8. We give some concluding remarks and discuss the prospects of future work in Section 9.

\section{Stellar properties of HD 40307}

We list the basic stellar properties of HD 40307 in Table \ref{properties}. This K2.5 V star is a nearby dwarf with a Hipparcos parallax of 77.95 $\pm$ 0.53 mas, which implies a distance of 12.83 $\pm$ 0.09 pc. It is somewhat smaller \citep[M$_{\rm \star} = 0.77 \pm 0.05$ M$_{\oplus}$; ][]{sousa2008} and less luminous \citep[$\log$ L$_{\rm star}/$L$_{\odot} = -0.639\pm0.060$; ][]{ghezzi2010} than the Sun. The star is quiescent \citep[$\log R'_{\rm HK} < -4.99$; ][]{mayor2009a} and relatively metal-poor with $[$Fe/H$] =$ -0.31$\pm$0.03 \citep{sousa2008}. It also lacks massive planetary companions, which makes it an ideal target for high-precision RV surveys aiming at finding low-mass planets. According to the calibration of \citet{barnes2007}, HD 40307 likely has an age similar to that of the Sun ($\sim$ 4.5 Gyr).

\begin{table}
\caption{Stellar properties of HD 40307.}\label{properties}
\begin{center}
\begin{tabular}{lrl}
\hline \hline
Parameter &  Estimate & Reference \\
\hline
Spectral Type & K2.5 V & \citet{gray2006} \\
$\log R'_{\rm HK}$ & -4.99 & \citet{mayor2009a} \\
$\pi$ [mas] & 77.95$\pm$0.53 & \citet{vanleeuwen2007} \\
log L$_{\rm star}/$L$_{\odot}$  & -0.639$\pm$0.060 & \citet{ghezzi2010} \\
log $g$ & 4.47$\pm$0.16 & \citet{sousa2008} \\
M$_{\rm star}$ [M$_\odot$] & 0.77$\pm$0.05 & \citet{sousa2008} \\
T$_{\rm eff}$ [K] & 4956$\pm$50 & \citet{ghezzi2010} \\
$[$Fe/H$]$ & -0.31$\pm$0.03 & \citet{sousa2008} \\
$v$ sin $i$ [kms$^{-1}$] & $<$1 & \citet{mayor2009a} \\
$P_{\rm rot}$ [days] & $\sim$ 48 & \citet{mayor2009a} \\
Age [Gyr] & $\sim$ 4.5 & \citet{barnes2007} \\
\hline \hline
\end{tabular}
\end{center}
\end{table}

\section{Statistical analyses}

\subsection{Statistical models}

We modelled the HARPS RVs using a statistical model with a moving average (MA) term and two additional Gaussian white noise components consisting of two independent random variables.

The choice of an MA approach instead of binning is based on the results of \citet{tuomi2012c} and accounts for the fact that uncertainties of subsequent measurements likely correlate with one another at time-scales of an hour in an unknown manner. We limit our analysis to MA models of third order (MA(3) models) because higher order choices did not improve the noise model significantly. Effectively, the MA(3) component in our noise model corresponds to binning. However, unlike when binning measurements and artificially decreasing the size of the data set, this approach better preserves information on possible signals in the data. The two Gaussian components of the noise model are the estimated instrument noise with zero-mean and known variance (nominal uncertainties in the RVs) and another with zero-mean but unknown variance corresponding to all excess noise in the data. The latter contains the white noise component of the stellar surface, usually referred to as stellar ``jitter'', and any additional instrumental systematic effects not accounted for in the nominal uncertainties. Keplerian signals and white noise component were modelled as in \citet{tuomi2011b}.

In mathematical terms, an MA($p$) model is implemented on measurement $m_{i}$ as
\begin{equation}\label{moving_average}
  m_{i} = r_{k}(t_{i}) + \gamma + \epsilon_{i} + \sum_{j=1}^{p} \phi_{j} \big[ m_{i-j} - r_{k}(t_{i-j}) - \gamma \big] \exp \big( t_{i-j} - t_{i} \big),
\end{equation}
where $r_{k}(t_{i})$ is the superposition of $k$ Keplerian signals at epoch $t_{i}$ and $\gamma$ is the reference velocity. The random variable $\epsilon_{i}$ is the Gaussian white noise component of the noise model with zero-mean and variance $\sigma^{2} = \sigma_{i}^{2} + \sigma_{J}^{2}$, where $\sigma_{i}^{2}$ is the (fixed) nominal uncertainty of the $i$th measurements and $\sigma_{J}^{2}$ is a free parameter describing the magnitude of the jitter component. Finally, the free parameters of the MA($p$) model are denoted as $\phi_{j}, j = 1, ..., p$ -- they describe the amount of correlation between the noise of the $i$th and $i-j$th measurements. The exponential term in Equation (\ref{moving_average}) ensures that correlations in the noise are modelled on the correct time-scale. Specifically, using hours as units of time, the exponential term (that is always $< 1$ because $t_{i} > t_{i-j}$) vanishes in few hours.

To demonstrate the impact of binning on this data set, Section \ref{sec:nightbins} shows the analyses of binned RVs with the common assumption that the excess noise is purely Gaussian. This corresponds to using the nightly average as the individual RV measurements and setting $\phi_{j} = 0$ for all $j$ in Eq. (\ref{moving_average}).

\subsection{Bayesian analyses and detection thresholds}\label{sec:bayes}

To estimate the model parameters and, especially, their uncertainties as reliably as possible, we drew random samples from the parameter posterior densities using posterior sampling algorithms. We used the adaptive Metropolis algorithm of \citet{haario2001} because it can be used to receive robust samples from the posterior density of the parameter vector when applied to models with multiple Keplerian signals \citep[e.g.][]{tuomi2011b,tuomi2012}. This algorithm is simply a modified version of the famous Metropolis-Hastings Markov chain Monte Carlo (MCMC) algorithm \citep{metropolis1953,hastings1970}, which adapts the proposal density to the information gathered from the posterior density. We performed samplings of models with $k = 0, 1, ... 7$ Keplerian signals.

The samples from the posterior densities were then used to perform comparisons of the different models. We used the one-block Metropolis-Hastings (OBMH) method \citep{chib2001,clyde2007} to calculate the relative posterior probabilities of models with differing numbers of Keplerian signals \citep[e.g.][]{tuomi2009,tuomi2011,tuomi2011b,tuomi2012}. We performed several samplings using different initial values of the parameter vector and calculated the means and the corresponding deviations as measures of uncertainties of our Bayesian evidence numbers $P(m | \mathcal{M}_{k})$, where $m$ is the measurement vector and $\mathcal{M}_{k}$ denotes the model with $k$ Keplerian signals.

The prior probability densities in our analyses were essentially uniform densities. As in \citet{tuomi2012}, we adopted the priors of the RV amplitude $\pi(K_{i}) = U(0, a_{RV})$, reference velocity $\pi(\gamma) = U(-a_{RV}, a_{RV})$, and jitter $\pi(\sigma_{J}) = U(0, a_{RV})$, where $U(a,b)$ denotes a uniform density in the interval $[a,b]$. Since the observed peak-to-peak difference in the raw RVs is lower than 10 m s$^{-1}$, the hyperparameter $a_{RV}$ was conservatively selected to have a value of 20 ms$^{-1}$. The priors of the longitude of pericentre ($\omega$) and the mean anomaly ($M_{0}$) were set to $U(0, 2\pi)$, in accordance with the choice of \citet{ford2007} and \citet{tuomi2012}. We used the logarithm of the orbital period as a parameter of our model because, unlike the period as such, it is a scale-invariant parameter. The prior of this parameter was set uniform such that the two cut-off periodicities were $T_{min}$ and $T_{max}$. These hyperparameters were selected as $T_{min} = 1.0$ days and $T_{max} = 10 T_{obs}$ because we did not expect to find signals with periods less than 1 day. Also, we did not limit the period space to the length of the baseline of the HARPS time series ($T_{obs}$), because signals in excess of that can be detected in RV data \citep{tuomi2009a} and because there might be long-period signals apparent as a trend with or without curvature in the data set.

Unlike in traditional Bayesian analyses of RV data, we did not use uniform prior densities for the orbital eccentricities. Instead, we used a semi-Gaussian as $\pi(e_{i}) \propto \mathcal{N}(0, \sigma_{e}^{2})$ with the corresponding normalisation, where the hyperparameter $\sigma_{e}$ was chosen to have a value of 0.3. This value decreases the posterior probabilities of very high eccentricities in practice, but still enables them if they explain the data better than lower ones \citep{tuomi2012,tuomi2012b}.

For the MA components $\phi_{j}$, we selected uniform priors as $\pi(\phi_{j}) = U(-1, 1)$, for all $j=1, 2, 3$. This choice was made to ensure that the MA model was stationary, i.e. time-shift invariant -- a condition that is satisfied exactly when the values are in the interval [-1, 1].

Finally, we did not use equal prior probabilities for the models with differing numbers of Keplerian signals. Instead, following \citet{tuomi2012}, we set them as $P(\mathcal{M}_{k}) = 2P(\mathcal{M}_{k+1})$, which means that the model with $k$ Keplerian signals was always twice as probable prior to the analyses than the model with $k+1$ Keplerian signals. While this choice makes our results more robust in the sense that a posterior probability that exceeds our detection threshold is actually already underestimated with respect to equal prior probabilities, there is a physical motivation as well. We expect that the dynamical interactions of planets in any given system make the existence of an additional planet less probable because there are fewer dynamically stable orbits. This also justifies the qualitative form of our prior probabilities for the eccentricities.

Our criterion for a positive detection of $k$ Keplerian signals is as follows. First, we require that the posterior probability of a $k$-Keplerian model is at least 150 times greater than that of the $k-1$-Keplerian model \citep{kass1995,tuomi2011,tuomi2012,tuomi2011b,feroz2011}. Second, we require that the radial velocity amplitudes of all signals are statistically significantly different from zero. Third, we also require that the periods of all signals are well-constrained from above and below. These criteria were also applied in \citet{tuomi2012}.

We describe the parameter posterior densities using three numbers, namely, the maximum \emph{a posteriori} (MAP) estimate and the limits of the corresponding 99\% Bayesian credibility sets (BCSs) or intervals in one dimension \citep[e.g.][]{tuomi2009}.

\subsection{Periodogram analysis}

As is traditionally the case when searching for periodic signals in time series, we used least-squares periodograms \citep{lomb1976,scargle1982} to probe the next most significant periods left in the data. In particular, we used the least-squares periodograms described in \citet{cumming2004}, which adjust for a sine wave and an offset at each test period and plot each test period against the F-ratio statistic (or power) of the fit. While strong powers likely indicate the existence of a periodic signal (though strong powers may be caused by sampling-related features in the data as well), the lack of them does not necessarily mean that there are no significant periodicities left \citep[e.g.][]{tuomi2012}. This is especially so in multi-Keplerian fits due to strong correlations and aliases between clearly detected signals and yet-undetected lower amplitude companions \citep[e.g.][]{anglada2010}. The reason is that residuals must necessarily be calculated with respect to a model that is assumed to be correct, which is clearly not the case when adding additional degrees of freedom, i.e. additional planetary signals, to the model. Therefore, determining the reliability of a new detection based on goodness-of-fit comparisons is prone to biases, which effectively reduces the sensitivity and reliability of these detections \citep{tuomi2012}.

While periodograms of the residuals are very useful, they do not properly quantify the significance of the possibly remaining periodicities and, therefore, we used them as a secondary rather than a primary tool to assess the significance of new signals. The analytic false-alarm probability (FAP) thresholds as derived by \citet{cumming2004} are provided in the figures as a reference and for illustrative purposes only. We used the same periodogram tools to assess the presence of periodicities in the time series of a few activity indicators.

\section{Analysis of the RV data}

The 345 measurements taken on 135 separate nights were obtained over a baseline of $\sim$ 1900 days. In contrast to the discussion in \citet{mayor2009a}, we could not confirm a long-period trend using our new RVs and we did not detect evidence for a trend in the new CCF RVs obtained using the HARPS-DRS. As shown in \citet{anglada2012a} (Fig. 3), changes in the continuum flux accross each echelle order (also called blaze function) induce RV shifts of several ms$^{-1}$ if not properly accounted for. This effect was reported to affect HARPS measurements in \citet{pepe2010} and appears to have been fixed by HARPS-DRS v3.5 based on the release notes\footnote{www.eso.org/sci/facilities/lasilla/instruments/harps/tools/drs.html [www.eso.org]} (issued on 29 October, 2010). With respect to HD 40307 in particular, we found that when RVs were derived without blaze function correction, a strong positive drift ($\sim$ 2 ma$^{-1}$yr$^{-1}$) was left in the Doppler time series. We speculate that the trend reported earlier may be caused by this blaze function variability. A notable feature of the HD 40307 data set is that the epochs have a long gap of 638 days between 3055 - 3693 [JD-2450000]. This feature can effectively decrease the phase-coverage of the data on longer periods and complicates the interpretation of periodograms due to severe aliases.

\subsection{Analysis of binned data}\label{sec:nightbins}

In a first quicklook analysis, we worked with the nightly averages of the radial velocities as obtained from HARPS-TERRA using the standard setup for K dwarfs. In this setup, all HARPS echelle apertures are used and a cubic polynomial is fitted to correct for the blaze function of each aperture. After this correction, the weighted means $\hat{v}_e$ of all RV measurements within each night $e$ are calculated. As a result, we obtain internal uncertainties of the order of 0.3-0.4 ms$^{-1}$ for the HARPS-TERRA RVs. Because of stellar and/or instrumental systematic errors, we observed that these individual uncertainties are not representative of the real scatter within most nights with five or more measurements. Using three of those nights we estimate that at least 0.6 m$s^{-1}$ must be added in quadrature to each individual uncertainty estimate. After this, the uncertainty of a given epoch is obtained as $\sigma_{e}^{-1}= \sum^{N_e}_i (\sigma^e_{i})^{-1}$, where the sum is calculated over all exposures obtained during a given night. Finally, based on their long-term monitoring of inactive stars, \citet{pepe2011} inferred a noise level of 0.7 ms$^{-1}$ to account for instrumental and stellar noise. After some tests, we found that adding 0.5 ms$^{-1}$ in quadrature to the uncertainties of the nightly averages ensured that none of the epochs had uncertainties below the 0.7 ms$^{-1}$ level. The typical uncertainties of a single night derived this way were of the order of 0.8 ms$^{-1}$. These corrections are basically only well-educated guesses based on the prior experience with RV data and reported stability of the instrument. Therefore, one must be especially careful not to over-interpret the results derived from them (e.g., powers in periodograms and significance of the signals). In the fully Bayesian approach, we treat the excess noise as a free parameter of the model, therefore the Bayesian estimates of the noise properties should in principle also be more reliable.

\begin{figure}
\center
\includegraphics[width=0.45\textwidth,clip]{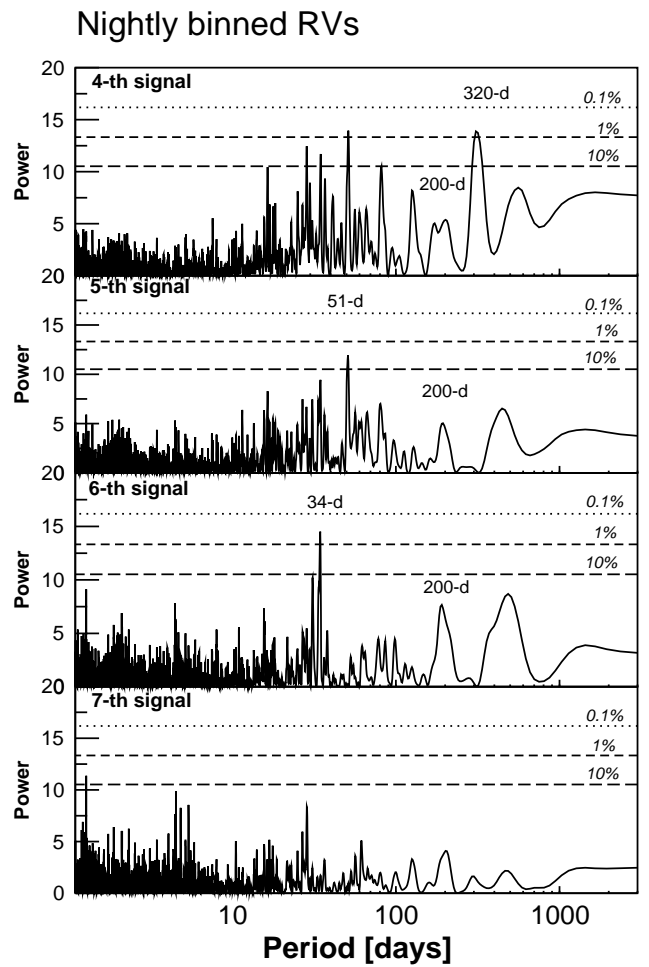}
\caption{Least-squares periodograms of the binned HD 40307 radial velocities for the residuals of the models with three (top) to six (bottom) periodic signals. The analytic 10\%, 1\%, and 0.1\% FAPs are shown as horizontal lines.}\label{periodograms_bin}
\end{figure}

First, we re-analysed the nighly binned RVs to see whether we could independently reproduce the results of \citet{mayor2009a} when HARPS-TERRA measurements and our Bayesian methods were used. Assuming an unknown Gaussian noise parameter \citep[e.g.][]{tuomi2011b} in addition to the estimated measurement uncertainties, the posterior samplings and the corresponding model probabilities easily revealed the three strong signals corresponding to periods of 4.3, 9.6, and 20.4 days. The residual periodogram of the three-Keplerian model revealed additional strong periodicities exceeding the 1\% FAP level (Fig. \ref{periodograms_bin}, top panel) and we tested more complicated models with up to six Keplerian signals. Especially, we tested whether the additional power present in the three-Keplerian model residuals (Fig. \ref{periodograms_bin}, top panel) at periods of 28.6, 34.8, 51.3, and 308 days, peaking above the 10\% FAP level, are statistically significant by starting our MCMC samplings at nearby seed periods.

The global four-Keplerian solution was found to correspond to the three previously known super-Earth signals and an additional signal with an MAP period of 320 days. This period was bounded from above and below and its amplitude was strictly positive -- in accordance with our detection criteria. The corresponding posterior probability of the four-Keplerian model was 1.9$\times 10^{5}$ times greater than that of a three-Keplerian one, making the 320-day signal significant. In addition to this signal, we could identify a 51-day periodicity (Fig. \ref{periodograms_bin}, second panel) that satisfied the detection criteria as well. Including this fifth signal in the model further increased the model probabilitity by a factor of 6.6$\times 10^{6}$. We could furthermore identify a sixth signal with our six-Keplerian model, correponding to a period of 34.4 days. However, even though the samplings converged well and the solution looked well-constrained, the six-Keplerian model was only five times more probable than the five-Keplerian one and would not be detected using our criteria in Section \ref{sec:bayes}. Both of the two new significant signals had MAP estimates of their radial velocity amplitudes slightly lower than 1.0 ms$^{-1}$ -- the signals at 51 and 320 days had amplitudes of 0.70 [0.31, 1.09] ms$^{-1}$ and 0.75 [0.38, 1.12] ms$^{-1}$, respectively, where the uncertainties are denoted using the intervals corresponding to the 99\% BCSs. We note that the periodogram of sampling does not have strong powers at the periods we detect \citep[see Fig. 1 in][]{mayor2009a}.

Given the uncertain nature of the signal at 34.4 days and the potential loss of information when using the nightly averaged RVs (artificial reduction of the number of measurements), we performed a complete Bayesian reanalysis of the full dataset (345 RVs), now including the aforementioned moving average approach to model the velocities.

\subsection{Analysis using all RV measurements}\label{sec:bayesian_full}

The analyses of the unbinned data immediately showed the three previously announced signals \citep{mayor2009a} with periods of 4.3, 9.6, and 20.4 days. Modelling the data with the superposition of $k$ Keplerian signals and an MA(3) noise model plus the two Gaussian white noise components, our posterior samplings and periodogram analyses identified these signals very rapidly, enabling us to draw statistically representative samples from the corresponding parameter densities.

\begin{figure}
\center
\includegraphics[width=0.45\textwidth,clip]{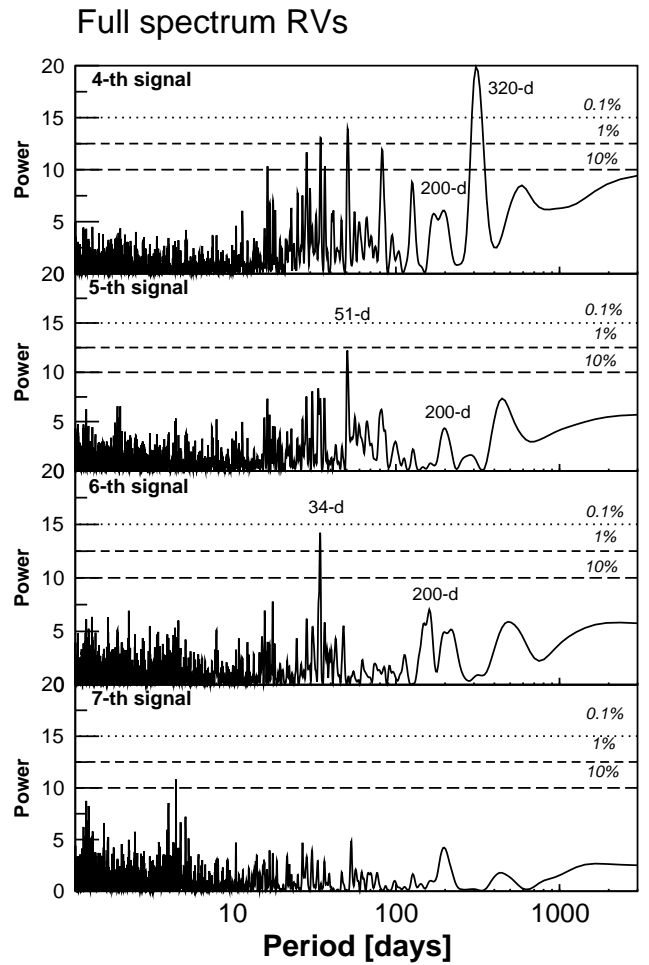}
\caption{Least-squares periodograms of all 345 RVs of HD 40307 for the residuals of the models with three (top) to six (bottom) periodic signals together with the analytic 10\%, 1\%, and 0.1\% FAPs.}\label{periodograms_b}
\end{figure}

The residual periodogram of this model (three Keplerians and MA(3) components of the noise removed) revealed some significant powers exceeding the 0.1\% and 1\% FAP level at 320 and 50.8 days, respectively (Fig. \ref{periodograms_b}, top panel). Samplings of the parameter space of a four-Keplerian model indicated that the global solution contained the 320-day periodicity as the fourth signal and yielded a posterior probability for the four-Keplerian model roughly $1.5 \times 10^{6}$ times higher than for the three-Keplerian one. The nature of this signal and its relation to the stellar activity (Section \ref{sec:activity}) is discussed in Section \ref{sec:chromatic}.

We continued by calculating the periodogram of the residuals of our four-Keplerian model (Fig. \ref{periodograms_b}, second panel) and observed a periodogram power that almost reached the 1\% FAP level at a period of 50.8 days. Including this fifth signal further increased the posterior probability of our model by a factor of $6.4 \times 10^{5}$.

The residuals of the five-Keplerian model contained a periodicity at 34.7 days (Fig. \ref{periodograms_b}, third panel) exceeding the 1\% FAP level. The corresponding six-Keplerian model with this candidate received the highest posterior probability -- roughly $5.0 \times 10^{8}$ times higher than that of the five-Keplerian model. Since the parameters of this sixth candidate were also well-constrained, we conclude that including the 34.7-day signal in the statistical model is fully justified by the data.

We also attempted to sample the parameter space of a seven-Keplerian model but failed to find a clear probability maximum for a seventh signal (see also the residual periodogram of the six-Keplerian model in Fig. \ref{periodograms_b}, bottom panel). Although the periodicity of the seventh signal did not converge to a well-constrained probability maximum, all periodicities in the six-Keplerian model at 4.3, 9.6, 20.4, 34.7, 50.8, and 320 days were still well-constrained, i.e. their radial velocity amplitudes were statistically distinguishable from zero and their periods had clear probability maxima. Because we were unable to constrain the parameters of a seventh signal using the posterior samplings, we cannot be sure whether the corresponding Markov chains had converged to the posterior density and cannot reliably calculate an estimate for the posterior probability of the seven-Keplerian model. Therefore we stopped looking for additional signals.

From this analysis, we can state confidently that there are six significant periodicities in the HARPS-TERRA radial velocities of HD 40307 when the whole spectral range of HARPS is used. As we show in the next section, one of them has the same period as the chromospheric activity indicator (S-index) and requires more detailed investigation. The analysis of all 345 RVs indicates that for these data binning appears to be a retrograde step in extracting periodic signals from the RV data. We infer that binning serves to alter measurement uncertainties and damp the significance levels of the periodicities in the data.

\section{Stellar activity}\label{sec:activity} 

We examined the time series of two activity indicators derived from the cross correlation function properties as provided by the HARPS-DRS. They are the bisector span (BIS) and the full-width at half-maximum (FWHM) of the CCF. These indices monitor different features of the average stellar line. Briefly, BIS is a measure of the stellar line asymmetry and should correlate with the RVs if the observed offsets are caused by spots or plages rotating with the star \citep{queloz2001}. The FWHM is a measure of the mean spectral line width. Its variability (when not instrumental) is usually associated with changes in the convective patterns on the surface of the star. A third index, the so-called S-index in the Mount Wilson system \citep{baliunas1995}, is automatically measured by HARPS-TERRA on the blaze-corrected one-dimensional spectra provided by the HARPS-DRS. The S-index is a measure of the flux of the CaII H and K lines ($\lambda_H = 3933.664$ \AA ~ and $\lambda_K = 3968.470$ \AA, respectively) relative to a locally defined continuum \citep{lovis2011} and is an indirect measurement of the total chromospheric activity of the star. For simplicity, the analysis of the activity indicators was performed throughout for the 135 nightly averaged values using sequential least-squares fitting of periodic signals that are each described by a sine-wave model (period, amplitude and phase).

\subsection{Analysis of the FWHM and BIS}

\begin{figure}[tb] \centering
\includegraphics[width=0.45\textwidth,clip]{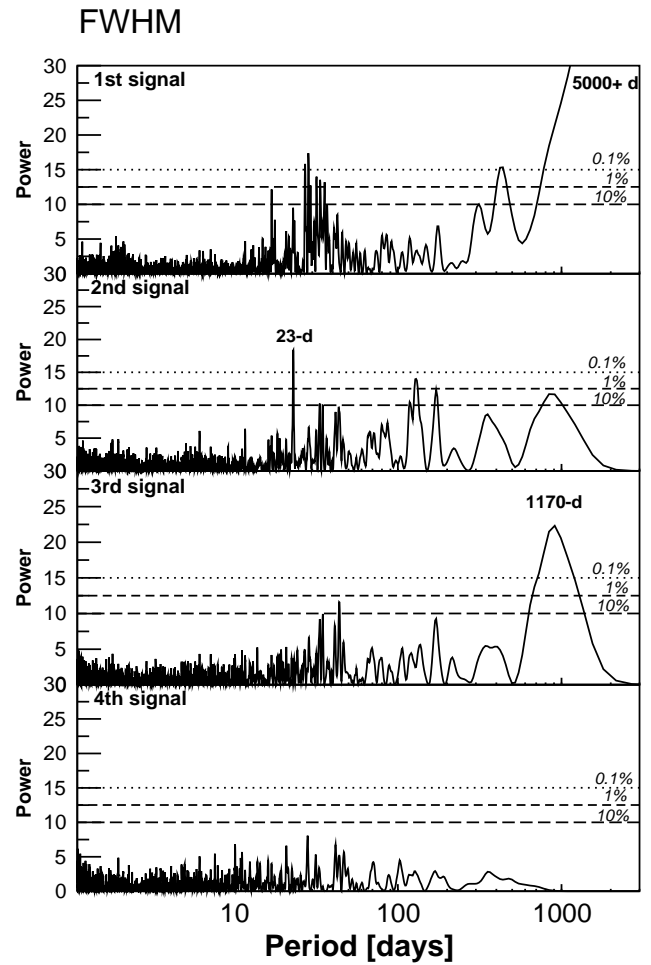} 
\caption{Periodogram series of the signals detected in the FWHM, from most significant to less significant (top to bottom).}\label{fig:fwhm}
\end{figure}

\begin{figure}[tb] \centering
\includegraphics[width=0.45\textwidth,clip]{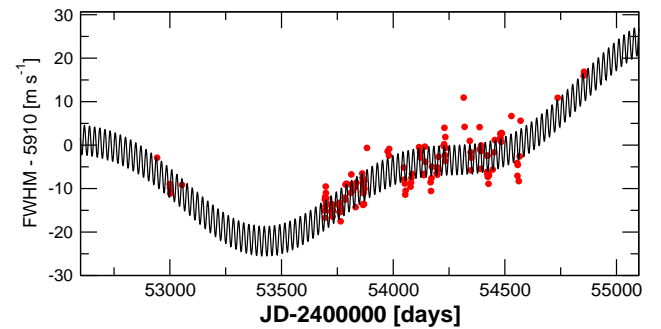}
\caption{Time series of the FWHM activity index. The solid black line represents the best fit to a model containing three sinusoids (periods of 5000, 23 and 1170 days).}\label{fig:fwhm_curve}
\end{figure}

The BIS was remarkably stable (RMS $\sim 0.5$ ms$^{-1}$) and the periodogram of its time series did not show any significant powers. Visual inspection of the time series for the FWHM already shows a very significant trend of 5.3 ms$^{-1}$ yr$^{-1}$. The 345 measurements of the FWHM are listed in Table \ref{tab:rvs}. A sinusoidal fit to this trend suggested a period of 5000 days or more (see top panel in Fig. \ref{fig:fwhm}). After removing the trend, two more signals strongly show up in the residuals. The first one was found at 23 days and had an analytic FAP of 0.005\%. After fitting a sinusoid to this signal and calculating the residuals, an extremely significant peak appeared at 1170 days with an analytic FAP of 0.002\%. After including this in a model with three sinusoids, no additional signals could be seen in the periodogram of the residuals with analytic FAP estimates lower than 10\% (Fig. \ref{fig:fwhm}, bottom panel). We also show the FWHM values together with the fitted periodic curves in Fig. \ref{fig:fwhm_curve}. While the signals in the FWHM were significant, we did not clearly detect their counterparts in the RVs. Given that BIS does not show any obvious signals either, we suspect that the periodicities in the FWHM might be caused by instrumental effects, e.g. tiny changes of the focus inside the spectrograph, rather than intrinsic variability of the stellar lines. The instrumental origin would reconcile the absence of correspondent drifts in the BIS and in the RVs. A similar indication of drifts and sensitivity of the FWHM to instrumental issues has been reported in e.g. \citet{lovis2011}. While this adds some caveats on the long-term stability of the HARPS instrumental profile (and therefore its long-term precision), unless it is found to have similar periods, we see no reason to suspect that any of the signals in the RVs are spuriously induced by changes in the FWHM (intrinsic or instrumental).

\subsection{Analysis of the S-index}

\begin{figure}[tb] \centering
\includegraphics[width=0.45\textwidth,clip]{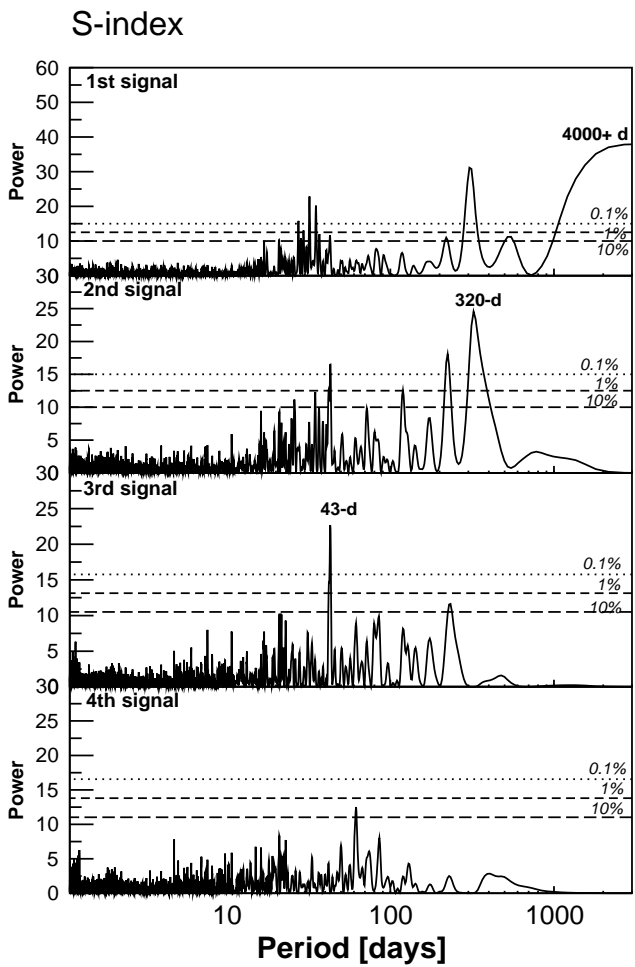}
\caption{Periodogram series of the signals detected in the S-index, from most significant to less significant (top to bottom).}\label{fig:sper}
\end{figure}

\begin{figure}[tb] \centering
\includegraphics[width=0.45\textwidth,clip]{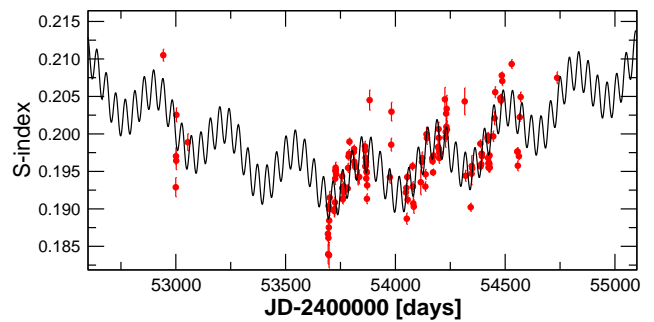}
\caption{Time-series of the S-index. The solid black represents the best fit to a model containing three sinusoids (periods of 4000, 320, and 43 days). Long-period trend and two signals at 320 and 43 days are clearly detected in the time-series of the S-index.}
\label{fig:sindex}
\end{figure}

The S-index also shows strong coherent variablity. The full set of 345 measurements of the S-index are provided in Table \ref{tab:rvs}. Again, this analysis was performed for the nightly binned measurements using least-squares periodograms and the sequential inclusion of sinusoidal signals. The last two S-index measurements were well above the average and could not be reproduced by any smooth function. Even if they are representative of a physical process, such outlying points cannot be easily modelled by a series of a few sinusoids because these would add many ambiguities to the interpretation of the results. When these two points were removed, the periodograms looked much cleaner and we could unambiguously identify clear periodicities. Therefore, the analysis discussed here is based on the first 133 nightly averages only. As illustrated in the periodograms of Fig. \ref{fig:sper}, this analysis showed strong evidence for three signals in the S-index. From most significant to less significant, we found these to consist of a long-period quadratic trend (with a period exceeding the data baseline), a signal with P$\sim$ 320 days and a third signal at 43 days. Fig. \ref{fig:sindex} shows the best fit to a parabolic trend together with the two periodicities at 320 and 43 days. As has been reported for several other stars \citep{dumusque2011}, we find it likely that the long-period trend is the signature of the stellar cycle that is not yet fully covered by the baseline of the observations. This trend might also be related to the trend observed in the FWHM, perhaps through an increase of the average magnetic field.

Using more epochs, \citet{lovis2011} reached the conclusion that the stellar magnetic cycle of HD 40307 is about 5000 days -- a result that is compatible with ours. However, the nature of the 320-day signal is less clear. This period seems to be too long to be caused by rotation because it is more than a factor of three longer than the longest currently known rotation periods which are about 100 days \citep{irwin2011}. Long periods like this would be difficult to explain by angular momentum evolution \citep{barnes2012}. Indeed, the third signal in the S-index (43 days) appears to be a much better match to the expected stellar rotation of HD 40307 \citep{mayor2009b}. The S-index also shows a few other tentative periodicities at 26, 63, and 167 days (FAP $\sim$ 2-5\%).

HD 40307 is a very quiet star \citep[log R$^\prime_{HK}\sim-4.99$; ][]{mayor2009a} and it is in the effective temperature range where the correlation between magnetic activity and RV jitter shows the smallest correlation \citep[$\sim$ 4300-5000 K, see Fig. 2 in ][]{dumusque2011b}. Therefore, apart from the two most obvious signals in the S-indices (long-period trend and 320-day period), the other tentative periodicities are unlikely to produce detectable effects on the RVs. Indeed, no RV counterpart was identified at 43 days (or any of the potential fourth signals at periods of 26, 63, or 167 days), not even at low significance. Several studies have demonstrated a clear correlation of the S-index with observed RV offsets. For example, \citet{dumusque2011b} performed a comprehensive analysis of simultaneous HARPS RVs and chromospheric emission measurements of G and K dwarfs, showing that the correlation between the chromospheric emission and RV offsets can be clearly traced and quantified as a function of spectral type. \citet{pepe2011} used this correlation to detrend the RV measurements of HD 85512 (K6 V) and reported the detection of a $m \sin i \sim 2$M$_\oplus$ candidate with an orbital period of $\sim 58$ days. More recently, \citet{anglada2012a} showed that long-period signals correlated with Doppler signals can be strongly wavelength dependent even within the limited HARPS wavelength range. Since Keplerian signals cannot be wavelength depedent, this provides a new set of diagnostics to distinguish activity-induced signals from Keplerian signals. Instead of using the detrending technique employed by \citet{pepe2011}, we investigate in the next section if any of the signals are wavelength dependent.

\subsection{Wavelength-dependent signals}\label{sec:chromatic}

\citet{anglada2012a} showed that there is a correlation between the S-indices and RV measurements of HD 69830 (G8 V) and that an apparent long-period trend in the RVs completely disappeared when one obtains the velocities using the redmost half of the HARPS spectra only. Following \citet{anglada2012a}, we call this effect
\emph{chromatic jitter}. Chromatic jitter can cause coherent RV variability (e.g., periodic blueshifts following the magnetic cycle) but can also behave randomly, effectively adding uncertainty and correlated noise to the measurements. At present, we do not sufficiently understand the underlying physics to be able to predict the existence and nature of the wavelength dependence of activity-related signals. Despite this lack of detailed knowledge, the diminished sensitivity of redder wavelengths to activity has been shown to be an efficient method in verifying or falsifying the existence of proposed candidates. For example, CRIRES/VLT RV measurements \citep{huelamo2008,figueira2010} in the H-band ($\sim 1.5 \mu$m) were used to rule out the existence of a gas giant candidate around TW Hya that was initially detected in the optical RVs by \citet{setiawan2008}. Another example was the brown dwarf candidate around LP 944-20. This candidate was first detected at optical wavelengths (2 kms$^{-1}$ semi-amplitude) and then ruled out by the same group using nIR NIRSPEC/Keck observations \citep{martin2006}. The chromatic nature of activity-induced signals has also abundant qualitative support from theory and simulations \citep[e.g.][]{reiners2010,barnes2011}. Below we explore the wavelength dependence of the 320-day signal (and all the others) to investigate its (their) nature in more detail. 

\begin{figure}[tb]
\center
\includegraphics[width=0.40\textwidth,clip]{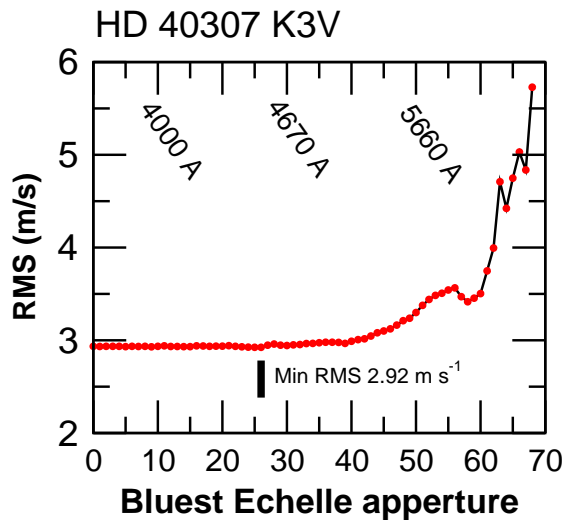}
\caption{RMS as a function of the bluest echelle aperture used. The corresponding $\lambda_{BC}$ are illustrated at the top of the panel. A shallow minimum is reached at aperture 26 ($\lambda_{BC} = 433$ nm), indicating activity induced signals/jitter at the bluest wavelengths.}
\label{fig:minorder} 
\end{figure}

As suggested by \citet{anglada2012a}, the first qualitative way of assessing the presence of chromatic jitter consists of obtaining radial velocity measurements for all epochs using the stellar spectrum from 680 nm (the longest wavelength available at HARPS) to some blue cut-off wavelength $\lambda_{BC}$, and plot the RMS of the resulting RVs as a function of $\lambda_{BC}$. For a perfectly stable star, the RMS should decrease monotonically as more echelle apertures are used to obtain the velocity data. However, \citet{anglada2012a} showed that even for quiet G and K dwarfs without reported planets, a minimum in this RMS was typically reached when $\lambda_{BC}$ was chosen to be between 450-550 nm. For M dwarfs, this cut-off was found to be closer to 600 nm. In Fig. \ref{fig:minorder}, we show the RMS of the radial velocities of HD 40307 as a function of $\lambda_{BC}$ and note that there appears to be a shallow minimum at $\lambda_{BC}= 453$ nm. Although this minimum is not very deep (no Keplerian solution was subtracted from the RVs), its existence already suggests the presence of wavelength-dependent noise that acts stronger in the blue part of the spectrum.

\begin{figure}[tb]
\centering
\includegraphics[width=0.45\textwidth,clip]{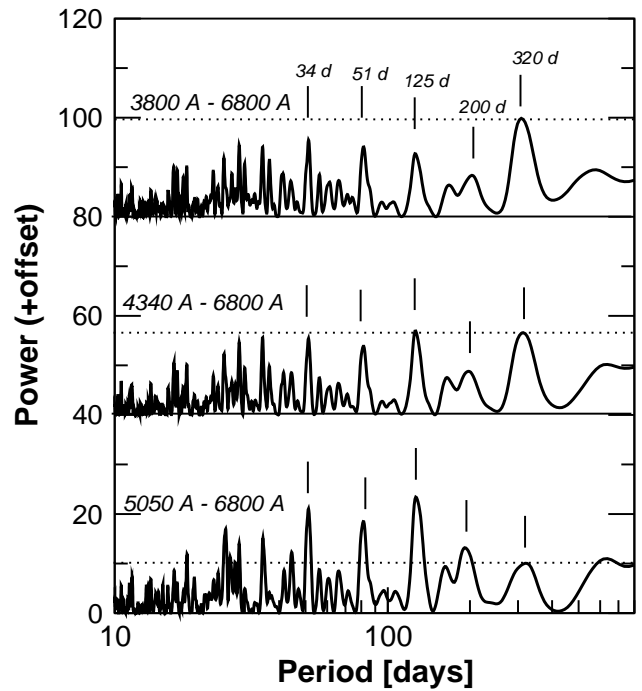}
\caption{Residual periodogram of the three-planet solution as a function of the blue cut-off wavelength $\lambda_{BC}$. The top periodogram is obtained using the full-spectrum RVs. The bottom periodogram is obtained using the aperture 40 (505.0 nm) as $\lambda_{BC}$. All signals discussed in the text are marked with narrow vertical lines. The horizontal line represents the peak value of peak closer to 320 days. When $\lambda_{BC} = 433$ nm (central periodogram), the emerging 125-day peak is already as high as the activity-induced signal. The 120-day peak is an alias of the most likely period of 200 days for planet candidate d.} \label{fig:periodogram_series}
\end{figure}

The next test consists of assessing if a signal observed using the full spectrum is also present when examining a subsection of it. For simplicity, we restrict this test to the nightly binned measurements and only use periodogram tools. In agreement with \ref{sec:bayesian_full}, when the full spectrum was used, the fourth dominant signal was found at a period of $\sim$ 320 days. Any activity-induced signals only present at the bluest wavelengths should lose their significance as one shifts $\lambda_{BC}$ towards the red wavelengths. As illustrated in Fig. \ref{fig:periodogram_series}, this behaviour is clearly found for the 320-day signal. Interestingly, when $\lambda_{BC}$ is set to 430 nm (20th HARPS aperture), the 320-day peak becomes as high as a new one emerging around 120 days. When the $\lambda_{BC}$ is set to 503 nm (40th HARPS aperture), the 120-day signal clearly dominates the periodogram. It is noteworthy that the powers of both the 34- and 51-day peaks remain mostly constant, which shows that their significance is not seriously affected by the choice of the cut-off wavelength.

\begin{figure}[tb]
\centering
\includegraphics[width=0.45\textwidth,clip]{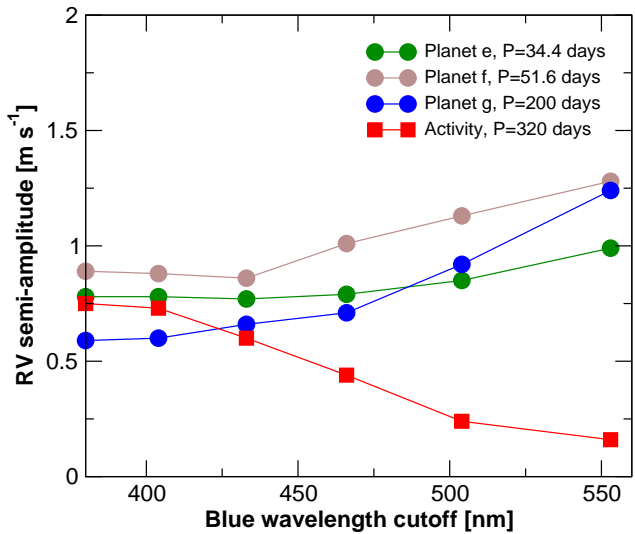}
\caption{Amplitudes of the four signals under discussion as a function of the blue cut-off $\lambda_{BC}$. While all proposed candidates keep a constant (or slightly increase) the derived semi-amplitude, the 320-day signal is almost nonexistent (the period actually becomes unconstrained) when wavelengths redder than 480 nm are used to derived a seven-planet orbital solution.}
\label{fig:K_vs_wave}
\end{figure}

As we show in the next section, the 120-day signal is a yearly alias of the most favoured period at 200 days, which also appears conspicuously in the bottom periodogram of Fig. \ref{fig:periodogram_series}. To further illustrate the wavelength dependence of the 320-day signal, we show the fitted RV amplitudes of the four candidate periodicities (periods of 34, 51, 200, and 320 days) as a function of $\lambda_{BC}$ (Fig. \ref{fig:K_vs_wave}). These amplitudes are obtained by simultaneous least-squares fitting of seven circular orbits to the seven periods of interest (periods are also allowed to adjust). While the planet candidates at 34 and 51 days show roughly constant amplitudes, the 320-day signal almost completely vanishes when $\lambda_{BC}$ is redder than 450 nm (K$<$ 0.4 ms$^{-1}$). It is also worth noting that the 200-day signal increases its amplitude as the 320-day signal disappears. The 200-day signal is seriously affected by yearly aliases which, effectively, correlates its amplitude with the signal at 320 days. The 320-day period is so suppressed at red wavelengths that for any $\lambda_{BC}$ redder than 450 nm, we were forced to keep the period fixed at 320 days to estimate its amplitude and to prevent it from converging to a very different periodicity.

\begin{figure}[tb]
\centering
\includegraphics[width=0.45\textwidth,clip]{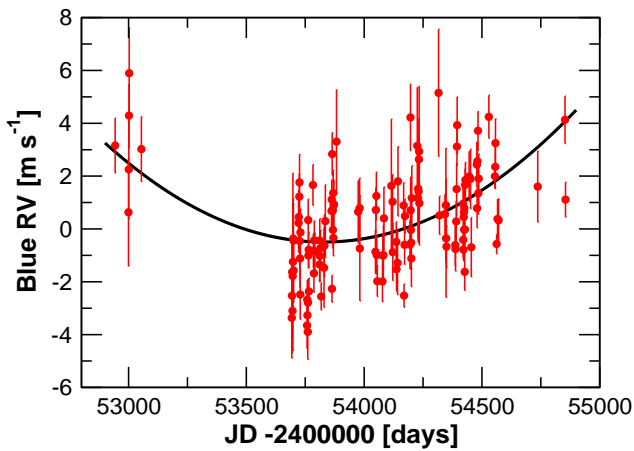}
\caption{Residuals of the three-planet fit when only the bluemost part of the spectra were used to extract the Doppler measurements. A quadratic trend similar to the one detected in the S-index is the most obvious signal left in these residuals.} 
\label{fig:blueRV}
\end{figure}

At this point, it is fair to ask why the long-term variability in the S-index (and in the FWHM) does not show in the RV data as well. We can show that it actually does, very prominently, when calculating the RVs using the bluemost part of the spectrum only. To illustrate this, we obtained the RV measurements using the wavelength range from 380 nm to 440 nm (20 bluest echelle apertures). After subtracting the signals of the three planets reported by \citet{mayor2009a}, a parabolic shape is clearly left in the RV residuals. Fig. \ref{fig:blueRV} illustrates how well the parabolic trend matches the long-period trend also observed in the S-index (compare Figs. \ref{fig:sindex} and \ref{fig:blueRV}). No blue RV counterparts of the other tentative periodicities in the S-index were identified in the data. 

As a conclusion to this section, we identified the chromatic nature of the $\sim$ 320-day signal. We note that the internal errors of the individual spectra, given in the third column of Table \ref{tab:rvs}, have a median value of 0.4-0.5 ms$^{-1}$. This number is significantly smaller than the measured intranight scatter and also smaller than the best RMS we could receive from any six- or seven-planet fit. This means that systematic noise (stellar and/or instrumental) is a dominant source of uncertainty and that accuracy is not lost by ignoring the Doppler information at the bluest wavelengths. A careful optimisation of $\lambda_{BC}$ might provide even more significant detections but, given the unknown nature of this chromatic jitter, choosing $\lambda_{BC}= 490.0 nm$ (37th HARPS aperture) appears to provide a fair compromise. We suspect that the observed behaviour of HD 40307 RVs \citep[also HD 69830, HD 85512, and several other nominally stable stars, see ][]{anglada2012a} is a manifestation of the same processes observed in magnetically active stars but occuring at lower RV amplitude levels. The correlation between the S-index and the occurrence of the RV period at 320 days allows us to establish a very likely connection between this activity-related signal and a spurious RV signal and rule out the Keplerian nature of the corresponding candidate. We are in the process of analysing the chromatic dependence of stellar jitter using a more representative sample \citep[i.e. the sample analysed in ][]{dumusque2011}, the results of which will be presented in a future publication. Given the substantial changes in the potential signals in the system (e.g., appearance of the candidate at a period of 125 or 200 days), we perform a complete orbital re-analysis of the redmost half of the spectrum only (490-680 nm) in the next section.

\begin{table}
\center
\caption{Overview of the significant periodicities found in two activity indicators}
\label{tab:activity}
\begin{tabular}{lcc}
\hline \hline
Period & Quantity & RV counterpart \\
(days) &          &                \\
\hline \hline
22.97  & FWHM     & -              \\
43.0   & S-index  & -              \\
320    & S-index  & Yes (Blue)     \\
1220   & FWHM     & -              \\
$>$4000   & S-index  & (Blue?)        \\
$>$5000   & FWHM     & (Blue?)        \\
\hline\hline
\end{tabular}
\end{table}

\section{Analysis of the radial velocities from the 490-680 nm spectra}

First, we performed a rapid analysis of the RVs obtained using the redmost part of the spectra using least-squares periodograms. As was found to be the case when analysing the RVs from full spectra, three additional periodicities compared to the \citep{mayor2009a} solutions could be extracted from the data if all orbital eccentricities were fixed to zero. Two of the signals had the same periods (34 and 51 days) and the third one showed at $\sim$ 200 days (and on its yearly alias at 125 days). Although a model with circular orbits always looks more appealing aesthetically, we admit that this assumption cannot be justified using a pure least-squares approach without the use of prior information and/or physical constrains (e.g., orbital dynamics). As before, and in order to obtain a robust detection of all the candidates, we performed a full Bayesian analysis of the 345 red RV measurements using the methods and models described in Section \ref{sec:bayes}.

The parameter spaces of models with $k = 0, ..., 6$ were easy to sample using the adaptive Metropolis algorithm and the Markov chains converged rapidly to the posterior densities. Again, the three candidates presented by \citet{mayor2009a} with periods of 4.3, 9.6, and 20.4 days were extremely significant and easy to find using our MCMC samplings.

\begin{figure}
\center
\includegraphics[width=0.45\textwidth,clip]{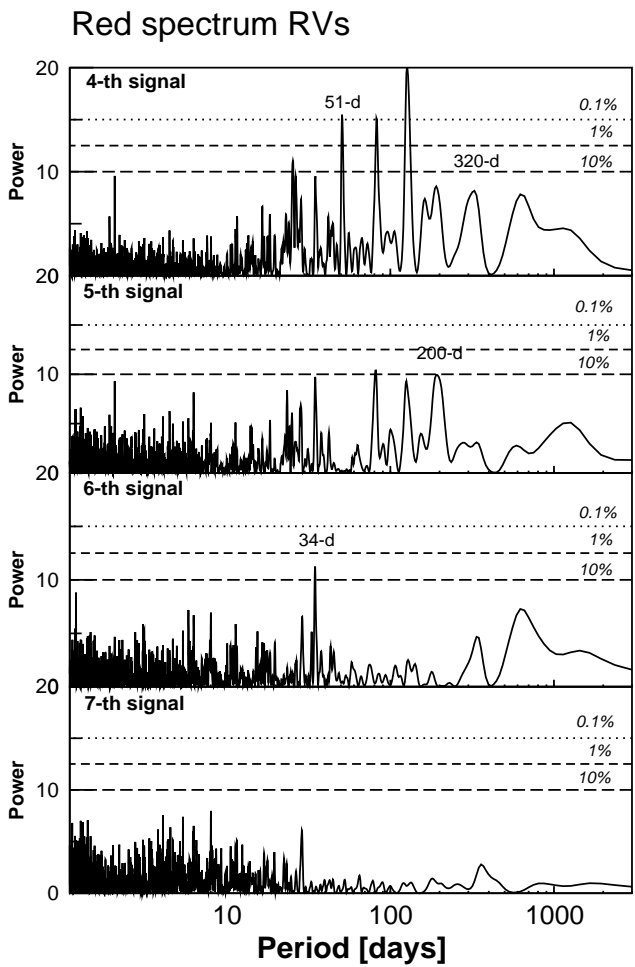}
\caption{As in Fig. \ref{periodograms_b} but for the unbinned radial velocities of the 490-680 nm spectrum.}\label{periodograms}
\end{figure}

After removing the MA(3) component of the noise and the signals in the three-Keplerian model, the power spectrum of the residuals showed very strong peaks at periods of 50.8, 82.1, and 128.3 days (Fig. \ref{periodograms}, top panel) exceeding 0.1\% FAP levels. Even though the highest peak corresponds to the 125-day period, the MCMC samplings showed that the 50.8-day periodicity is the global solution by providing higher posterior probabilities to the four-Keplerian model. We note that when calculating the residual periodogram of the three-Keplerian model, it is necessary to assume a fixed 3-Keplerian solution (and noise model) for the data. In contrast, the MCMC automatically accounts for correlations by also adjusting the noise parameters simultaneously to find the most probable areas of the parameter space (not only minimising the least-squares sum). After several MCMC samplings starting in the vicinity of the different possible periods suggested by the periodogram, this 50.8-day signal was always found to correspond to the global four-Keplerian solution and made the four-Keplerian model $3.1 \times 10^{12}$ times more probable than the three-Keplerian one.

The residual periodograms of four- and five-Keplerian models had moderate powers at 34.6, 81, 125, and 197 days (Fig. \ref{periodograms}, panels two and three). We performed similar samplings of the parameter spaces of five- and six-Keplerian models using the favoured periodogram powers as starting points of the Markov chains. As a result, we found that the period space near the 197- and 34.6-day powers provided the global maxima and improved the posterior model probabilities by factors of $3.6 \times 10^{11}$ and $2.5 \times 10^{9}$ when moving from $k=4$ to $k=5$ and from $k=5$ to $k=6$, respectively, where $k$ is the number of Keplerian signals in the model. Removing all six signals and the MA(3) components of the noise from the data left the periodogram of the residuals without any considerable powers (Fig. \ref{periodograms}, bottom panel). Similarly as for the full spectrum, the period of the seventh signal in a seven-Keplerian model did not converge to a clear probability maximum, thus failing to satisfy our requirement that the period of a signal has to be well-constrained for a positive detection. 

\begin{table}
\center
\caption{Relative posterior probabilities of models with $k=0, ..., 6$ Keplerian signals ($\mathcal{M}_{k}$) and the periods ($P_{s}$) of the signals added to the model when increasing the number of signals in the model by one. $P(d | \mathcal{M}_{k})$ is the Bayesian evidence and RMS denotes the common root mean square value of the residuals for comparison.}\label{probabilities}
\begin{tabular}{lcccc}
\hline \hline
$k$ & $P(\mathcal{M}_{k} | d)$ & $\log P(d | \mathcal{M}_{k})$ & RMS [ms$^{-1}$] & $P_{s}$ [days] \\
\hline
0 & 1.6$\times10^{-127}$ & -765.92 $\pm$ 0.06 & 2.20 & -- \\
1 & 8.5$\times10^{-106}$ & -715.20 $\pm$ 0.04 & 1.88 & 9.6 \\
2 & 7.7$\times10^{-74}$ & -640.92 $\pm$ 0.13 & 1.50 & 20.4 \\
3 & 3.5$\times10^{-34}$ & -548.90 $\pm$ 0.27 & 1.16 & 4.3 \\
4 & 1.1$\times10^{-21}$ & -519.43 $\pm$ 0.14 & 1.05 & 50.8 \\
5 & 4.1$\times10^{-10}$ & -492.12 $\pm$ 0.05 & 0.99 & 198 \\
6 & $\sim$ 1 & -469.40 $\pm$ 0.12 & 0.92 & 34.7 \\
\hline \hline
\end{tabular}
\end{table}

\begin{figure*}
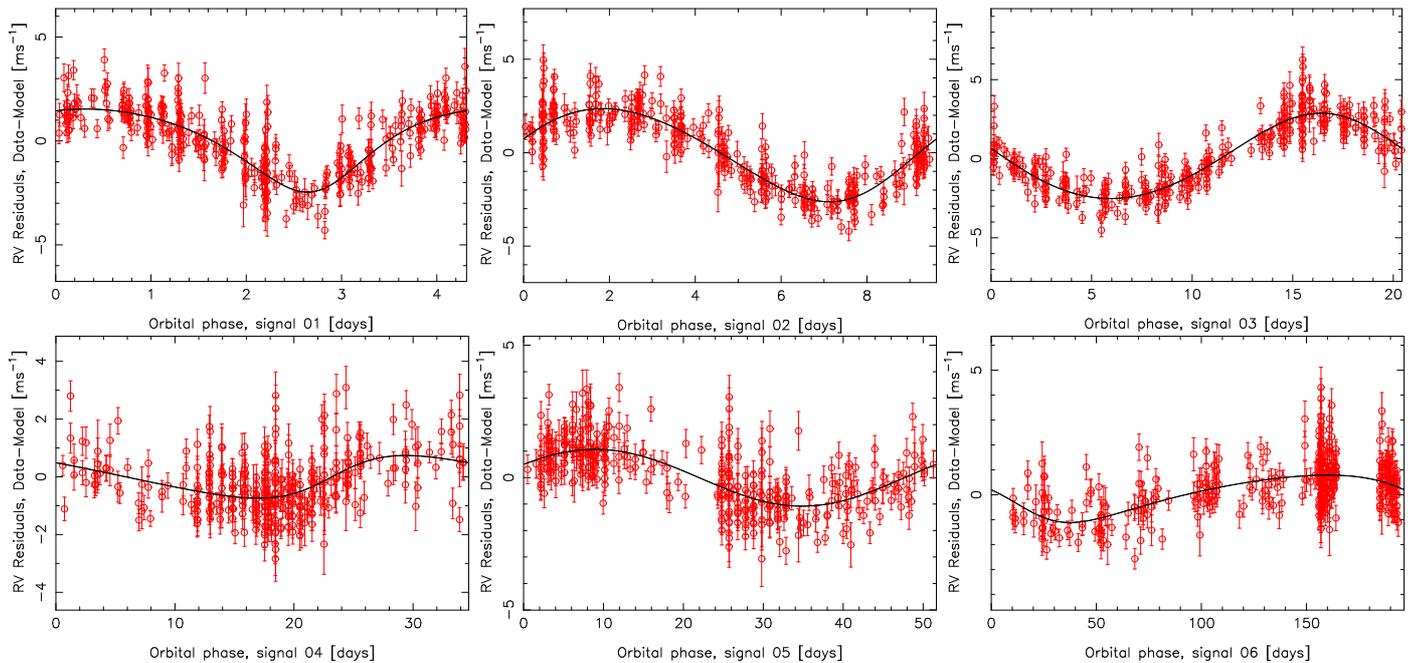

\center
\includegraphics[angle=-90,width=0.33\textwidth]{rvdist06_scresidc_rv_HD40307e_1.ps}
\includegraphics[angle=-90,width=0.33\textwidth]{rvdist06_scresidc_rv_HD40307e_2.ps}
\includegraphics[angle=-90,width=0.33\textwidth]{rvdist06_scresidc_rv_HD40307e_3.ps}

\includegraphics[angle=-90,width=0.33\textwidth]{rvdist06_scresidc_rv_HD40307e_4.ps}
\includegraphics[angle=-90,width=0.33\textwidth]{rvdist06_scresidc_rv_HD40307e_5.ps}
\includegraphics[angle=-90,width=0.33\textwidth]{rvdist06_scresidc_rv_HD40307e_6.ps}
\caption{Phase-folded MAP Keplerian signals of the six-planet solution with the other five signals and the MA(3) components removed from each panel.}\label{orbits}
\end{figure*}

To summarise, the posterior probabilities (and the corresponding Bayesian evidences) of models with $k = 0, 1, 2, ...$ Keplerian signals were found to be heavily in favour of $k = 6$ (Table \ref{probabilities}). This was the case although the selected prior probabilities penalise the models more heavily as the number of signals increases. We do not show the results for a seven-Keplerian model in Table \ref{probabilities} because the posterior samplings did not converge to a clear maximum (nor several maxima) in the seven-Keplerian model and therefore we cannot be sure whether the OBMH estimate yielded trustworthy values for the marginal integrals needed to assess the model probabilities. We plot the phase-folded MAP estimated orbits of the six Keplerian signals in Fig. \ref{orbits} after removing the MA(3) components.

\begin{figure*}
\center
\includegraphics[angle=-90,width=0.24\textwidth]{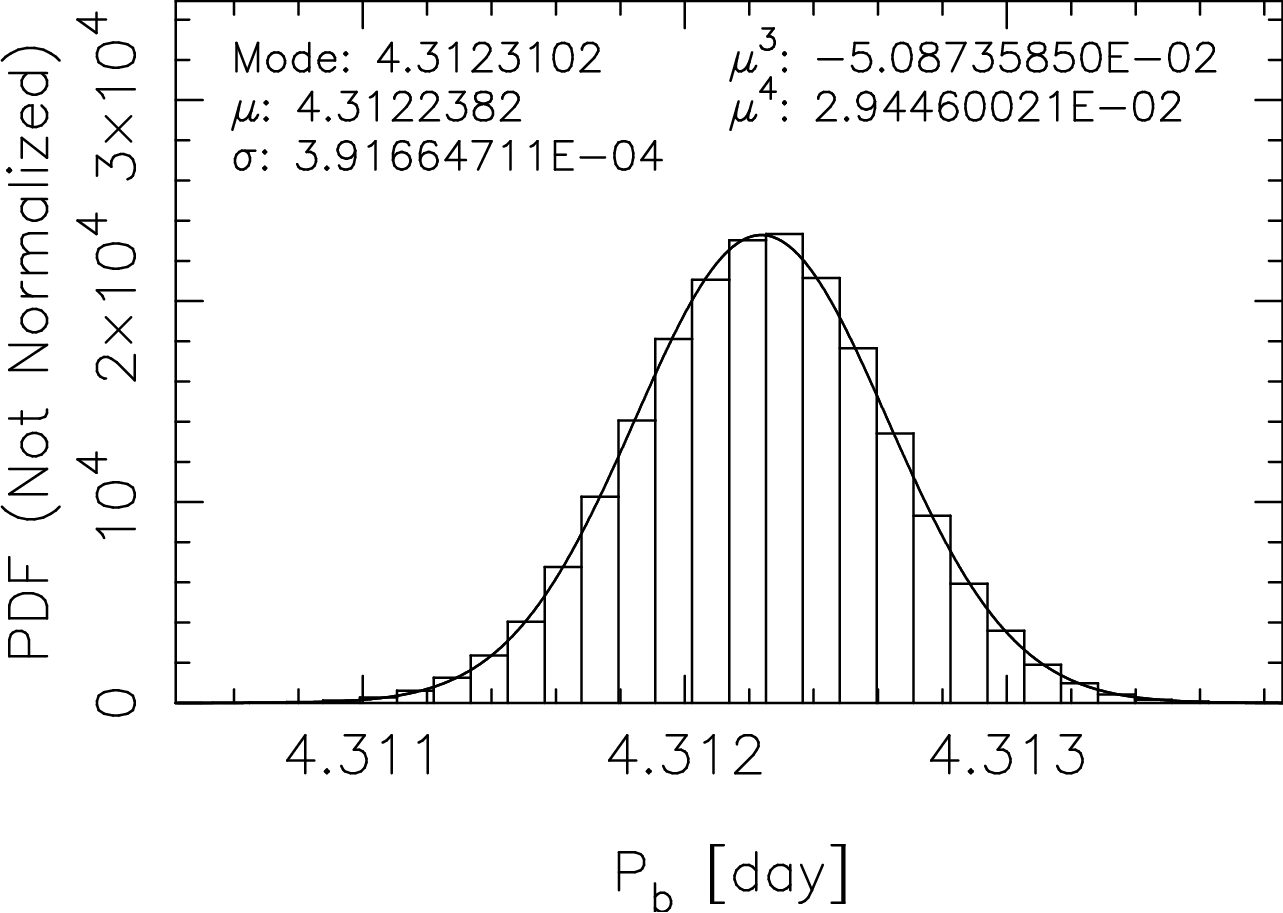}
\includegraphics[angle=-90,width=0.24\textwidth]{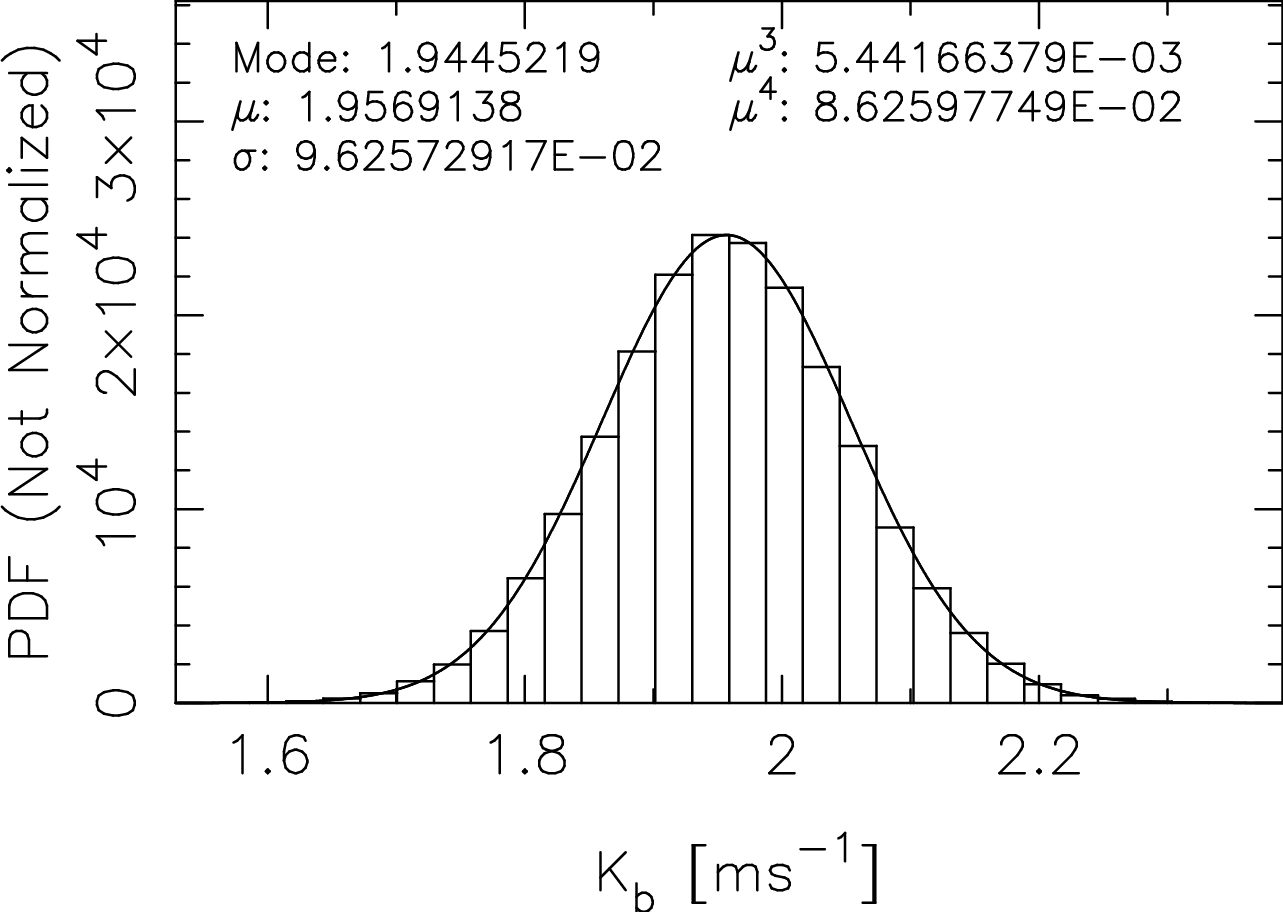}
\includegraphics[angle=-90,width=0.24\textwidth]{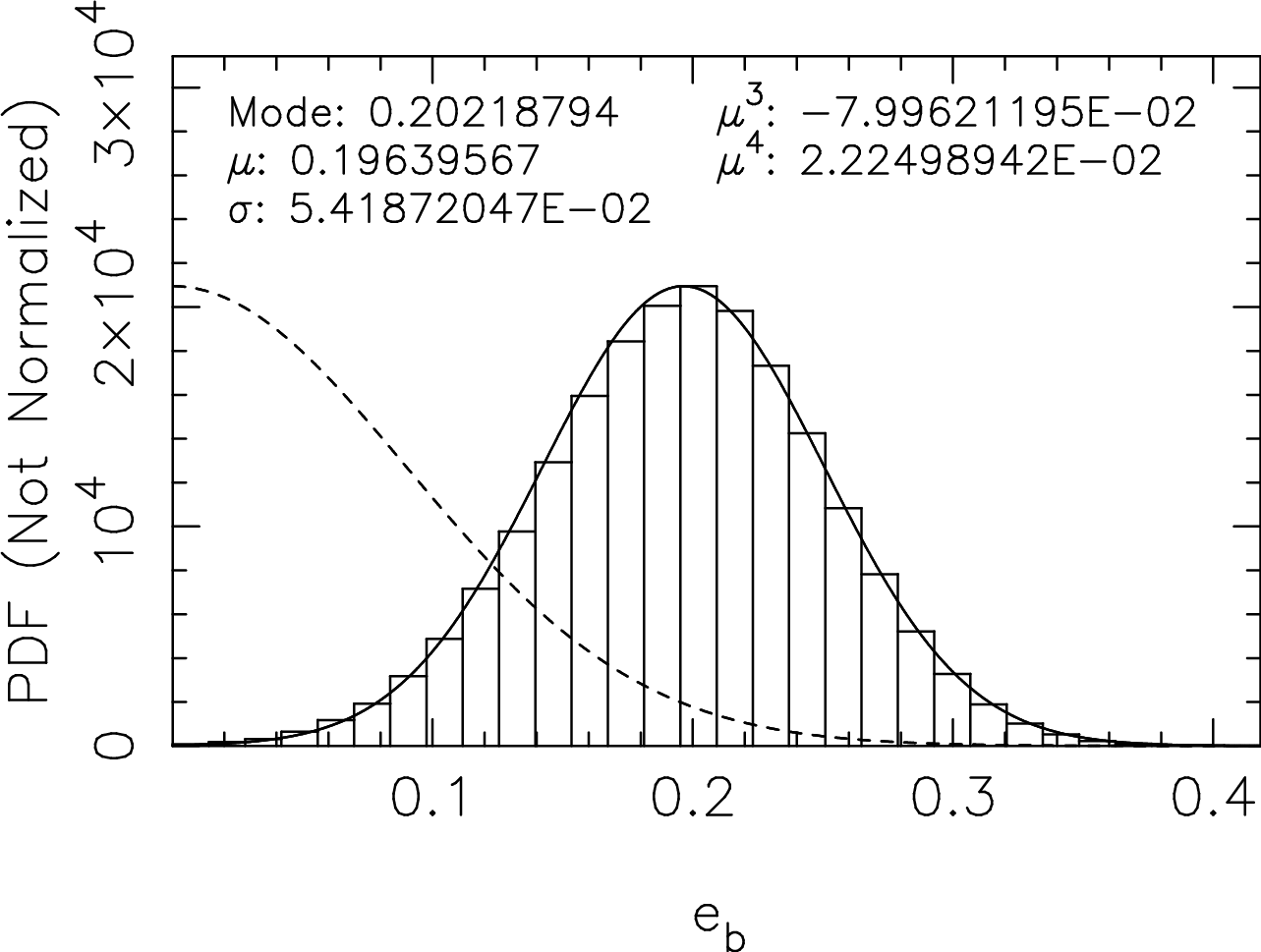}

\includegraphics[angle=-90,width=0.24\textwidth]{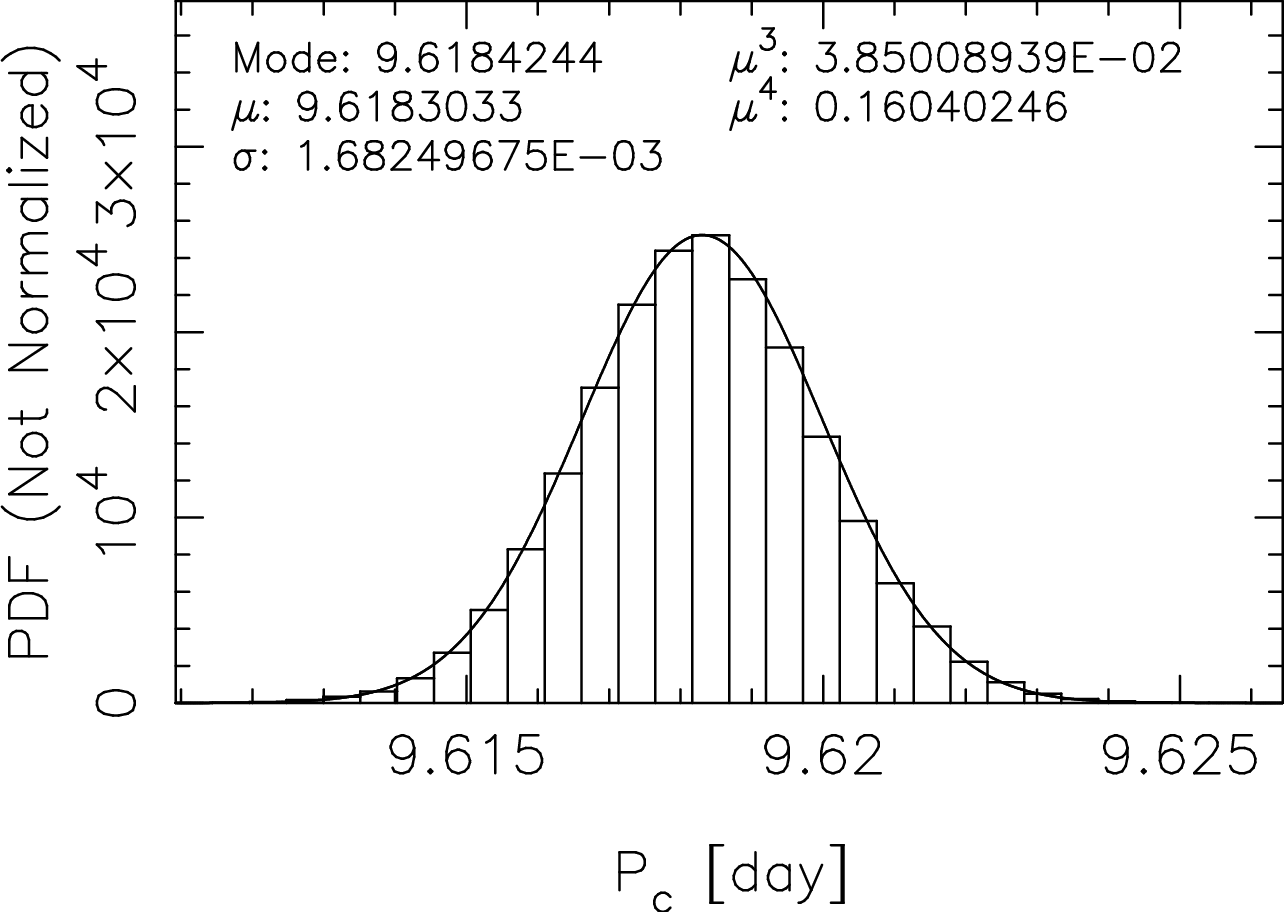}
\includegraphics[angle=-90,width=0.24\textwidth]{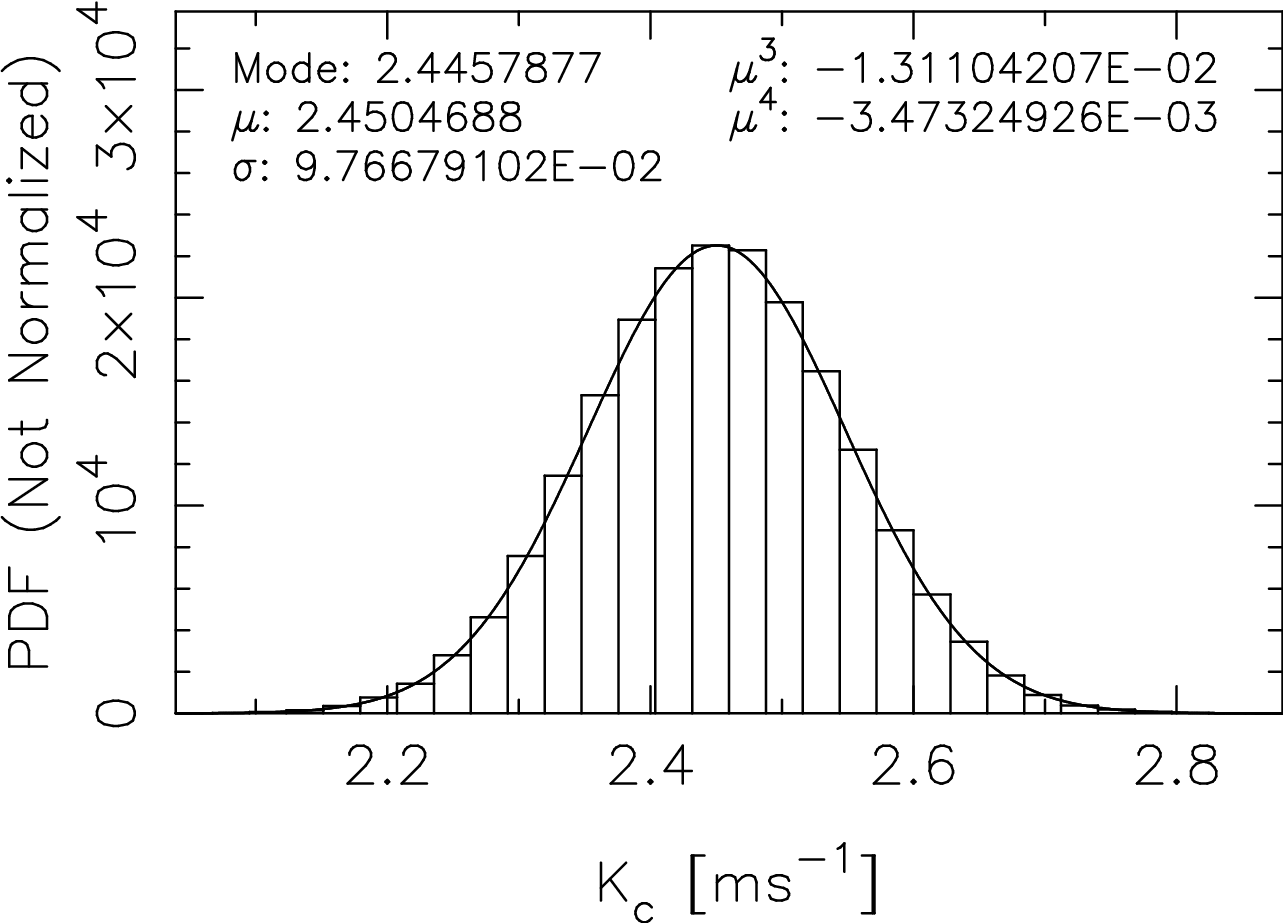}
\includegraphics[angle=-90,width=0.24\textwidth]{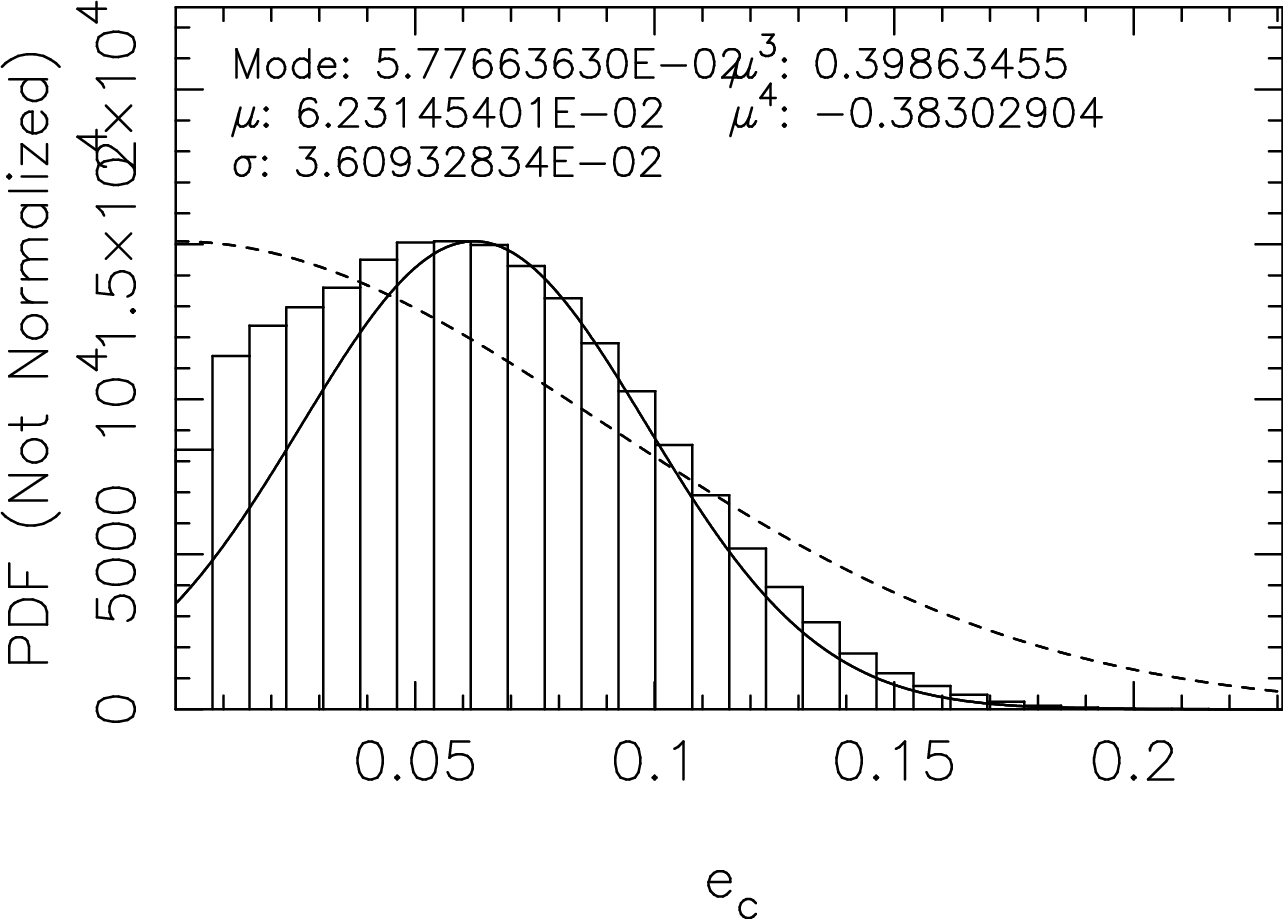}

\includegraphics[angle=-90,width=0.24\textwidth]{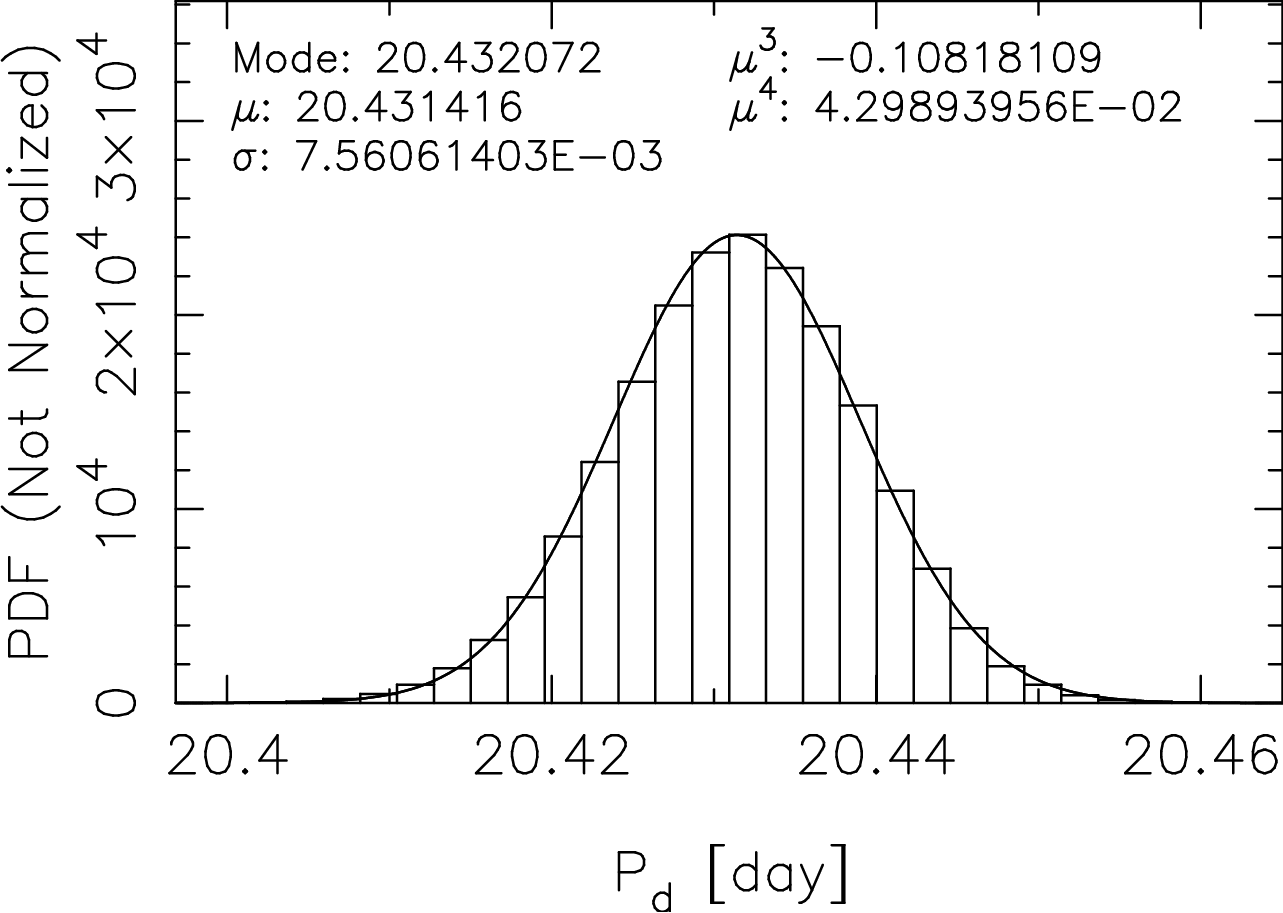}
\includegraphics[angle=-90,width=0.24\textwidth]{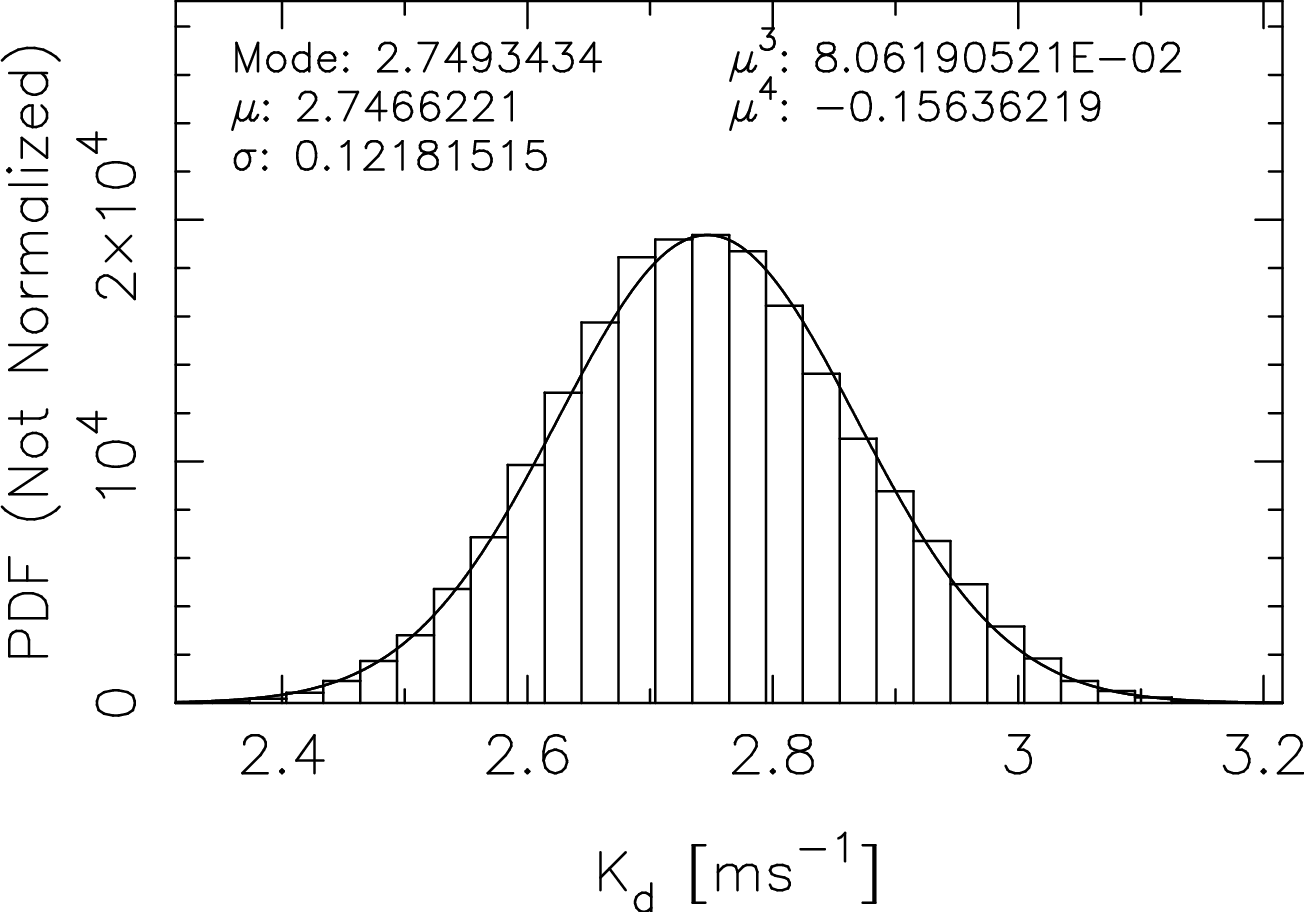}
\includegraphics[angle=-90,width=0.24\textwidth]{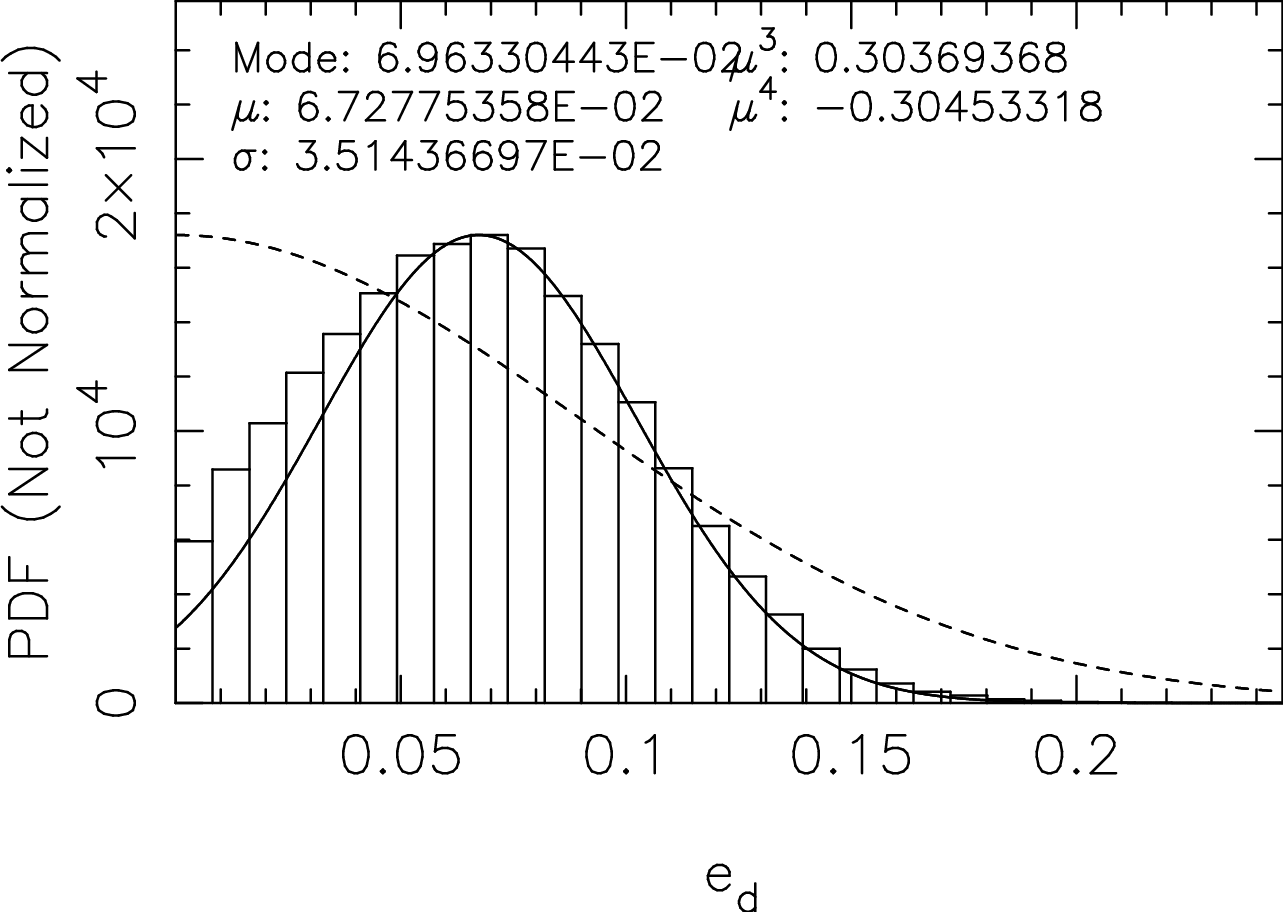}

\includegraphics[angle=-90,width=0.24\textwidth]{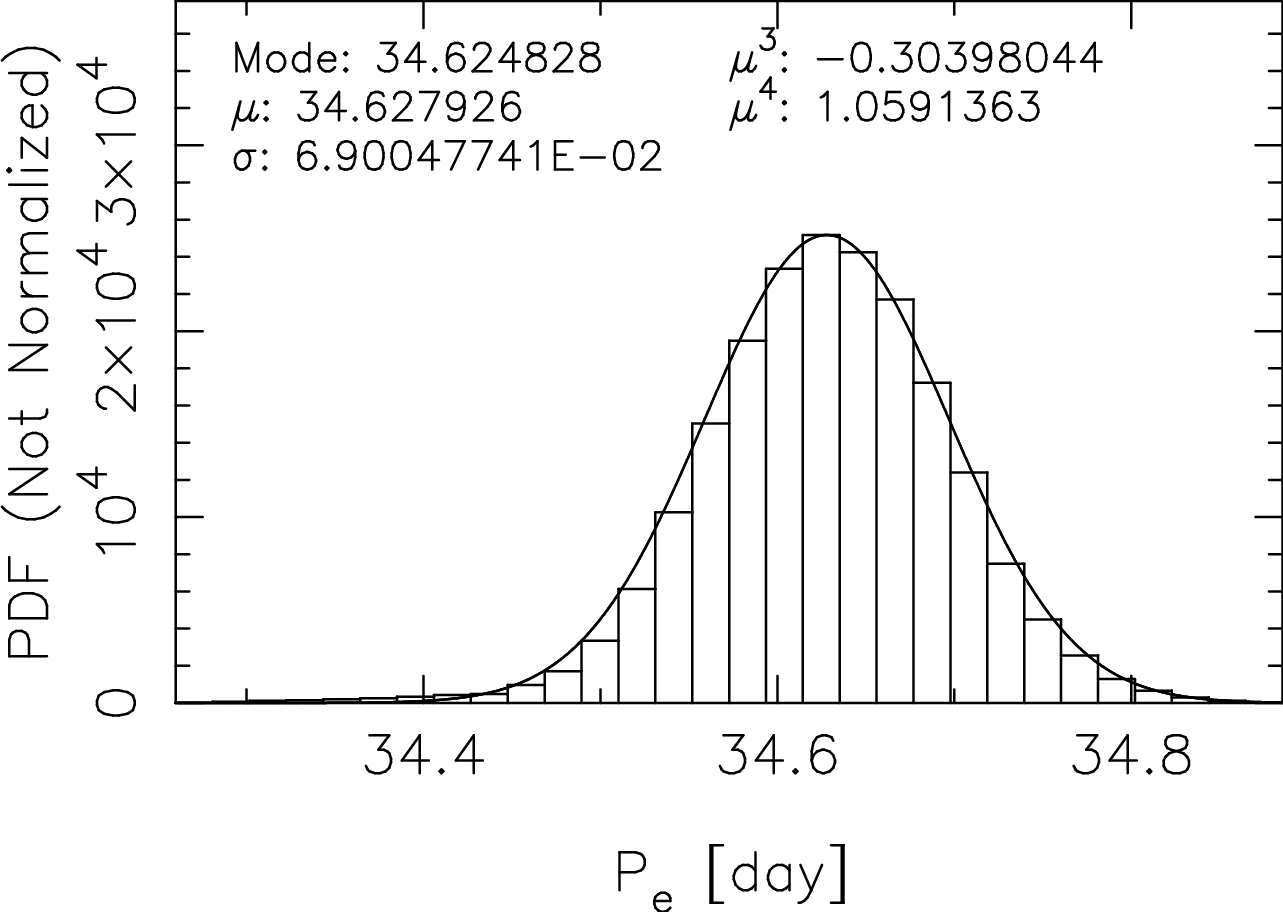}
\includegraphics[angle=-90,width=0.24\textwidth]{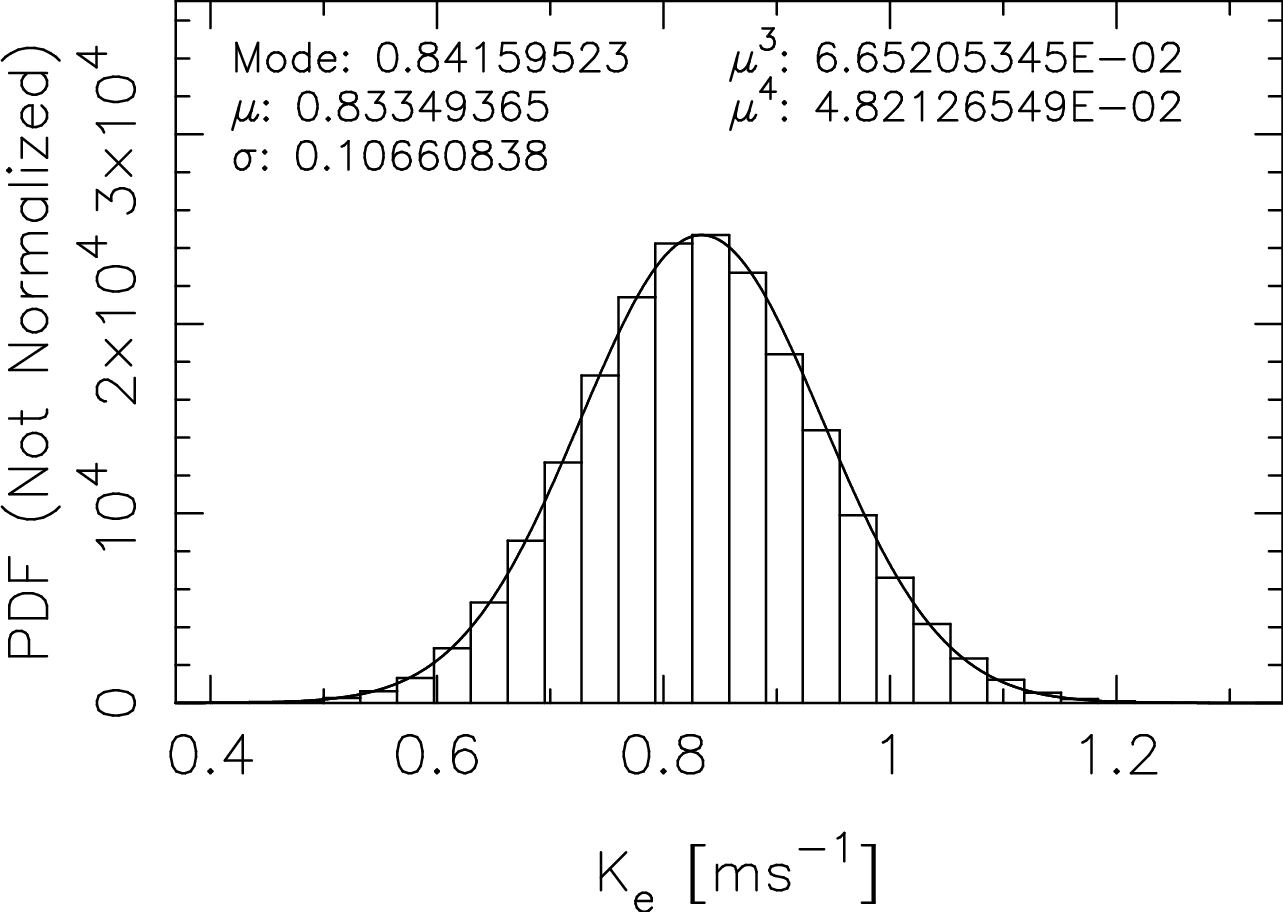}
\includegraphics[angle=-90,width=0.24\textwidth]{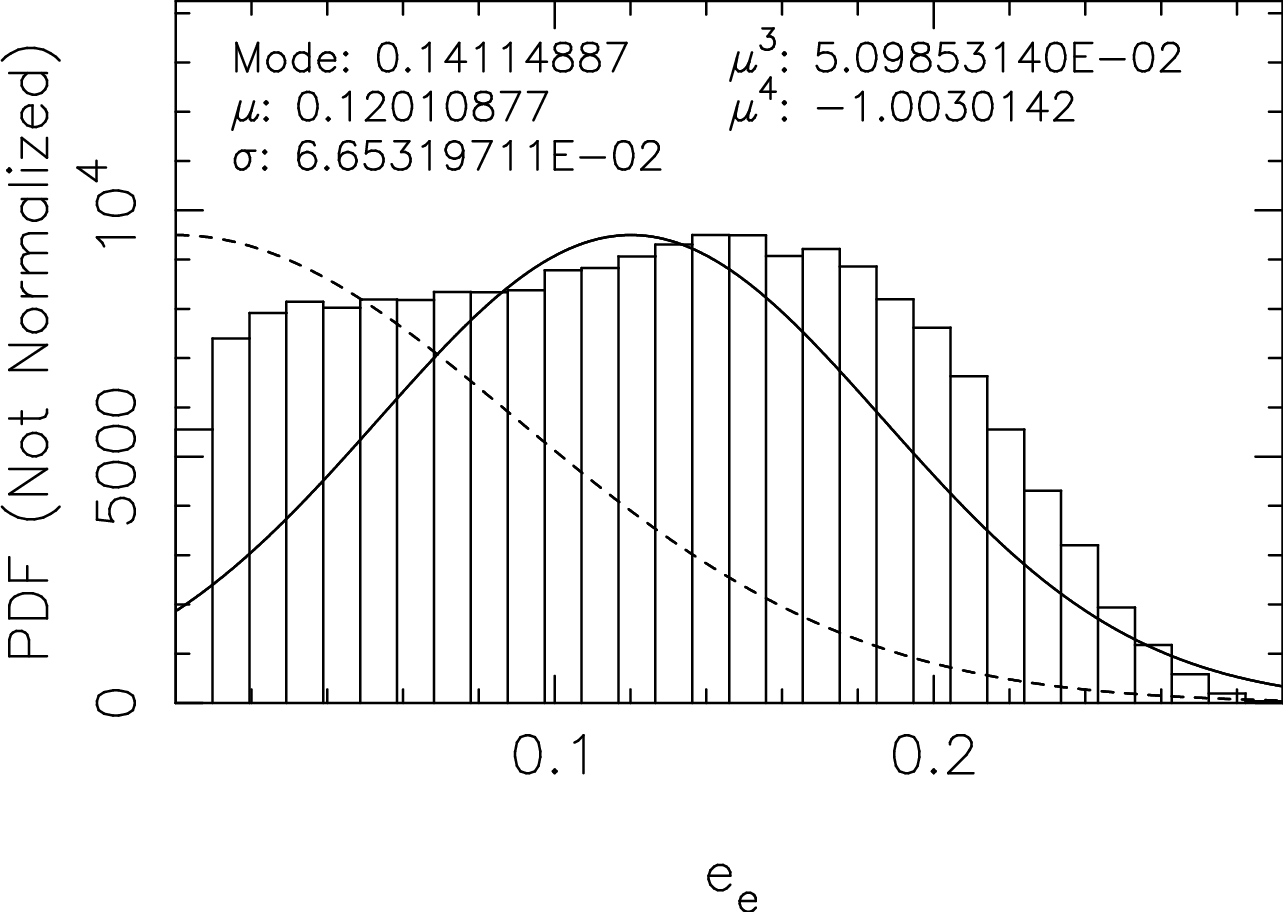}

\includegraphics[angle=-90,width=0.24\textwidth]{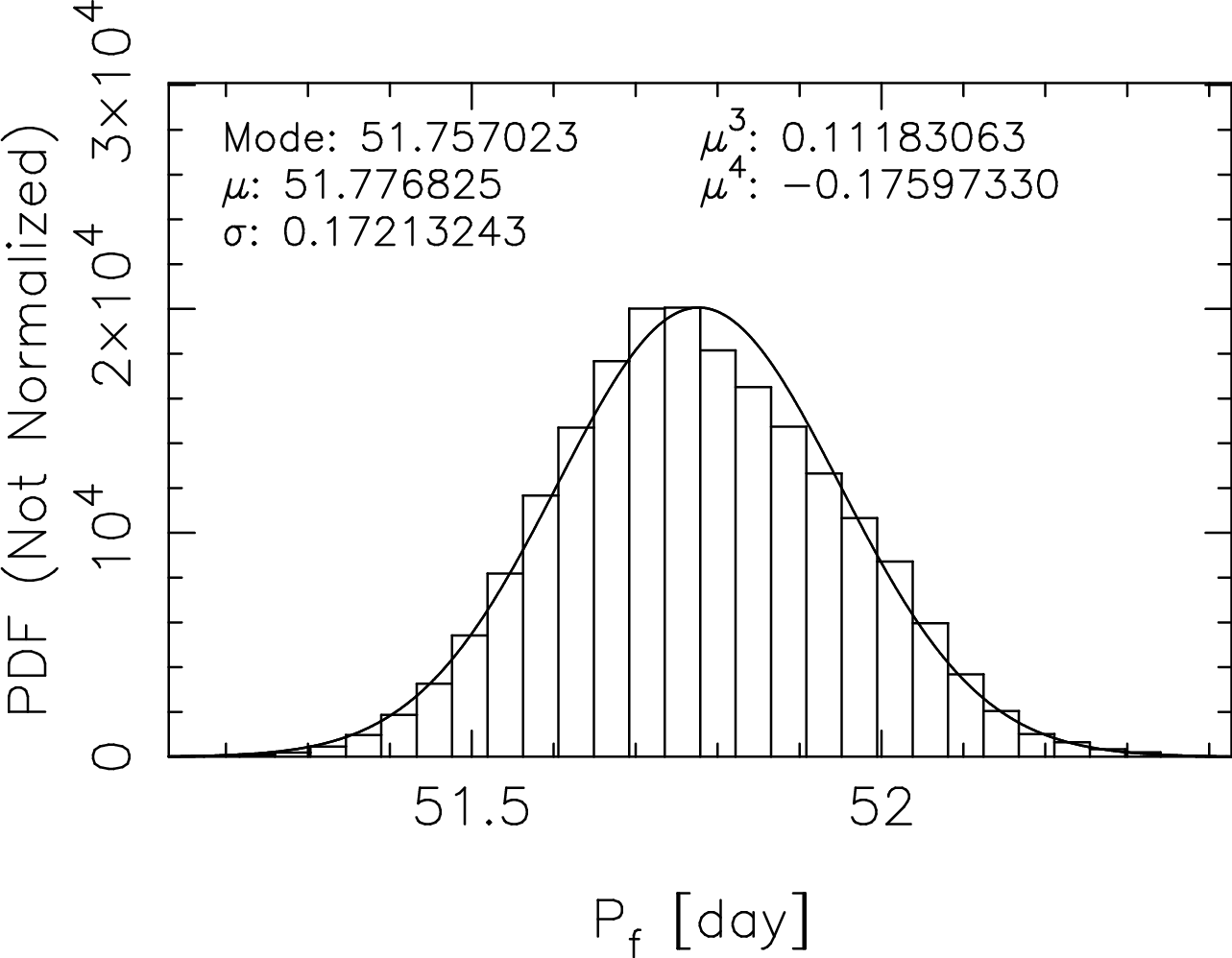}
\includegraphics[angle=-90,width=0.24\textwidth]{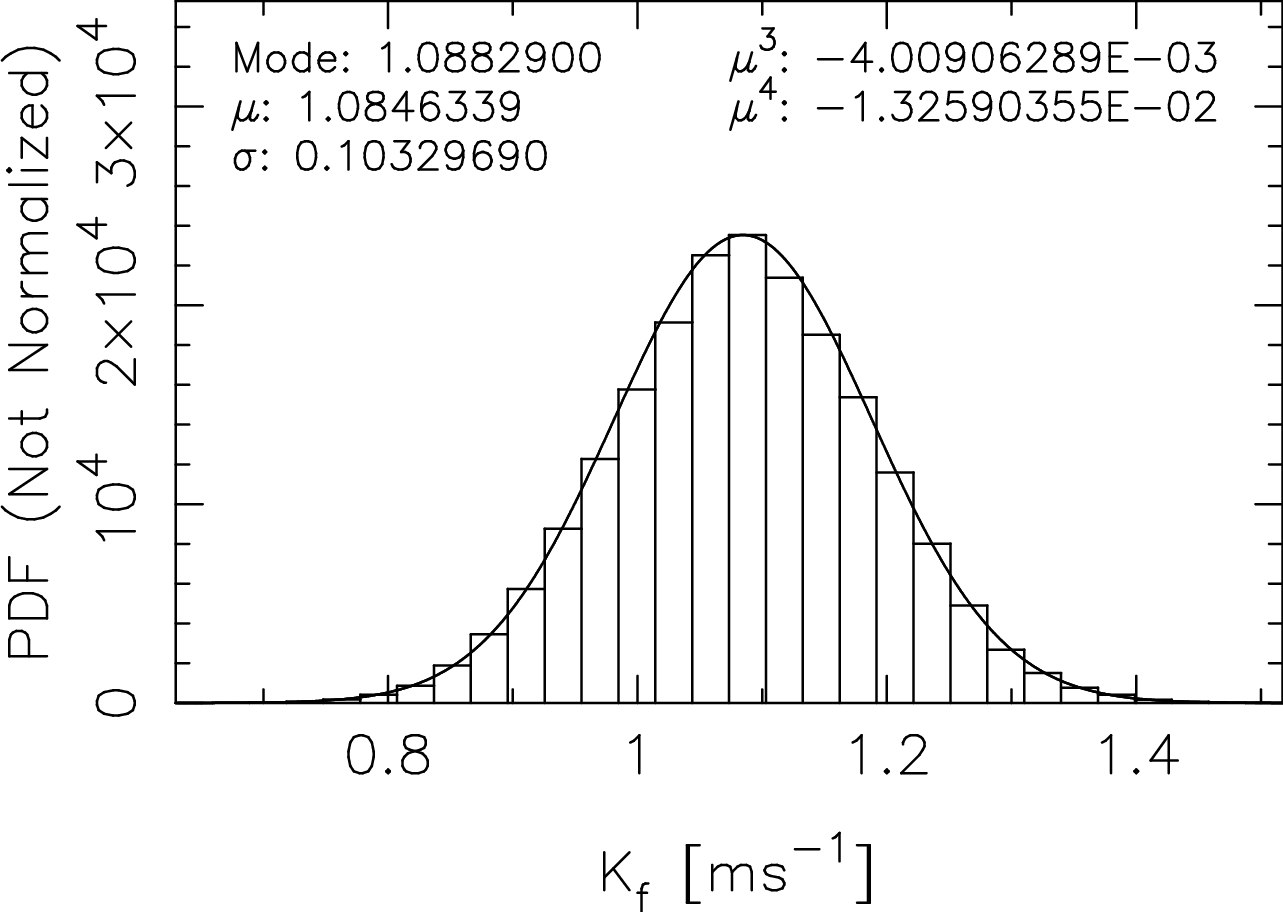}
\includegraphics[angle=-90,width=0.24\textwidth]{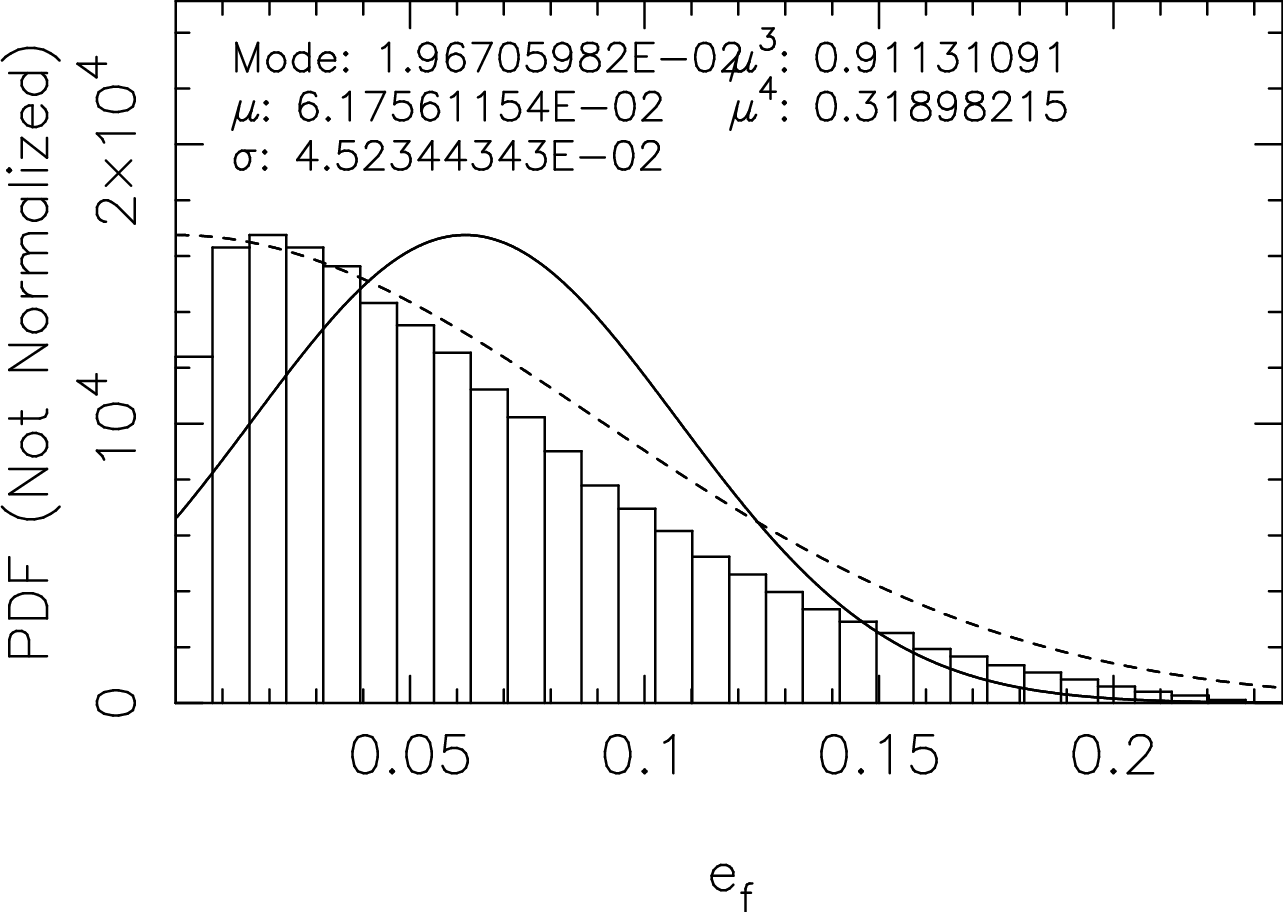}

\includegraphics[angle=-90,width=0.24\textwidth]{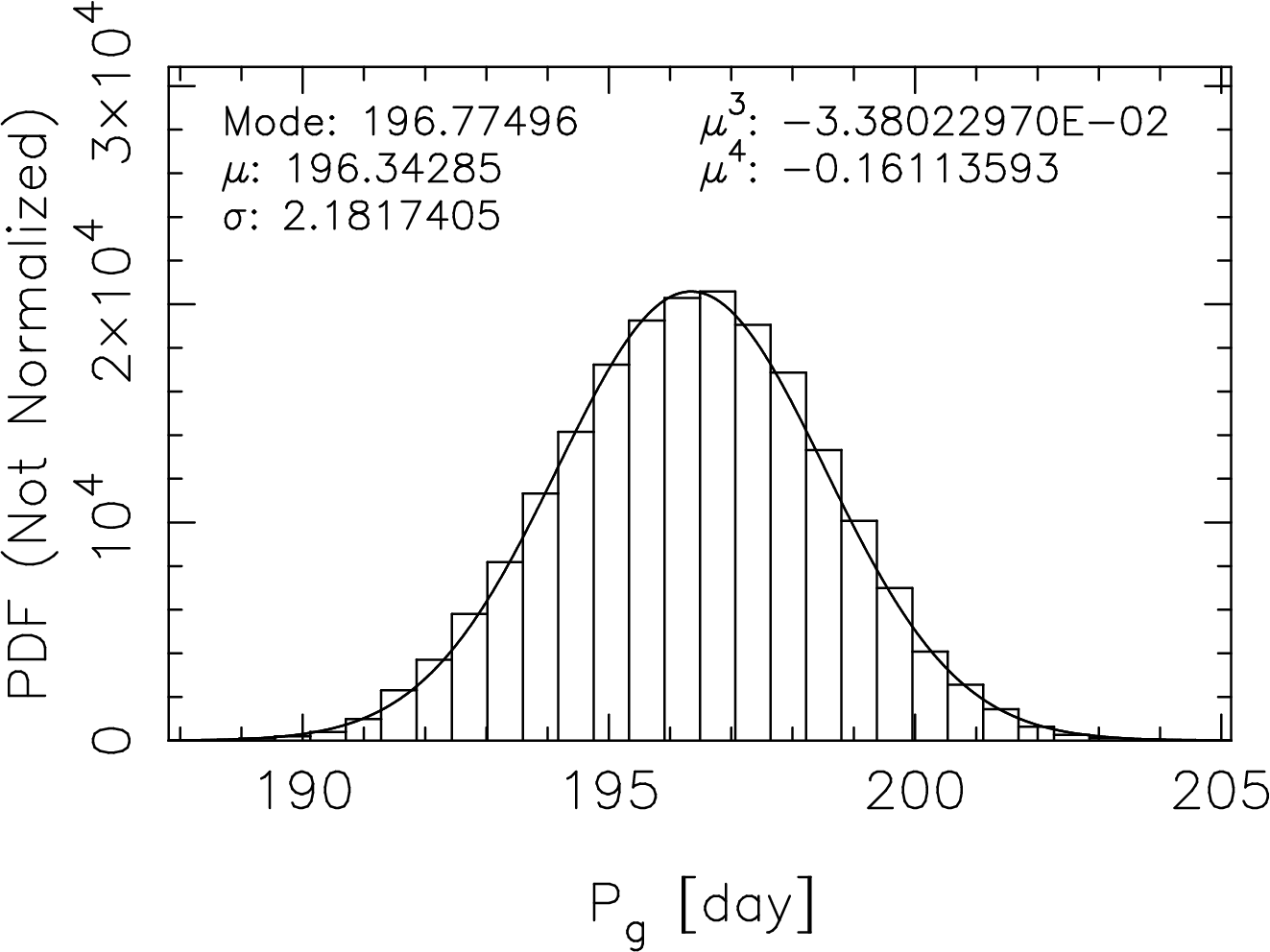}
\includegraphics[angle=-90,width=0.24\textwidth]{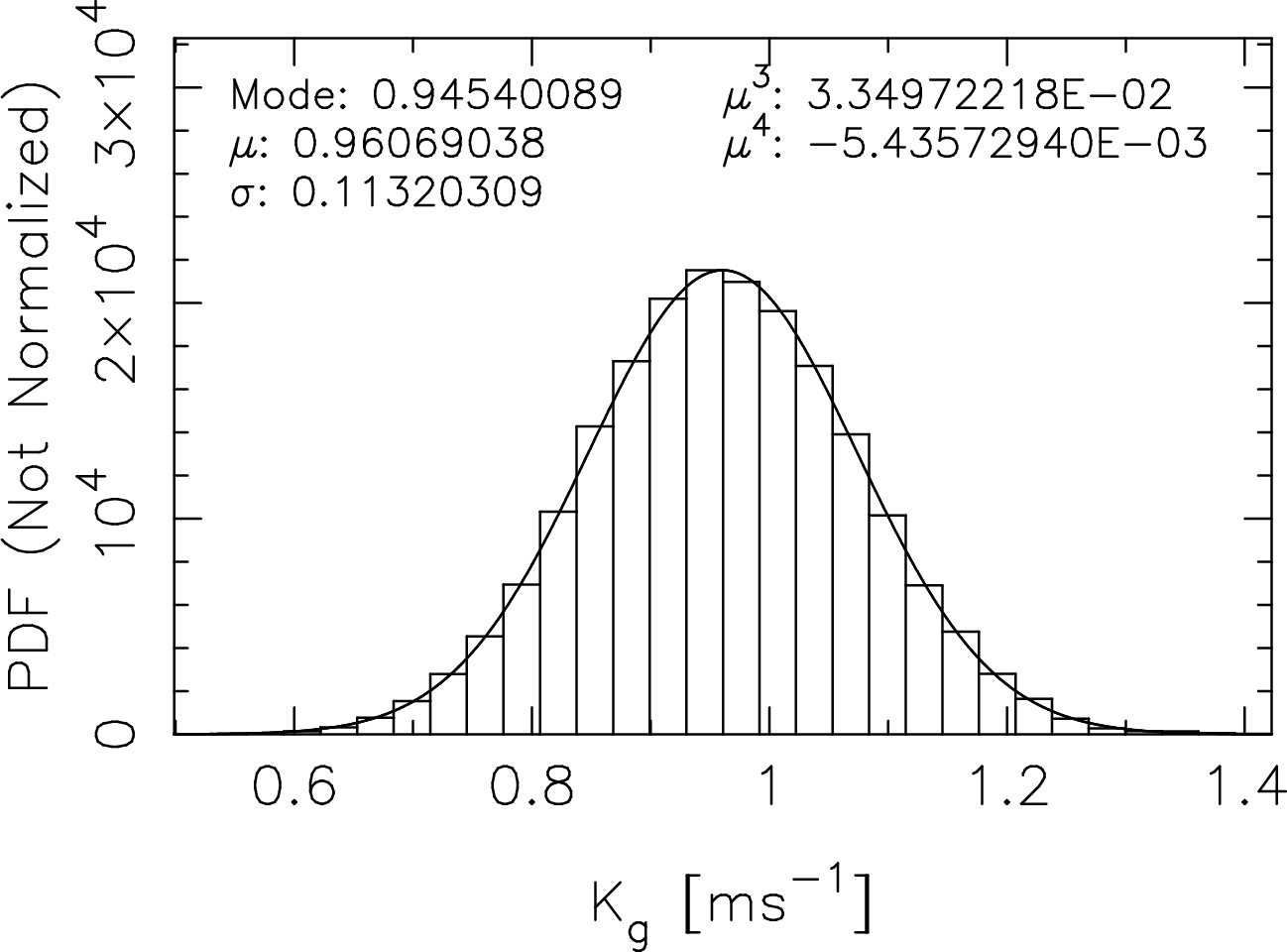}
\includegraphics[angle=-90,width=0.24\textwidth]{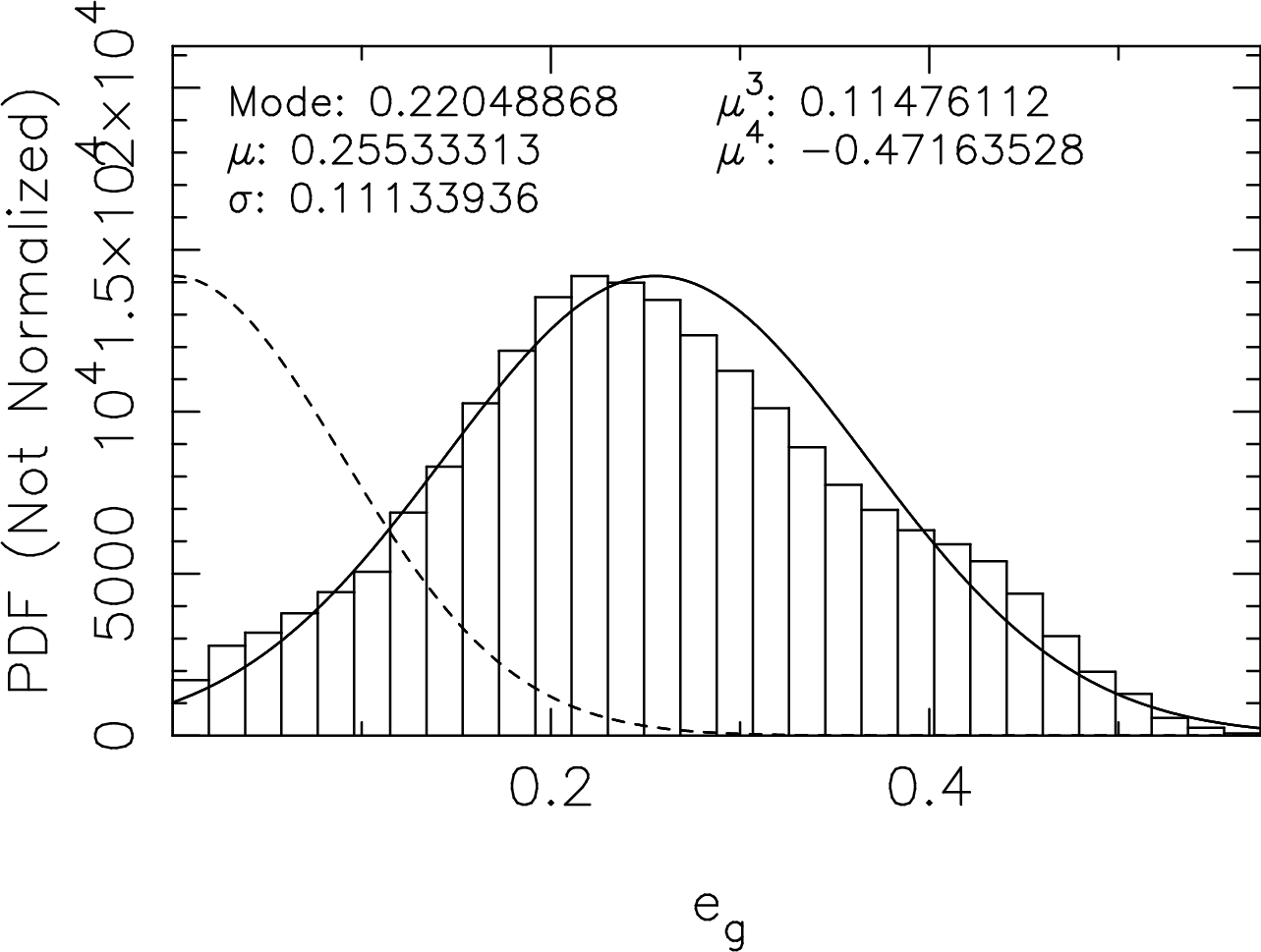}
\caption{Distributions estimating the posterior densities of orbital periods ($P_{x}$), eccentricities ($e_{x}$), and radial velocity amplitudes ($K_{x}$) of the six Keplerian signals. The dashed lines show the prior densities for comparison for the orbital eccentricities. The solid curves are Gaussian densities with the same mean ($\mu$) and variance ($\sigma^{2}$) as the parameter distributions. Additional statistics, mode, skewness ($\mu^{3}$), and kurtosis ($\mu^{4}$) of the distributions are also shown.}\label{densities}
\end{figure*}

One of our detection criteria for Keplerian signals in the radial velocities is that their amplitudes should be statistically significantly different from zero. The MAP estimates and the corresponding 99\% BCSs of the six-Keplerian model parameters are shown in Table \ref{parameters6}. The BCSs of the RV amplitudes clearly indicate that all six amplitude parameters indeed differed significantly from zero in accordance with the model probabilities that imply the existence of six Keplerian signals in the data. This was clearly the case despite the 34.7- and 198-day signals having amplitudes with MAP estimates below 1.0 ms$^{-1}$. Also, the periods of all signals were well-constrained and had clear MAP values and close-Gaussian densities (Table \ref{parameters6}; Fig. \ref{densities}). The distributions of orbital eccentricities are also shown in Fig. \ref{densities}. In agreement with the quicklook detection using least-squares periodograms assuming circular orbits, these distributions demonstrate that all signals had eccentricities consistent with close-circular orbits.

We also analysed the RVs derived from two independent blocks of the red part of the spectrum to assess if any of the signals was only present at certain wavelength ranges. The first block comprised the range between 465 and 553 nm (from now on Red1) and the second one covered the range between 553 to 690 nm (Red2). Red1 contains the richest part of a HARPS spectrum of a mid-K dwarf in terms of Doppler information (higher signal-to-noise, but also higher density of lines). The mean uncertainty of Red1 is approximately 0.6 ms$^{-1}$. As a result, we found that the same six signals were directly spotted even with periodograms only. The amount of Doppler information in Red2 is lower and nominal uncertainties were found to be approximately 1.0 ms$^{-1}$ on average. Using sequential periodogram subtraction of the signals we could recover the same six candidates except that the 125-day alias was preferred against the 200-day signal for HD 40307 g. However, the Bayesian samplings including the complete model for the uncertainties still converged to the same six-planet solution (P$_{\rm g} \sim$ 200 days). Given that there is no \emph{a priori} reason to prefer one or the other (125 versus 200 days, no activity signals appear to be related to any of them) and because the MCMC samplings insist on favouring the 200-day period, we conclude that the solution in Table \ref{parameters6} is the one preferred by the data. The significances and model probabilities using Red1 and Red2 are obviously lower than when using the whole red half of the spectrum. Yet, with the exception of the 34-day candidate in Red2 (chains converged to the same solution, but model probability ratio was close to the 1:150 threshold), all signals could be confidently identified even using two subsets of the HARPS data (each contains 1/3 of the full HARPS wavelength coverage). These tests give us further confidence in the reality of the reported candidate signals.

As a final consistency check, we calculated the model probabilities using different information criteria and compared them to the Bayesian model probabilities derived using our approximation for the marginal integrals. First, we estimated the relative posterior probabilities using the Akaike information criterion \citep[AIC; see e.g.][]{liddle2007,burnham2002}, which is known to provide only rough approximations for the marginal integrals. The AIC (actually its small-sample version, AICc) yielded posterior probabilities for the different models that did not differ by more than a factor of 5 from the probabilities received using the OBMH estimate (Table \ref{probabilities}). Using the AIC in model selection yielded the same qualitative results the OBMH estimate did. A common criticism of the AIC is that does not penalise the more complex models as much as it should in all applications \citep[e.g.][ and references therein]{burnham2002}. To obtain a more reliable comparison, we repeated the model selection procedure using the Bayesian information criterion (BIC), sometimes called the Schwarz information criterion \citep{schwarz1978}, which is known to penalise more complicated models, e.g. models with more Keplerian signals in this particular case, more heavily than the AIC. Even according to the more restictive BIC, the six-Keplerian model had the highest posterior probability and was 1.3$\times 10^{5}$ times more probable than the five-Keplerian model. These results provide further confidence in the conclusion that six periodic signals explain the variations in the data.

\begin{table*}
\center
\caption{Six-planet solution of HD 40307 radial velocities. MAP estimates of the parameters and their 99\% BCSs. Parameter $e_{\rm dyn}$ denotes the estimates for eccentricity after excluding solutions corresponding to dynamically instability in $10^{6}$ year time scales.}\label{parameters6}
\begin{tabular}{lccc}
\hline \hline
Parameter & HD 40307 b & HD 40307 c & HD 40307 d \\
\hline
$P$ [days] & 4.3123 [4.3111, 4.3134] & 9.6184 [9.6135, 9.6234] & 20.432 [20.408, 20.454] \\
$e$ & 0.20 [0.04, 0.34] & 0.06 [0, 0.17] & 0.07 [0, 0.18] \\
$e_{\rm dyn}$ & 0.16 [0.04, 0.25] & 0.05 [0, 0.13] & 0.05 [0, 0.13] \\
$K$ [ms$^{-1}$] & 1.94 [1.67, 2.25] & 2.45 [2.17, 2.75] & 2.75 [2.40, 3.10] \\
$\omega$ [rad] & 3.4 [2.4, 4.2] & 4.1 [-] & 0.3 [-] \\
$M_{0}$ [rad]& 4.3 [2.5, 5.9] & 5.3 [-] & 5.6 [-] \\
$m_{p} \sin i$ [M$_{\oplus}$] & 4.0 [3.3, 4.8] & 6.6 [5.6, 7.7] & 9.5 [8.0, 11.2] \\
$a$ [AU] & 0.0468 [0.0445, 0.0492] & 0.0799 [0.0759, 0.0839] & 0.1321 [0.1255, 0.1387] \\
\hline
& HD 40307 e & HD 40307 f & HD 40307 g \\
\hline
$P$ [days] & 34.62 [34.42, 34.83] & 51.76 [51.30, 52.26] & 197.8 [188.8, 203.5] \\
$e$ & 0.15 [0, 0.28] & 0.02 [0, 0.22] & 0.29 [0, 0.60] \\
$e_{\rm dyn}$ & 0.06 [0, 0.18] & 0.03 [0, 0.13] & 0.22 [0, 0.45] \\
$K$ [ms$^{-1}$] & 0.84 [0.53, 1.16] & 1.09 [0.77, 1.37] & 0.95 [0.65, 1.27] \\
$\omega$ [rad] & 5.3 [-] & 6.2 [-] & 1.6 [-] \\
$M_{0}$ [rad]& 6.2 [-] & 0.5 [-] & 5.6 [-] \\
$m_{p} \sin i$ [M$_{\oplus}$] & 3.5 [2.1, 4.9] & 5.2 [3.6, 6.7] & 7.1 [4.5, 9.7] \\
$a$ [AU] & 0.1886 [0.1782, 0.1969] & 0.247 [0.233, 0.258] & 0.600 [0.567, 0.634] \\
\hline
$\gamma$ [ms$^{-1}$] & -0.47 [-0.71, -0.25] \\
$\sigma_{j}$ [ms$^{-1}$] & 0.73 [0.59, 0.87] \\
\hline \hline
\end{tabular}
\end{table*}

\subsection{Noise properties}

In addition to ruling out wavelength-dependent signals, the Bayesian analysis of the RVs at different wavelength ranges allows the comparison of the noise properties. As described in Section \ref{sec:bayes}, we modelled the radial velocity noise using a third-order MA model (three free parameters) together with a Gaussian white noise component (one free parameter). The posterior distributions and MAP values for these parameters quantify the significance and the magnitude of each component of the noise (white vs correlated).

The MA parameters $\phi_{j}$, $j=1, 2, 3$, which describe the correlation between the error of the $i$-th and ($i-j$)th measurement, are clearly positive for the RVs from full spectra (Table \ref{noise_parameters}). The noise of the measurements appears to correlate positively the strongest with the noise of the previous measurement and the effect decreases for the measurements obtained before that. However, this correlation decreases for the RVs obtained from the redmost part of the spectra (490-680 nm). According to our results, this decrease is balanced by a corresponding increase in the magnitude of the Gaussian white noise component (Table \ref{noise_parameters}). Therefore, the noise in the radial velocities becomes ``whiter`` when excluding the bluest half of the spectra from the analyses. This result means that in addition to removing long-term activity-related signals, using redder wavelengths also helps reducing correlations in the noise that might cause biases to any estimates of the Keplerian signals if not accounted for.

\begin{table}
\center
\caption{Noise parameters obtained when using the RVs from the full spectra and the red part (490-680 nm) of the spectra only as denoted using the MAP estimates and the 99\% BCSs.}\label{noise_parameters}
\begin{tabular}{lcc}
\hline \hline
Parameter & Full & Red \\
\hline
$\phi_{1}$   & 0.54 [0.28, 0.77]   & 0.27 [0.03, 0.50] \\
$\phi_{2}$   & 0.23 [-0.13, 0.58]  & 0.10 [-0.25, 0.42] \\
$\phi_{3}$   & 0.28 [-0.06, 0.66]  & 0.25 [-0.07, 0.54] \\
$\sigma_{j}$ [ms$^{-1}$] & 0.53 [0.41, 0.65] & 0.73 [0.59, 0.87] \\
\hline \hline
\end{tabular}
\end{table}

We note that the six signals in the velocities were rather independent of the exact MA model used. We obtained consistent solutions with MA(2) and MA(4) models as well, though these noise models were found to have slightly lower posterior probabilities. This indicates, as can be seen in Table \ref{noise_parameters}, that the MA components MA($p$) with $p > 1$ do not affect the solution much because their 99\% BCSs imply that they are consistent with zero.

\section{Orbital stability}\label{sec:stability}

We investigated the long-term stability of the planetary system corresponding to our six-Keplerian solution (Table \ref{parameters6}). This analysis was performed in three steps. First, we checked the Lagrange stability criteria, which provide a first qualitative assessement of the allowed orbital configurations. Second, we numerically integrated orbits directly drawn from the MCMC samplings and assessed the fraction corresponding to stable orbits at the Myr time scale. As a last test, we analysed the stability of the six-planet configuration by testing different eccentricities and arguments of the node of each planet.

\subsection{Lagrange stability criteria}

\begin{figure}
\center
\includegraphics[angle=-90, width=0.49\textwidth]{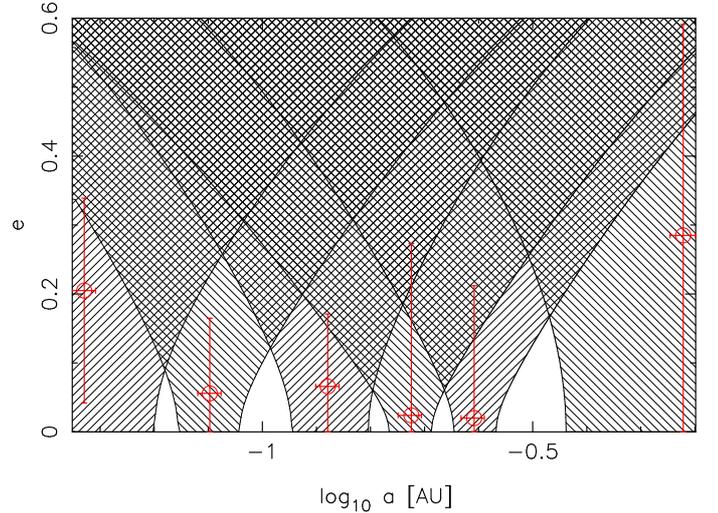}
\caption{Approximate Lagrange stability thresholds (shaded areas) between each two planets (red circles) and the MAP orbital parameters of the six-planet solution (Table \ref{parameters6}).}\label{barnes_stability}
\end{figure}

As a rough initial test, we checked whether the subsequent planets in the system satisfied the approximate Lagrange stability criterion expressed analytically in
\citet{barnes2006} and applied for instance in \citet{tuomi2012}. We plotted the resulting areas of instability in Fig. \ref{barnes_stability} -- if the $i-1$th or $i+1$th planet has orbital parameters within the shaded areas surrounding the $i$th planet, the system is likely unstable over the long term. While this criterion is but a rough approximation of reality because it does not take the stabilising and/or destabilising effects of orbital resonances into account and is only valid for two planets at a time, it provides simple guidelines on the prospects of stability in the system.

According to the Lagrange stability criterion and the corresponding analytical thresholds in Fig. \ref{barnes_stability}, the six planet candidates orbiting HD 40307 form a stable system if their orbital eccentricities are close to or below their MAP estimates. Alternatively, it can be stated that if the planets exist and have masses approximately equal to the minimum masses in Table \ref{parameters6}, they are all likely on close-circular orbits. Especially, the outer companion is unlikely to have orbtital eccentricity in excess of 0.4 (Fig. \ref{barnes_stability}).

\subsection{Numerical integration of MCMC samples}

To further test the dynamical stability of the system, we integrated the planetary orbits for $10^{6}$ years, which is appropriate to exclude 'almost all' unstable configurations \citep{barnes2004}. In our integrations, we used the Bulirsch-Stoer integrator \citep{bulirsch1966} that has been used before when integrating planetary orbits \citep[e.g.][]{tuomi2009a,tuomi2011}. We found that most of the unstable configurations corresponded to one (or more) of the eccentricities of the companions being higher than its MAP estimate and the system turned out to be unstable in $10^{4} - 10^{5}$ years. Specifically, we drew a sample from the posterior density of the orbital parameters and used each vector in this sample as an initial state of orbital integrations. We increased the sample size such that there were 100 initial states that could not be shown to correspond to unstable orbital configurations. This subsample was found to correspond to 4\% of the total sample.

\begin{figure*}
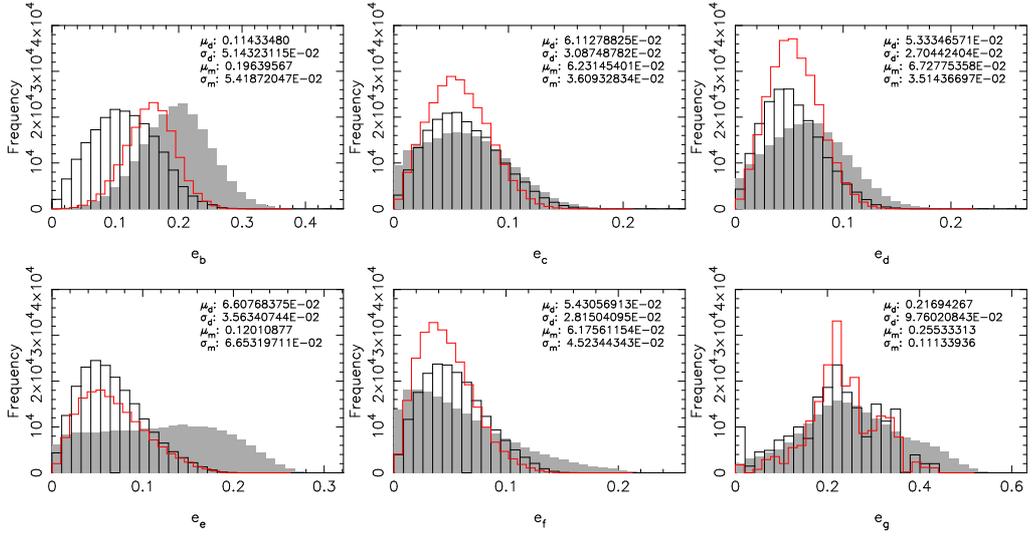

\center
\includegraphics[angle=-90,width=0.24\textwidth]{rvdist06_rv_HD40307e_nat_hist_eccs2_b.ps}
\includegraphics[angle=-90,width=0.24\textwidth]{rvdist06_rv_HD40307e_nat_hist_eccs2_c.ps}
\includegraphics[angle=-90,width=0.24\textwidth]{rvdist06_rv_HD40307e_nat_hist_eccs2_d.ps}

\includegraphics[angle=-90,width=0.24\textwidth]{rvdist06_rv_HD40307e_nat_hist_eccs2_e.ps}
\includegraphics[angle=-90,width=0.24\textwidth]{rvdist06_rv_HD40307e_nat_hist_eccs2_f.ps}
\includegraphics[angle=-90,width=0.24\textwidth]{rvdist06_rv_HD40307e_nat_hist_eccs2_g.ps}
\caption{Distributions estimating the orbital eccentricities given measurements alone (grey histogram; mean $\mu_{m}$ and standard deviation $\sigma_{m}$), showing the values the eccentricities had during orbital integrations of $10^{6}$ years that could not be shown to be unstable (white histogram; mean $\mu_{d}$ and standard deviation $\sigma_{d}$), and their combination (red histogram).}\label{dynamical_densities}
\end{figure*}

In Fig. \ref{dynamical_densities} we plot the posterior densities of the planetary eccentricities together with the densities obtained from the subsample of solutions that were not found to be unstable. Fig. \ref{dynamical_densities} states essentially the same thing as Fig. \ref{barnes_stability}. That is, it shows that the stable configurations mostly correspond to orbital eccentricities closer to zero than those derived from the Bayesian MCMC analysis. The corresponding updated estimates for orbital eccentricities are shown in Table \ref{parameters6} and denoted using $e_{\rm dyn}$. While this experiment shows that a non-negligible fraction of the solutions allowed by the data corresponds to stable configurations, we aimed to explore the source of the dynamical instabilities and the role of potential dynamical resonances in the system as well.

\subsection{Sources of (in)stability}

Owing to the dense packing of the planets in the HD 40307 system, one still has to rely on systematic numerical integrations to correctly investigate any aspects of its stability. For example, from a direct numerical integration of the MAP six-planet solution given in Table \ref{parameters6}, one quickly finds that the orbital evolution leads to collisional trajectories within a timescale of $10^{3}$ years. 

As was already shown above, it is likely that the planets have close-circular orbits. This configuration is also supported by the confidence intervals given for the individual eccentricities in Table \ref{parameters6}, which essentially state that circular orbits cannot be ruled out by the data. Therefore, to slightly simplify the dynamical analysis, we performed a new fit to the observations, for which we allowed only circular orbits. For this nominal solution (S1 hereafter) we analysed the impact of changing the initial eccentricity $e$ and initial mean anomaly $M$ of individual planets on the stability of the whole system.

\begin{figure*}[ht]
  \centering
  \includegraphics[width=0.49\textwidth]{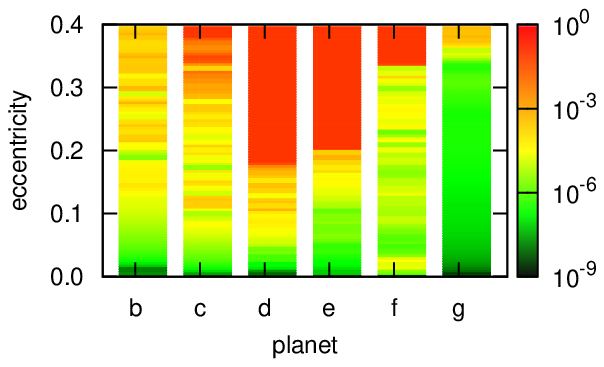} 
  \includegraphics[width=0.49\textwidth]{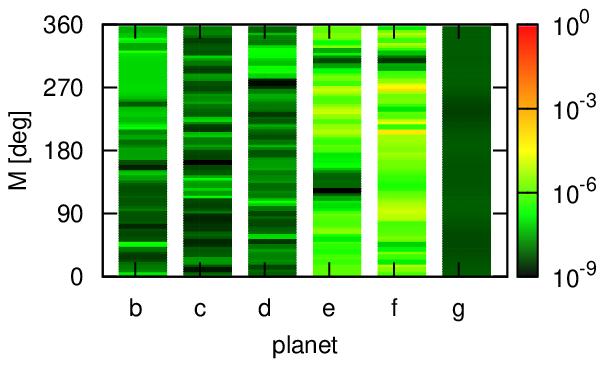} 
  \caption{\label{fig:1}Stability analysis of the circular six-planet solution (S1). The panels show the influence of changing the initial eccentricity (left) and initial mean anomaly (right) on the stability of the complete planetary system. "Red" corresponds to highly unstable orbits while "black" denotes regular motion. Each of the 100 initial conditions per planet were integrated numerically for 1 kyr, for which we then computed the stability criterion $D$ (Eq. \ref{eq:1}).}
\end{figure*}

Starting from S1 we changed only the initial $e$ or $M$ of an individual planet. We then integrated the obtained system for 1000 years using the symplectic SABA2 integrator \citep{laskar2001} with a time step $\tau=0.1$ days. The stability of the system was afterwards estimated using a frequency analysis \citep{laskar1993}. For this we divided the solution into two parts and computed the mean motions $n$ for all planets for each part separately. To obtain the measure of stability ($D$), we calculated the maximum relative change in the difference between both mean motions as
\begin{equation}\label{eq:1}
 D = \max \left| \frac{n^i_1 - n^i_2}{n^i_1}\right|,\qquad i=b,\ldots,g.
\end{equation}
For a stable system, the individual mean motions will not vary with time, i.e. $n_1^i\approx n_2^i$. The results are shown in Fig. \ref{fig:1}. Here "red" denotes strongly chaotic orbits, while "black" corresponds to very regular ones. 

The left panel of Fig. \ref{fig:1} confirms that the stability of the system is crucially dependent on the individual eccentricities of the planets. Only for initial eccentricities very close to zero regular motion can be found ($D\sim 10^{-9}$). Using the right panel of Fig.~\ref{fig:1} it is also possible to identify the reasons for dynamical stability in HD 40307. While the system is not sensitive at all for example to the initial mean anomaly of planet $g$, it crucially depends on $M_e$ and $M_f$. In each column of these planets one finds in Fig. \ref{fig:1} initial conditions that lead to very regular motion ($D\sim 10^{-9}$ for e.g. $M_f(t_0)\approx 310^\circ$) and also conditions that lead to chaotic motion ($D\sim 10^{-2}$ for e.g. $M_f(t_0)\approx 267^\circ$). When analysing the results of the respective integrations, we found that the orbital periods of candidates e and f need to have a ratio of 3:2 to maintain stable motion. This resonant configuration helps to avoid close encounters between the planets due to the libration of the resonance angle $\theta = -2\lambda_e + 3\lambda_f-\omega_e$. We note that this configuration is also maintained over long time intervals. Integrations up to 1 Myr using these initial conditions did not show any signs of chaotic motion.

In contrast, for highly unstable initial conditions the angle $\theta$ circulates, which leads very soon to collisions between some planets or the ejection of one of them.

In conclusion to the discussion on the dynamics of the system we find that -- despite the presence of many planet candidates in the system -- there exists a substantial number of dynamically stable orbits allowed by the data. In particular, orbits with eccentricities close to zero are dynamically favoured due to the existence of the stabilising 3:2 mean motion resonance between candidates e and f. We note that we have only tested solutions based on the minimum masses of the candidates. It is possible that increasing the masses of all planets by a small factor (e.g., $\times$ 1.2) makes the system unstable and could therefore impose interesting upper limits to the masses. Additional RV measurements are necessary to better constrain the eccentricities of the several candidates and derive more robust conclusions on the dynamical features of the system.

\section{Habitability and prospective follow-up}\label{sec:habitability}

\begin{figure*}
\center
\includegraphics[width=0.80\textwidth,clip]{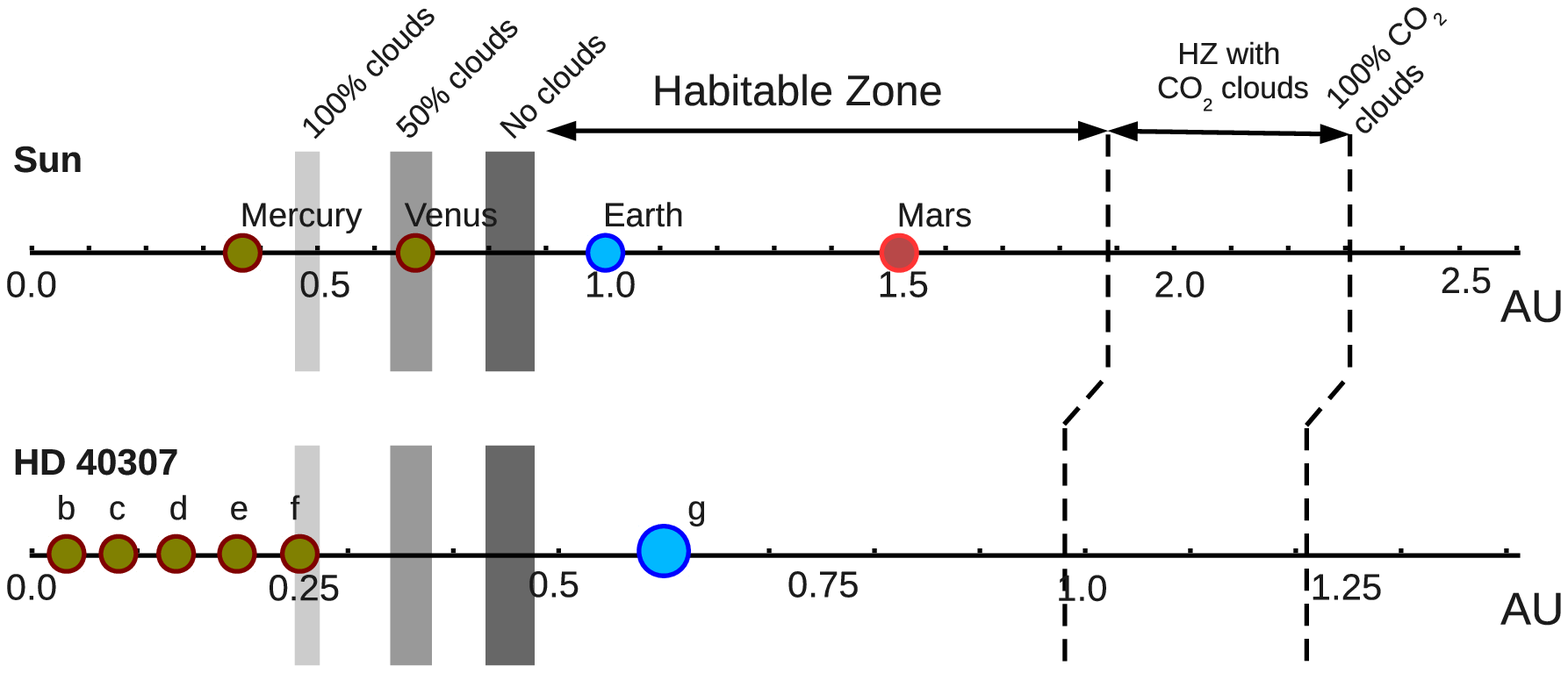}
\caption{Representation of the liquid water habitable zone around HD 40307 as described by \citet{selsis2007} compared to our solar system. Note the compactness of the orbital configuration of the five inner planet candidates around HD 40307 compared to the planets in the inner solar system. }\label{fig:hz}
\end{figure*}

The so-called liquid water habitable zone (HZ) is the area around a star where an Earth-like planet could support liquid water on its surface. For HD 40307, this HZ lies between $0.43$ and $0.85$ AU (see Fig. \ref{fig:hz}). This interval was computed using the recipes given in \citet{selsis2007} and is a function of stellar luminosity and effective temperature. We used the luminosity and temperature estimates given by \citet{ghezzi2010}. With an orbital semi-major axis of $0.60$ AU, the planet candidate HD 40307 g receives $\sim$ 62\% of the radiation the Earth receives from the Sun and lies safely within the HZ of its stellar host. Even though the radiation is somewhat low compared to that received by the Earth, we note that the Earth lies actually reasonably close to the inner boundary of the Sun's HZ (Fig. \ref{fig:hz}).

As is the case with the previously reported planets in the HZs of nearby stars, additional observational information is necessary -- on top of confirming the planetary nature of the three new signals we report in this work -- to decide if HD 40307 g indeed can support water on its surface. In the meantime, and given that it would be a rather massive object for a telluric planet, detailed climatic \citep[e.g.][]{barnes2012} and planetary interior simulations \citep[e.g.][]{korenaga2010,stein2011} can be used to estimate its suitability for supporting liquid water and, perhaps, life. Unlike the other previously reported candidate habitable planets around nearby M dwarfs (e.g., GJ 581 and GJ 667C), HD 40307 g is located farther away from the central star and, like the habitable zone planet candidate Kepler 22b \citep{borucki2012}, it should not suffer from tidal locking either \citep{kasting1993,barnes2009b}.

Given its relatively long-period orbit, the transit probability of planet candidate HD 40307 g is very low \citep[$\sim$0.6\% ][]{charbonneau2007}. Also, previous attempts to detect the transits of the inner candidate HD 40307 b (P $\approx 4.3$ days) have been unsuccessful \citep{gillon2010}. Therefore, if coplanarity is a common feature of planetary systems, the fact that HD 40307 b is not transiting its star bodes ill for the transit chances of any outer companion. HD 40307 is a nearby star (13 pc) compared to the typical target of the Kepler mission (e.g., Kepler 22 is located at $\sim$ 200 pc). Because of this, the star-planet angular separation can be as large as $\sim$ 46 mas. Space-based direct-imaging missions such as Darwin/ESA \citep{cockell2009} or NASA/TPF \citep{lawson2006} aim to reach the sensitivity to detect Earth-sized planets at an inner working angle of $\sim$ 40 mas. Therefore, HD 40307 g would be the first habitable-zone candidate in the super-Earth mass regime that could be targeted by the planned direct imaging observatories.

\section{Discussion}

After examining the wavelength dependence of the signals in the RVs of HD 40307 and their relation to the chromospheric activity indicator (S-index), we easily detect the three super-Earth candidates previously reported by \citet{mayor2009a} using independent data analysis methods. In addition to these, we also confidently detect three new periodic signals in the RVs of this K2.5 V dwarf star. Our analyses of the HARPS RVs indicates the existence of a system with six super-Earth candidates around this star (Table \ref{probabilities}; Fig. \ref{periodograms}) and our dynamical analyses show that long-term stable orbits are allowed within our credibility sets of the parameters. Our results make HD 40307 one of the few known planetary systems with more than five planet candidates: HD 10180 \citep{lovis2011,tuomi2012}, the Kepler-11 six-planet system \citep{lissauer2011}, and the solar system.

A handful of potentially habitable planets have been reported to date. GJ 581 d \citep{mayor2009b} and HD 88512 b \citep{pepe2011} orbit at the extended HZ where thick clouds of CO$_2$ or water would be required to sustain wet atmospheres. As is the case with Kepler-22 b \citep{borucki2012} and GJ 667C c \citep{anglada2012b}, HD 40307 g orbits well within the more classic and conservative definition of the HZ. With a radius of 2.4 $R_\oplus$, Kepler-22 b is likely to be a small-scale version of Neptune rather than a rocky planet with a solid surface \citep{borucki2012}. However, the precise natures of GJ 667C c and HD 40307 g are unknown and they could also be scaled-down versions of Neptune-like planets with thick atmospheres and solid cores. Compared to the candidates orbiting nearby low-mass stars (GJ 581 d, GJ 667C c, HD 85512 b), HD 40307 g is not likely to suffer from tidal-locking -- which improves its chances of hosting an Earth-like climate. Like for the other HZ candidates, it is not yet possible to determine its physical and geochemical properties and a direct-imaging mission seems to be the most promising way of obtaining observational information on these properties.

For stable stars, the HARPS data can be readily used to detect Keplerian periodicities in RVs with amplitudes below 1.0 ms$^{-1}$ \citep{pepe2011}. This accuracy can be additionally improved by obtaining RV measurements using optimized data-reduction techniques \citep[e.g. HARPS-TERRA; ][]{anglada2012a} and analysing the wavelength dependence of the signals. Also, detections should not rely on criteria solely based on the analyses of the model residuals -- this procedure is prone to biases and the weakest signals in the data can easily escape detection \citep[compare e.g. the results of ][]{lovis2011,feroz2011,tuomi2012}.

The ability to detect such low-amplitude signals also introduces a practical problem. The orbits of the planets can no longer be readily compared with large numbers of data points using phase-folded curves (Fig. \ref{orbits}) because -- while the signals may be significant -- their existence cannot be so readily verified by eyesight from such figures. Therefore, when discovering low-mass planets with the radial velocity method, it is necessary to demonstrate the significance of the signals using some other means. The detection criterion we describe, namely, that the radial velocity amplitude is statistically distinguishable from zero, provides an easy way of demonstrating this. The distributions of all amplitude parameters in Fig. \ref{densities} clearly satisfy this criterion and show that all signals we report are indeed significantly present in the data.

Additional measurements are needed to verify the existence and better constrain the orbits and minimum masses of the three new planet candidates reported here. Moreover, given that all signals are of planetary origin, detailed dynamical analyses can additionally constrain the orbits and masses by excluding unstable configurations, and especially, constraining the upper limits of the planetary masses that could then be used to determine the inclinations of their orbits. In this sense, and given the relatively dense dynamical packing of the system, the planet candidates around HD 40307 are likely to have masses close to their minimum values.

The outer planet candidate (HD 40307 g) has a minimum mass of $\sim$ 7 M$_{\oplus}$ and orbits the star within the liquid water HZ. Given its priviledged position comfortably within the HZ, it is likely to remain inside it for the full orbital cycle even if its eccentricity is not strictly zero. A more detailed characterization of this candidate is very unlikely using ground based studies because it is very inlikely to transit the star, and a direct imaging mission seems the most promising way of learning more about its possible atmosphere and life-hosting capabilities. The planetary system around HD 40307 has an architecture radically different from that of the solar system (see Figure \ref{fig:hz}), which indicates that a wide variety of formation histories might allow the emergence of roughly Earth-mass objects in the habitable zones of stars.

\begin{acknowledgements}

M. Tuomi is supported by RoPACS (Rocky Planets Around Cool Stars), a Marie Curie Initial Training Network funded by the European Commission's Seventh Framework Programme. G. Anglada-Escud\'e is supported by the German Ministry of Education and Research under 05A11MG3. E. Gerlach would like to acknowledge the financial support from the DFG research unit FOR584. A. Reiners acknowledges research funding from DFG grant RE1664/9-1. S. Vogt gratefully acknowledges support from NFS grant AST-0307493. This work is based on data obtained from the ESO Science Archive Facility under request number GANGLFGGCE173161. This research has made extensive use of the SIMBAD database, operated at CDS, Strasbourg, France; and the NASA's Astrophysics Data System. We acknowledge the significant efforts of the HARPS-ESO team in improving the instrument and its data reduction pipelines, and obtaining the observations that made this work possible.

\end{acknowledgements}


\appendix

\section{Radial velocities and activity indices}

\longtab{2}{
\begin{longtable}{lccccccc}

\caption{\label{tab:rvs} HARPS-TERRA Differential radial velocities of HD 40307 from the full spectra (F) and the red part (490-680 nm) of the spectra (R). These velocities are in the solar system barycentric reference frame and are corrected for the perspective acceleration effect. The S-index measurements in the Mount Wilson system and corresponding uncertainties (photon noise) are also provided. The rightmost column contains the measurements of the FWHM of the cross correlation function as provided by the HARPS-DRS. Since no uncertainties for the FWHM are provided by the HARPS-DRS, we use the measured standard deviation over all 135 nightly averaged measurements (6.5 m s$^{-1}$) as a measure of the uncertainty of each nightly average used in the analysis.}
\\
\hline\hline
Time & Velocity (F) & Unc. (F) & Velocity (R) & Unc. (R)    & S-index & Unc.       & FWHM        \\
JD & [ms$^{-1}$] & [ms$^{-1}$] & [ms$^{-1}$]  & [ms$^{-1}$] & [--]    & [10$^{-3}$]& [kms$^{-1}$] \\
\hline
\endfirsthead
\caption{Continued.}\\
\hline\hline
Time & Velocity (F) & Unc. (F) & Velocity (R) & Unc. (R)    & S-index & Unc.       & FWHM        \\
JD & [ms$^{-1}$] & [ms$^{-1}$] & [ms$^{-1}$]  & [ms$^{-1}$] & [--]    & [10$^{-3}$]& [kms$^{-1}$] \\
\hline
\endhead
\hline\hline
\endfoot
2452942.82149 & 3.06 & 0.65 & 1.87 & 0.63   & 0.21054 & 0.80   & 5.9071 \\   
2452999.74728 & -2.03 & 0.90 & -1.86 & 0.63 & 0.19181 & 1.34   & 5.9011 \\     
2452999.77714 & 0.37 & 1.00 & -2.45 & 0.67  & 0.19406 & 1.22   & 5.9009 \\    
2453000.76055 & 3.41 & 0.92 & 3.48 & 0.68   & 0.19707 & 1.18   & 5.9001 \\   
2453001.66845 & 5.93 & 0.84 & 5.34 & 0.77   & 0.19646 & 1.19   & 5.8988 \\   
2453002.66863 & 5.44 & 0.86 & 2.99 & 0.80   & 0.20258 & 0.93   & 5.8992 \\   
2453054.59115 & 0.52 & 0.77 & -1.24 & 0.64  & 0.19893 & 1.14   & 5.9008 \\    
2453692.71322 & 0.80 & 0.77 & 2.23 & 0.58   & 0.18624 & 1.09   & 5.8972 \\   
2453692.71565 & 1.49 & 0.65 & 1.04 & 0.50   & 0.18699 & 1.05   & 5.8930 \\   
2453692.71802 & 2.22 & 0.69 & 3.79 & 0.65   & 0.18699 & 1.09   & 5.8948 \\   
2453692.72053 & 2.23 & 0.69 & 4.08 & 0.54   & 0.18726 & 1.05   & 5.8967 \\   
2453692.72300 & 1.31 & 0.55 & 1.25 & 0.52   & 0.18872 & 1.02   & 5.8950 \\   
2453692.72541 & 2.21 & 0.64 & 3.18 & 0.60   & 0.18726 & 1.01   & 5.8956 \\   
2453692.72783 & 0.95 & 0.67 & 1.01 & 0.56   & 0.18277 & 1.00   & 5.8945 \\   
2453692.73029 & 0.95 & 0.64 & 1.66 & 0.65   & 0.18538 & 0.99   & 5.8948 \\   
2453692.73273 & 1.06 & 0.60 & 2.35 & 0.51   & 0.18755 & 1.02   & 5.8939 \\   
2453692.73513 & 2.14 & 0.69 & 2.79 & 0.56   & 0.18738 & 1.02   & 5.8955 \\   
2453692.73762 & 1.60 & 0.65 & 2.30 & 0.41   & 0.19030 & 1.07   & 5.8943 \\   
2453692.74008 & 2.18 & 0.72 & 2.88 & 0.56   & 0.18666 & 1.05   & 5.8954 \\   
2453692.74250 & 1.94 & 0.65 & 2.70 & 0.64   & 0.18378 & 1.05   & 5.8956 \\   
2453692.74496 & 2.48 & 0.74 & 2.06 & 0.67   & 0.18525 & 1.08   & 5.8953 \\   
2453692.74735 & 0.41 & 0.67 & 1.86 & 0.61   & 0.18848 & 0.96   & 5.8923 \\   
2453692.74982 & 0.28 & 0.56 & 1.26 & 0.58   & 0.18663 & 0.96   & 5.8952 \\   
2453692.75226 & 0.21 & 0.63 & 2.53 & 0.61   & 0.18659 & 0.94   & 5.8938 \\   
2453693.61842 & 0.74 & 0.85 & 0.17 & 0.69   & 0.18743 & 1.21   & 5.8961 \\   
2453693.62016 & 2.19 & 1.02 & 1.59 & 0.68   & 0.18634 & 1.20   & 5.8983 \\   
2453693.62191 & 2.20 & 1.05 & 2.31 & 0.70   & 0.18454 & 1.23   & 5.8957 \\   
2453693.62371 & 3.48 & 1.01 & 2.23 & 0.81   & 0.18290 & 1.29   & 5.8990 \\   
2453693.62544 & 1.65 & 0.91 & 2.73 & 0.62   & 0.18509 & 1.29   & 5.8968 \\   
2453693.62717 & 2.08 & 0.95 & 3.29 & 0.71   & 0.18034 & 1.27   & 5.8943 \\   
2453693.62891 & 3.69 & 1.16 & 3.46 & 0.71   & 0.18243 & 1.31   & 5.9009 \\   
2453693.63068 & 0.52 & 1.13 & 2.20 & 0.78   & 0.18539 & 1.26   & 5.8954 \\   
2453693.63240 & 2.05 & 1.01 & 2.00 & 0.86   & 0.18637 & 1.23   & 5.8968 \\   
2453693.63414 & 0.89 & 0.88 & -0.09 & 0.58  & 0.18740 & 1.19   & 5.8964 \\    
2453693.63589 & 1.27 & 0.96 & 0.20 & 0.82   & 0.18430 & 1.22   & 5.8982 \\   
2453693.63762 & 3.08 & 1.10 & 3.56 & 0.84   & 0.18463 & 1.30   & 5.8968 \\   
2453693.63939 & 2.21 & 1.08 & 2.31 & 0.74   & 0.18293 & 1.35   & 5.8979 \\   
2453693.64113 & 1.25 & 1.06 & 0.82 & 0.71   & 0.18577 & 1.33   & 5.8962 \\   
2453693.64290 & 1.80 & 0.89 & 2.94 & 0.73   & 0.18543 & 1.23   & 5.8949 \\   
2453693.64463 & 1.44 & 0.94 & 1.80 & 0.70   & 0.18958 & 1.26   & 5.8931 \\   
2453693.64636 & 1.36 & 0.90 & 1.38 & 0.73   & 0.18340 & 1.26   & 5.8951 \\   
2453693.64812 & 2.99 & 1.08 & 2.47 & 0.67   & 0.18493 & 1.34   & 5.9004 \\   
2453693.64987 & 3.52 & 1.53 & 5.40 & 0.82   & 0.18230 & 1.44   & 5.8967 \\   
2453693.65159 & 0.81 & 1.49 & 1.66 & 0.80   & 0.17806 & 1.51   & 5.9013 \\   
2453693.65339 & 0.63 & 2.01 & 0.13 & 1.10   & 0.17905 & 1.57   & 5.8984 \\   
2453693.65512 & 2.11 & 1.36 & 1.99 & 0.71   & 0.18354 & 1.57   & 5.8990 \\   
2453693.65688 & 1.41 & 1.48 & 0.52 & 0.89   & 0.18541 & 1.55   & 5.8991 \\   
2453693.65862 & 3.47 & 1.48 & 3.65 & 1.10   & 0.18220 & 1.46   & 5.9005 \\   
2453693.66035 & 3.56 & 1.36 & 5.18 & 0.80   & 0.18024 & 1.41   & 5.8990 \\   
2453694.71446 & 3.68 & 0.53 & 3.92 & 0.49   & 0.18893 & 0.91   & 5.8978 \\   
2453694.71838 & 2.80 & 0.65 & 2.62 & 0.65   & 0.18649 & 0.94   & 5.8975 \\   
2453694.72221 & 3.34 & 0.52 & 3.11 & 0.45   & 0.18729 & 0.91   & 5.8989 \\   
2453694.72598 & 2.99 & 0.52 & 3.83 & 0.49   & 0.18586 & 0.97   & 5.8969 \\   
2453694.72992 & 3.04 & 0.54 & 3.06 & 0.49   & 0.18547 & 0.95   & 5.8977 \\   
2453694.73376 & 3.32 & 0.60 & 2.80 & 0.56   & 0.18541 & 0.96   & 5.8988 \\   
2453694.73761 & 2.94 & 0.62 & 2.90 & 0.57   & 0.18763 & 0.99   & 5.8981 \\   
2453694.74126 & 4.09 & 0.69 & 4.87 & 0.69   & 0.18405 & 1.11   & 5.9007 \\   
2453694.74522 & 3.83 & 1.07 & 5.65 & 0.66   & 0.18498 & 1.28   & 5.8992 \\   
2453694.74905 & 3.44 & 0.83 & 3.34 & 0.65   & 0.18557 & 1.22   & 5.8947 \\   
2453695.68438 & 4.15 & 0.42 & 3.92 & 0.51   & 0.19010 & 0.68   & 5.8950 \\   
2453695.68817 & 4.61 & 0.43 & 4.97 & 0.35   & 0.19201 & 0.70   & 5.8920 \\   
2453695.69201 & 3.89 & 0.42 & 4.10 & 0.41   & 0.19170 & 0.70   & 5.8944 \\   
2453695.69592 & 4.00 & 0.39 & 4.50 & 0.35   & 0.18987 & 0.70   & 5.8958 \\   
2453695.69972 & 3.54 & 0.39 & 4.35 & 0.38   & 0.19025 & 0.67   & 5.8916 \\   
2453695.70350 & 4.89 & 0.40 & 5.82 & 0.34   & 0.19123 & 0.67   & 5.8919 \\   
2453695.70730 & 3.90 & 0.39 & 4.34 & 0.42   & 0.18864 & 0.68   & 5.8929 \\   
2453695.71114 & 4.42 & 0.45 & 5.34 & 0.45   & 0.19049 & 0.69   & 5.8916 \\   
2453695.71509 & 4.68 & 0.37 & 4.87 & 0.38   & 0.18974 & 0.71   & 5.8936 \\   
2453696.68945 & 4.81 & 0.50 & 4.89 & 0.51   & 0.18766 & 0.80   & 5.8962 \\   
2453696.69326 & 4.42 & 0.47 & 4.65 & 0.54   & 0.18826 & 0.80   & 5.8962 \\   
2453696.69706 & 3.57 & 0.53 & 3.21 & 0.61   & 0.18783 & 0.83   & 5.8972 \\   
2453696.70094 & 2.82 & 0.58 & 3.66 & 0.56   & 0.18688 & 0.82   & 5.8932 \\   
2453696.70477 & 2.92 & 0.43 & 2.97 & 0.55   & 0.18649 & 0.78   & 5.8951 \\   
2453696.70861 & 3.63 & 0.49 & 4.14 & 0.47   & 0.18861 & 0.82   & 5.8952 \\   
2453696.71244 & 3.60 & 0.43 & 3.33 & 0.41   & 0.18813 & 0.82   & 5.8962 \\   
2453696.71629 & 2.84 & 0.47 & 2.38 & 0.59   & 0.18730 & 0.77   & 5.8987 \\   
2453696.72009 & 3.20 & 0.46 & 3.25 & 0.36   & 0.18662 & 0.78   & 5.8954 \\   
2453696.72395 & 3.90 & 0.37 & 3.79 & 0.34   & 0.18783 & 0.78   & 5.8953 \\   
2453697.71626 & -1.12 & 2.69 & -2.24 & 1.05 & 0.18662 & 1.98   & 5.9028 \\     
2453697.72027 & -0.35 & 2.63 & 0.43 & 0.98  & 0.18598 & 1.89   & 5.9054 \\    
2453697.72392 & -1.80 & 1.96 & -0.12 & 1.16 & 0.18833 & 1.64   & 5.9018 \\     
2453697.72793 & -0.86 & 1.84 & -0.13 & 0.89 & 0.18536 & 1.62   & 5.9035 \\     
2453697.73183 & 0.77 & 1.78 & 1.05 & 0.86   & 0.18136 & 1.58   & 5.9016 \\   
2453697.73547 & -0.27 & 1.42 & -0.81 & 0.82 & 0.18244 & 1.47   & 5.8981 \\     
2453697.73933 & 0.61 & 1.77 & 1.57 & 0.91   & 0.18004 & 1.53   & 5.8971 \\   
2453697.74308 & -2.48 & 2.31 & -0.45 & 1.15 & 0.18324 & 1.73   & 5.8946 \\     
2453697.74695 & 1.14 & 2.43 & 1.99 & 0.73   & 0.18138 & 1.82   & 5.8990 \\   
2453698.72214 & -3.81 & 0.69 & -3.27 & 0.64 & 0.18974 & 1.04   & 5.8982 \\     
2453698.72601 & -3.52 & 0.61 & -2.95 & 0.55 & 0.18798 & 1.02   & 5.8981 \\     
2453698.72978 & -3.70 & 0.60 & -3.84 & 0.59 & 0.18900 & 1.06   & 5.8955 \\     
2453698.73369 & -2.89 & 0.69 & -4.03 & 0.53 & 0.18913 & 1.09   & 5.8996 \\     
2453698.73753 & -2.12 & 0.69 & -1.74 & 0.55 & 0.18817 & 1.07   & 5.8998 \\     
2453698.74132 & -3.49 & 0.60 & -3.07 & 0.46 & 0.18917 & 1.02   & 5.9006 \\     
2453698.74516 & -2.97 & 0.65 & -3.20 & 0.57 & 0.18742 & 1.01   & 5.9005 \\     
2453698.74902 & -2.96 & 0.68 & -2.54 & 0.57 & 0.18875 & 1.14   & 5.8996 \\     
2453698.75274 & -2.41 & 0.74 & -0.18 & 0.67 & 0.18695 & 1.15   & 5.8985 \\     
2453699.72090 & -3.09 & 0.56 & -1.73 & 0.49 & 0.18900 & 0.93   & 5.8939 \\     
2453699.72453 & -3.19 & 0.43 & -2.43 & 0.43 & 0.19005 & 0.86   & 5.8982 \\     
2453699.72836 & -1.84 & 0.52 & -1.78 & 0.55 & 0.19065 & 0.87   & 5.8967 \\     
2453699.73223 & -3.11 & 0.48 & -2.28 & 0.55 & 0.19062 & 0.88   & 5.8955 \\     
2453699.73606 & -3.85 & 0.53 & -2.10 & 0.50 & 0.18777 & 0.91   & 5.8977 \\     
2453699.73995 & -2.79 & 0.44 & -2.21 & 0.36 & 0.18990 & 0.89   & 5.8967 \\     
2453699.74375 & -3.02 & 0.58 & -2.62 & 0.50 & 0.19189 & 0.92   & 5.8971 \\     
2453699.74760 & -2.77 & 0.48 & -1.93 & 0.44 & 0.19041 & 0.86   & 5.8969 \\     
2453699.75138 & -3.02 & 0.51 & -1.48 & 0.46 & 0.19000 & 0.92   & 5.8982 \\     
2453700.75276 & -3.18 & 0.52 & -2.35 & 0.52 & 0.18997 & 0.92   & 5.8975 \\     
2453700.75646 & -3.61 & 0.49 & -2.09 & 0.49 & 0.19447 & 0.86   & 5.8977 \\     
2453700.76034 & -3.57 & 0.56 & -2.20 & 0.50 & 0.18906 & 0.88   & 5.8982 \\     
2453700.76415 & -3.57 & 0.55 & -1.98 & 0.49 & 0.18985 & 0.87   & 5.8983 \\     
2453700.76799 & -3.88 & 0.50 & -4.00 & 0.51 & 0.19370 & 0.85   & 5.8994 \\     
2453700.77189 & -3.74 & 0.47 & -1.81 & 0.46 & 0.19120 & 0.84   & 5.8970 \\     
2453700.77569 & -2.89 & 0.48 & -1.25 & 0.38 & 0.19242 & 0.84   & 5.8975 \\     
2453700.77959 & -3.74 & 0.44 & -2.46 & 0.42 & 0.19187 & 0.81   & 5.8977 \\     
2453700.78341 & -3.79 & 0.45 & -2.33 & 0.45 & 0.19137 & 0.81   & 5.8961 \\     
2453721.72774 & -0.58 & 0.51 & -0.12 & 0.55 & 0.19012 & 0.76   & 5.8957 \\     
2453721.73151 & -0.02 & 0.40 & 0.13 & 0.41  & 0.18984 & 0.77   & 5.8944 \\    
2453721.73536 & 0.15 & 0.46 & 0.67 & 0.45   & 0.19185 & 0.75   & 5.8928 \\   
2453721.73921 & -0.61 & 0.45 & -0.60 & 0.48 & 0.19262 & 0.80   & 5.8958 \\     
2453721.74307 & -0.22 & 0.47 & 0.32 & 0.47  & 0.18883 & 0.83   & 5.8943 \\    
2453721.74696 & -0.06 & 0.46 & 1.34 & 0.39  & 0.18745 & 0.86   & 5.8951 \\    
2453721.75079 & -0.15 & 0.45 & 0.18 & 0.46  & 0.18779 & 0.85   & 5.8937 \\    
2453721.75459 & -0.75 & 0.55 & 2.14 & 0.52  & 0.19036 & 0.84   & 5.8949 \\    
2453721.75842 & -0.58 & 0.46 & -0.02 & 0.54 & 0.18910 & 0.80   & 5.8954 \\     
2453721.76230 & -1.00 & 0.46 & -0.48 & 0.40 & 0.18967 & 0.78   & 5.8946 \\     
2453722.72704 & 0.59 & 0.30 & 1.65 & 0.30   & 0.18982 & 0.78   & 5.8951 \\   
2453722.73077 & 1.55 & 0.46 & 1.91 & 0.44   & 0.18878 & 0.79   & 5.8967 \\   
2453722.73453 & 0.35 & 0.46 & 1.69 & 0.53   & 0.18990 & 0.82   & 5.8951 \\   
2453722.73839 & 0.33 & 0.46 & 1.14 & 0.46   & 0.19035 & 0.82   & 5.8979 \\   
2453722.74226 & 1.81 & 0.40 & 1.91 & 0.39   & 0.19051 & 0.80   & 5.8961 \\   
2453722.74610 & 0.01 & 0.44 & 0.70 & 0.49   & 0.19076 & 0.84   & 5.8970 \\   
2453722.74997 & 0.38 & 0.44 & 1.69 & 0.50   & 0.19003 & 0.82   & 5.8948 \\   
2453722.75377 & 1.66 & 0.45 & 1.47 & 0.42   & 0.19084 & 0.81   & 5.8982 \\   
2453722.75760 & 1.28 & 0.48 & 1.47 & 0.48   & 0.19038 & 0.81   & 5.8951 \\   
2453722.76143 & 1.91 & 0.44 & 3.53 & 0.39   & 0.18872 & 0.84   & 5.8963 \\   
2453724.68964 & -1.30 & 0.44 & -0.76 & 0.48 & 0.19170 & 0.76   & 5.8950 \\     
2453724.69350 & -2.69 & 0.41 & -1.51 & 0.38 & 0.19169 & 0.74   & 5.8960 \\     
2453724.69729 & -1.24 & 0.42 & -0.55 & 0.36 & 0.19028 & 0.73   & 5.8947 \\     
2453724.70115 & -1.64 & 0.41 & -1.66 & 0.47 & 0.19108 & 0.74   & 5.8947 \\     
2453724.70501 & -1.53 & 0.42 & -1.56 & 0.36 & 0.19259 & 0.72   & 5.8937 \\     
2453724.70875 & -2.23 & 0.48 & -0.74 & 0.48 & 0.18971 & 0.73   & 5.8956 \\     
2453724.71265 & -2.14 & 0.40 & -2.12 & 0.44 & 0.19014 & 0.72   & 5.8939 \\     
2453724.71648 & -1.42 & 0.36 & -1.32 & 0.36 & 0.18899 & 0.74   & 5.8934 \\     
2453724.72034 & -0.98 & 0.40 & -0.98 & 0.45 & 0.19086 & 0.76   & 5.8977 \\     
2453724.72410 & -1.83 & 0.44 & -1.13 & 0.58 & 0.19198 & 0.78   & 5.8951 \\     
2453725.69712 & -0.10 & 0.41 & -0.90 & 0.48 & 0.18874 & 0.77   & 5.8956 \\     
2453725.70098 & 0.12 & 0.52 & 1.06 & 0.55   & 0.19180 & 0.77   & 5.8958 \\   
2453725.70477 & 0.61 & 0.43 & 1.15 & 0.47   & 0.19195 & 0.75   & 5.8951 \\   
2453725.70855 & -0.07 & 0.43 & 0.62 & 0.48  & 0.19391 & 0.79   & 5.8961 \\    
2453725.71242 & 0.27 & 0.44 & 1.30 & 0.47   & 0.19424 & 0.80   & 5.8957 \\   
2453725.71626 & 0.26 & 0.44 & 2.37 & 0.47   & 0.19498 & 0.78   & 5.8966 \\   
2453725.72006 & 0.30 & 0.43 & 0.91 & 0.49   & 0.19641 & 0.78   & 5.8966 \\   
2453725.72392 & -0.09 & 0.48 & 0.68 & 0.32  & 0.19523 & 0.79   & 5.8957 \\    
2453725.72779 & 0.56 & 0.39 & 0.64 & 0.45   & 0.19569 & 0.79   & 5.8976 \\   
2453725.73161 & 0.24 & 0.41 & 0.72 & 0.43   & 0.19779 & 0.76   & 5.8972 \\   
2453726.69222 & -1.23 & 0.42 & -0.68 & 0.36 & 0.19735 & 0.75   & 5.8948 \\     
2453726.69605 & -0.39 & 0.52 & -0.56 & 0.60 & 0.19193 & 0.90   & 5.8988 \\     
2453726.69999 & -1.04 & 0.49 & -0.95 & 0.40 & 0.19540 & 0.85   & 5.8981 \\     
2453726.70385 & -0.90 & 0.44 & -0.01 & 0.46 & 0.19450 & 0.81   & 5.8965 \\     
2453726.70759 & -0.42 & 0.33 & -0.08 & 0.38 & 0.19633 & 0.79   & 5.8978 \\     
2453726.71152 & -1.51 & 0.44 & -0.81 & 0.38 & 0.19472 & 0.79   & 5.8968 \\     
2453726.71531 & -1.29 & 0.40 & -0.61 & 0.38 & 0.19476 & 0.77   & 5.8955 \\     
2453726.71915 & -1.78 & 0.37 & -1.10 & 0.37 & 0.19419 & 0.76   & 5.8978 \\     
2453726.72297 & -1.16 & 0.39 & -0.65 & 0.39 & 0.19675 & 0.73   & 5.8935 \\     
2453726.72682 & -1.51 & 0.41 & -1.61 & 0.40 & 0.19529 & 0.72   & 5.8939 \\     
2453727.67219 & -3.98 & 0.45 & -1.95 & 0.39 & 0.19549 & 0.72   & 5.8951 \\     
2453727.67589 & -4.74 & 0.47 & -4.11 & 0.38 & 0.19477 & 0.75   & 5.8960 \\     
2453727.67976 & -4.06 & 0.45 & -1.58 & 0.36 & 0.19628 & 0.79   & 5.8974 \\     
2453727.68357 & -3.77 & 0.33 & -2.71 & 0.28 & 0.19665 & 0.74   & 5.8950 \\     
2453727.68745 & -4.66 & 0.40 & -4.20 & 0.46 & 0.19388 & 0.73   & 5.8972 \\     
2453727.69132 & -4.52 & 0.46 & -3.41 & 0.42 & 0.19491 & 0.74   & 5.8950 \\     
2453727.69508 & -4.16 & 0.45 & -3.58 & 0.50 & 0.19505 & 0.72   & 5.8957 \\     
2453727.69891 & -3.62 & 0.41 & -2.66 & 0.44 & 0.19700 & 0.73   & 5.8965 \\     
2453727.70278 & -3.68 & 0.43 & -3.30 & 0.47 & 0.19510 & 0.74   & 5.8958 \\     
2453727.70654 & -4.41 & 0.39 & -4.00 & 0.36 & 0.19613 & 0.79   & 5.8973 \\     
2453728.75882 & -7.30 & 0.39 & -6.77 & 0.34 & 0.19571 & 0.67   & 5.8944 \\     
2453728.76261 & -6.96 & 0.41 & -5.91 & 0.46 & 0.19505 & 0.66   & 5.8930 \\     
2453728.76641 & -6.16 & 0.43 & -5.73 & 0.49 & 0.19493 & 0.69   & 5.8946 \\     
2453728.77026 & -6.29 & 0.41 & -5.55 & 0.47 & 0.19498 & 0.66   & 5.8923 \\     
2453728.77409 & -6.17 & 0.40 & -5.04 & 0.39 & 0.19540 & 0.68   & 5.8932 \\     
2453729.73763 & -2.36 & 0.39 & -1.76 & 0.39 & 0.19499 & 0.78   & 5.8962 \\     
2453729.74153 & -2.48 & 0.41 & -2.47 & 0.48 & 0.19439 & 0.76   & 5.8949 \\     
2453729.74533 & -2.47 & 0.42 & -2.91 & 0.45 & 0.19289 & 0.75   & 5.8951 \\     
2453729.74917 & -2.15 & 0.43 & -1.29 & 0.40 & 0.19458 & 0.73   & 5.8953 \\     
2453729.75302 & -2.08 & 0.34 & -1.45 & 0.34 & 0.19526 & 0.74   & 5.8935 \\     
2453729.75680 & -2.30 & 0.47 & -2.07 & 0.51 & 0.19429 & 0.73   & 5.8943 \\     
2453729.76067 & -2.53 & 0.46 & -1.94 & 0.41 & 0.19421 & 0.73   & 5.8953 \\     
2453729.76447 & -3.02 & 0.42 & -2.98 & 0.42 & 0.19556 & 0.72   & 5.8954 \\     
2453729.76830 & -2.92 & 0.39 & -1.79 & 0.38 & 0.19425 & 0.69   & 5.8945 \\     
2453729.77221 & -1.97 & 0.35 & -1.33 & 0.27 & 0.19620 & 0.72   & 5.8940 \\     
2453757.58857 & -2.31 & 0.45 & -2.99 & 0.47 & 0.19530 & 0.74   & 5.8963 \\     
2453757.59426 & -2.54 & 0.39 & -3.14 & 0.45 & 0.19347 & 0.68   & 5.8958 \\     
2453758.57622 & -5.29 & 0.31 & -5.14 & 0.38 & 0.19282 & 0.52   & 5.8966 \\     
2453759.58537 & -3.83 & 0.32 & -4.30 & 0.38 & 0.19282 & 0.49   & 5.8945 \\     
2453760.66690 & -0.17 & 0.39 & -0.06 & 0.34 & 0.19132 & 0.63   & 5.8967 \\     
2453761.60321 & -0.77 & 0.34 & -1.16 & 0.36 & 0.19147 & 0.51   & 5.8962 \\     
2453762.59317 & -2.54 & 0.34 & -3.68 & 0.31 & 0.19299 & 0.65   & 5.8957 \\     
2453763.59467 & -4.14 & 0.28 & -4.40 & 0.35 & 0.19228 & 0.46   & 5.8954 \\     
2453764.61673 & -2.63 & 0.27 & -2.38 & 0.30 & 0.19295 & 0.40   & 5.8924 \\     
2453765.60207 & -4.38 & 0.29 & -4.72 & 0.35 & 0.19303 & 0.42   & 5.8946 \\     
2453782.55901 & 2.57 & 0.32 & 1.67 & 0.38   & 0.19276 & 0.52   & 5.8978 \\   
2453784.58651 & -5.98 & 0.37 & -6.03 & 0.36 & 0.19551 & 0.62   & 5.8974 \\     
2453786.61145 & -4.44 & 0.33 & -5.15 & 0.42 & 0.19704 & 0.53   & 5.9009 \\     
2453788.60032 & -5.19 & 0.36 & -5.57 & 0.44 & 0.19740 & 0.56   & 5.9009 \\     
2453790.59110 & 1.27 & 0.31 & 0.92 & 0.32   & 0.19900 & 0.50   & 5.9011 \\   
2453810.58376 & -1.11 & 0.30 & -2.02 & 0.37 & 0.19762 & 0.53   & 5.8990 \\     
2453811.57335 & 1.20 & 0.29 & 1.35 & 0.30   & 0.19763 & 0.50   & 5.9006 \\   
2453811.59677 & 2.28 & 0.34 & 2.13 & 0.43   & 0.19845 & 0.52   & 5.9012 \\   
2453812.55900 & 2.83 & 0.38 & 2.37 & 0.44   & 0.19635 & 0.61   & 5.9032 \\   
2453814.60118 & -3.69 & 0.34 & -4.49 & 0.41 & 0.19592 & 0.54   & 5.9006 \\     
2453817.59376 & 2.64 & 0.26 & 2.81 & 0.39   & 0.19563 & 0.60   & 5.8995 \\   
2453829.51773 & 3.30 & 0.79 & 3.40 & 0.66   & 0.19146 & 1.10   & 5.9002 \\   
2453829.52316 & 2.46 & 0.54 & 2.04 & 0.54   & 0.19708 & 0.99   & 5.9003 \\   
2453831.51533 & -1.27 & 0.24 & -1.84 & 0.38 & 0.19487 & 0.73   & 5.8947 \\     
2453831.52040 & -1.61 & 0.45 & -1.33 & 0.47 & 0.19360 & 0.75   & 5.8967 \\     
2453835.51806 & 0.01 & 0.53 & -0.23 & 0.49  & 0.19214 & 0.87   & 5.9006 \\    
2453835.52259 & 0.52 & 0.62 & 0.00 & 0.59   & 0.19646 & 0.99   & 5.9021 \\   
2453861.46191 & 0.44 & 0.31 & 0.36 & 0.39   & 0.19838 & 0.67   & 5.9034 \\   
2453862.46764 & -3.31 & 0.32 & -4.20 & 0.32 & 0.19787 & 0.56   & 5.9024 \\     
2453863.46235 & -4.49 & 0.32 & -3.93 & 0.35 & 0.19568 & 0.52   & 5.8993 \\     
2453864.45691 & -1.39 & 0.34 & -2.25 & 0.36 & 0.19687 & 0.67   & 5.9012 \\     
2453865.45514 & -4.05 & 0.30 & -4.40 & 0.31 & 0.19588 & 0.48   & 5.8962 \\     
2453866.45643 & -4.12 & 0.33 & -4.67 & 0.39 & 0.19632 & 0.51   & 5.8964 \\     
2453867.47453 & -1.7 & 0.24 & -2.06 & 0.30  & 0.19410 & 0.53   & 5.8989 \\    
2453868.45499 & 1.50 & 0.35 & 0.77 & 0.34   & 0.19502 & 0.57   & 5.9005 \\   
2453869.45631 & -0.41 & 0.31 & -0.44 & 0.31 & 0.19488 & 0.48   & 5.8964 \\     
2453870.45490 & -2.84 & 0.41 & -2.84 & 0.42 & 0.19138 & 0.67   & 5.9018 \\     
2453871.45587 & -3.91 & 0.28 & -3.86 & 0.35 & 0.19317 & 0.57   & 5.8999 \\     
2453882.46166 & 0.88 & 0.97 & 2.20 & 0.57   & 0.20431 & 1.33   & 5.9084 \\   
2453882.46547 & 1.32 & 1.07 & 1.06 & 0.73   & 0.20471 & 1.33   & 5.9085 \\   
2453882.46940 & 2.58 & 0.91 & 3.30 & 0.72   & 0.20460 & 1.36   & 5.9110 \\   
2453974.92261 & 1.42 & 0.48 & 1.03 & 0.50   & 0.19426 & 0.80   & 5.9085 \\   
2453980.90950 & 4.83 & 0.55 & 5.02 & 0.52   & 0.19708 & 0.88   & 5.9069 \\   
2453980.91504 & 3.58 & 0.48 & 3.58 & 0.56   & 0.20014 & 0.85   & 5.9082 \\   
2453981.91896 & 3.66 & 0.84 & 3.17 & 0.63   & 0.20339 & 1.19   & 5.9069 \\   
2453981.92156 & 3.45 & 0.86 & 4.01 & 0.51   & 0.20369 & 1.27   & 5.9085 \\   
2453981.92423 & 4.09 & 1.06 & 4.23 & 0.71   & 0.20454 & 1.29   & 5.9108 \\   
2453981.92686 & 3.63 & 1.02 & 3.58 & 0.71   & 0.20040 & 1.25   & 5.9101 \\   
2454047.80472 & -4.61 & 0.42 & -5.02 & 0.56 & 0.19225 & 0.64   & 5.9048 \\     
2454049.80059 & 0.92 & 0.37 & 1.31 & 0.44   & 0.19283 & 0.58   & 5.9020 \\   
2454051.81528 & -1.00 & 0.44 & -0.88 & 0.51 & 0.18873 & 0.75   & 5.9011 \\     
2454053.82521 & -1.36 & 0.38 & -0.77 & 0.4  & 0.19429 & 0.56   & 5.8986 \\    
2454055.83011 & -5.22 & 0.30 & -4.90 & 0.43 & 0.19127 & 0.57   & 5.8995 \\     
2454077.73742 & -2.98 & 0.30 & -2.73 & 0.36 & 0.19573 & 0.52   & 5.9004 \\     
2454079.70672 & 6.02 & 0.37 & 5.88 & 0.45   & 0.19301 & 0.67   & 5.9014 \\   
2454081.71480 & 2.25 & 0.53 & 3.26 & 0.54   & 0.19077 & 0.84   & 5.9024 \\   
2454083.76354 & 3.32 & 0.50 & 4.14 & 0.54   & 0.19034 & 0.89   & 5.9033 \\   
2454115.61193 & -3.41 & 2.12 & -4.25 & 1.01 & 0.19494 & 1.81   & 5.9101 \\     
2454115.61771 & -1.12 & 1.23 & -0.51 & 0.73 & 0.19225 & 1.53   & 5.9089 \\     
2454120.67504 & 1.62 & 0.49 & 1.60 & 0.49   & 0.19653 & 0.80   & 5.9097 \\   
2454120.68114 & -0.10 & 0.47 & 0.00 & 0.47  & 0.19609 & 0.78   & 5.9071 \\    
2454121.64059 & 0.46 & 0.44 & 0.13 & 0.52   & 0.19680 & 0.85   & 5.9071 \\   
2454121.64644 & -0.58 & 0.43 & -0.74 & 0.36 & 0.19683 & 0.81   & 5.9071 \\     
2454136.60823 & 0.91 & 0.34 & 0.82 & 0.47   & 0.19397 & 0.66   & 5.9017 \\   
2454136.61374 & 0.92 & 0.34 & 1.31 & 0.39   & 0.19206 & 0.62   & 5.9044 \\   
2454137.59986 & -0.35 & 0.32 & -0.59 & 0.35 & 0.19550 & 0.65   & 5.9026 \\     
2454137.60532 & -0.51 & 0.39 & -0.90 & 0.41 & 0.19380 & 0.66   & 5.9041 \\     
2454141.60605 & -1.17 & 0.56 & -1.19 & 0.52 & 0.20137 & 1.00   & 5.9098 \\     
2454141.61179 & -1.12 & 0.37 & -1.02 & 0.47 & 0.19861 & 0.96   & 5.9094 \\     
2454143.55743 & 2.02 & 0.46 & 1.29 & 0.45   & 0.19866 & 0.85   & 5.9058 \\   
2454143.56322 & 1.79 & 0.49 & 2.09 & 0.47   & 0.20033 & 0.82   & 5.9065 \\   
2454167.54364 & 1.07 & 0.31 & -0.43 & 0.36  & 0.19678 & 0.50   & 5.9014 \\    
2454169.51793 & -2.88 & 0.23 & -3.25 & 0.28 & 0.19639 & 0.44   & 5.8994 \\     
2454171.53889 & -5.73 & 0.32 & -5.89 & 0.37 & 0.19491 & 0.45   & 5.9024 \\     
2454173.54371 & -4.21 & 0.35 & -4.44 & 0.45 & 0.19714 & 0.47   & 5.9050 \\     
2454194.49833 & -3.09 & 0.41 & -3.97 & 0.41 & 0.19940 & 0.78   & 5.9030 \\     
2454194.50370 & -1.94 & 0.46 & -2.26 & 0.47 & 0.19719 & 0.79   & 5.9040 \\     
2454196.49825 & 4.68 & 0.51 & 3.80 & 0.53   & 0.19934 & 0.91   & 5.9051 \\   
2454196.50338 & 3.95 & 0.55 & 2.79 & 0.52   & 0.20203 & 0.90   & 5.9041 \\   
2454197.49435 & -0.12 & 0.48 & 0.25 & 0.49  & 0.19868 & 0.84   & 5.9009 \\    
2454197.49969 & -0.76 & 0.45 & -0.68 & 0.42 & 0.20035 & 0.85   & 5.9054 \\     
2454198.50730 & -4.55 & 0.41 & -4.84 & 0.46 & 0.19857 & 0.75   & 5.9048 \\     
2454198.51264 & -4.27 & 0.50 & -4.64 & 0.57 & 0.19747 & 0.78   & 5.9049 \\     
2454199.49468 & -2.93 & 0.40 & -2.46 & 0.44 & 0.19753 & 0.71   & 5.9034 \\     
2454199.49995 & -2.45 & 0.40 & -2.39 & 0.44 & 0.19613 & 0.72   & 5.9054 \\     
2454200.48249 & -0.62 & 0.43 & -1.47 & 0.54 & 0.19802 & 0.75   & 5.9033 \\     
2454200.48765 & 0.70 & 0.48 & 0.74 & 0.46   & 0.19712 & 0.76   & 5.9055 \\   
2454202.49346 & -0.47 & 0.57 & -1.44 & 0.61 & 0.19852 & 0.96   & 5.9057 \\     
2454202.49888 & -0.01 & 0.58 & -0.71 & 0.60 & 0.19647 & 0.97   & 5.9089 \\     
2454225.48698 & 5.40 & 0.98 & 4.79 & 0.64   & 0.20354 & 1.39   & 5.9095 \\   
2454225.49172 & 5.38 & 1.43 & 2.80 & 0.84   & 0.20578 & 1.79   & 5.9100 \\   
2454228.47911 & -0.67 & 1.04 & -1.47 & 0.76 & 0.20000 & 1.47   & 5.9078 \\     
2454228.48399 & -0.99 & 0.68 & -0.71 & 0.51 & 0.19916 & 1.22   & 5.9125 \\     
2454229.48224 & -1.41 & 0.59 & -2.26 & 0.57 & 0.20195 & 1.29   & 5.9143 \\     
2454229.48703 & -0.94 & 0.89 & -2.22 & 0.63 & 0.19863 & 1.23   & 5.9135 \\     
2454231.46180 & 0.43 & 0.55 & 0.44 & 0.52   & 0.20424 & 1.03   & 5.9061 \\   
2454231.46674 & 1.26 & 0.58 & 0.90 & 0.56   & 0.20124 & 0.99   & 5.9092 \\   
2454232.47095 & 0.22 & 0.64 & 0.45 & 0.55   & 0.20078 & 1.15   & 5.9063 \\   
2454232.47594 & -1.01 & 0.65 & -0.76 & 0.49 & 0.20121 & 1.11   & 5.9063 \\     
2454233.45885 & 0.96 & 1.40 & -0.99 & 0.90  & 0.20153 & 1.61   & 5.9094 \\    
2454233.46394 & -0.30 & 2.47 & -0.49 & 1.07 & 0.20526 & 1.98   & 5.9142 \\     
2454234.48796 & 1.42 & 0.76 & 1.29 & 0.54   & 0.20107 & 1.21   & 5.9081 \\   
2454234.49299 & 1.05 & 0.71 & 2.35 & 0.56   & 0.20014 & 1.13   & 5.9105 \\   
2454315.88827 & -4.59 & 2.00 & -6.20 & 1.00 & 0.20437 & 1.80   & 5.9209 \\     
2454319.90884 & -2.15 & 0.44 & -3.79 & 0.56 & 0.19446 & 0.75   & 5.9142 \\     
2454342.87431 & 0.47 & 0.44 & -0.90 & 0.55  & 0.19026 & 0.54   & 5.9109 \\    
2454346.90515 & 5.00 & 1.31 & 5.48 & 0.87   & 0.19405 & 1.46   & 5.9092 \\   
2454346.90736 & 3.80 & 1.30 & 3.11 & 0.74   & 0.19493 & 1.41   & 5.9077 \\   
2454346.90954 & 4.11 & 1.36 & 3.34 & 0.84   & 0.19976 & 1.53   & 5.9083 \\   
2454346.91177 & 4.40 & 1.11 & 2.73 & 0.66   & 0.18904 & 1.44   & 5.9095 \\   
2454346.91396 & 5.16 & 1.19 & 4.18 & 0.65   & 0.19593 & 1.47   & 5.9069 \\   
2454347.88542 & 5.81 & 1.35 & 4.70 & 1.00   & 0.19801 & 1.40   & 5.9038 \\   
2454347.88751 & 5.61 & 1.21 & 6.49 & 0.79   & 0.19498 & 1.37   & 5.9003 \\   
2454347.88962 & 4.91 & 1.44 & 5.42 & 0.75   & 0.19431 & 1.48   & 5.9045 \\   
2454347.89172 & 5.52 & 1.45 & 5.82 & 0.82   & 0.19381 & 1.54   & 5.9079 \\   
2454347.89390 & 5.39 & 1.11 & 6.90 & 0.70   & 0.19764 & 1.31   & 5.9041 \\   
2454349.87735 & 2.77 & 0.39 & 2.51 & 0.48   & 0.19545 & 0.54   & 5.9071 \\   
2454386.74976 & 3.17 & 0.39 & 2.23 & 0.47   & 0.19873 & 0.53   & 5.9141 \\   
2454387.78961 & 2.92 & 0.39 & 1.95 & 0.44   & 0.19570 & 0.48   & 5.9090 \\   
2454390.79759 & 2.81 & 0.39 & 2.45 & 0.52   & 0.19600 & 0.55   & 5.9045 \\   
2454392.73519 & -2.61 & 0.45 & -4.07 & 0.54 & 0.19740 & 0.69   & 5.9087 \\     
2454393.76810 & -0.76 & 0.40 & -2.09 & 0.55 & 0.19737 & 0.48   & 5.9099 \\     
2454394.76002 & 1.91 & 0.41 & 1.82 & 0.44   & 0.19706 & 0.57   & 5.9086 \\   
2454419.79420 & -4.06 & 0.36 & -4.76 & 0.38 & 0.19949 & 0.56   & 5.9075 \\     
2454420.76710 & -4.00 & 0.36 & -4.65 & 0.38 & 0.19976 & 0.46   & 5.9038 \\     
2454421.72662 & -6.46 & 0.31 & -6.55 & 0.42 & 0.19920 & 0.41   & 5.9033 \\     
2454422.74119 & -6.11 & 0.38 & -7.16 & 0.40 & 0.19682 & 0.52   & 5.9076 \\     
2454423.76829 & -1.77 & 0.38 & -3.12 & 0.43 & 0.19610 & 0.57   & 5.9027 \\     
2454424.74178 & 2.29 & 0.41 & 1.47 & 0.44   & 0.19561 & 0.56   & 5.9035 \\   
2454425.73676 & 3.14 & 0.40 & 1.80 & 0.44   & 0.19552 & 0.60   & 5.9029 \\   
2454426.71643 & 0.42 & 0.35 & -0.05 & 0.41  & 0.19620 & 0.41   & 5.9011 \\    
2454427.74603 & 2.48 & 0.33 & 0.75 & 0.47   & 0.19600 & 0.47   & 5.9032 \\   
2454428.75546 & 4.51 & 0.42 & 4.09 & 0.51   & 0.19717 & 0.48   & 5.9030 \\   
2454429.72905 & 3.77 & 0.35 & 2.27 & 0.37   & 0.19554 & 0.56   & 5.9052	\\   
2454445.73146 & 4.71 & 0.40 & 3.94 & 0.38   & 0.19983 & 0.68   & 5.9044 \\   
2454445.73671 & 4.54 & 0.43 & 4.75 & 0.45   & 0.19962 & 0.71   & 5.9042 \\   
2454451.75723 & 1.71 & 0.64 & 1.88 & 0.60   & 0.20123 & 0.93   & 5.9077 \\   
2454451.76263 & 2.42 & 0.49 & 1.47 & 0.51   & 0.20303 & 0.85   & 5.9089 \\   
2454454.76782 & 3.87 & 0.52 & 3.77 & 0.51   & 0.20442 & 0.73   & 5.9114 \\   
2454454.77327 & 4.14 & 0.44 & 4.44 & 0.55   & 0.20680 & 0.75   & 5.9108 \\   
2454478.66543 & -6.25 & 0.37 & -6.75 & 0.31 & 0.20483 & 0.50   & 5.9114 \\     
2454479.62214 & -6.75 & 0.35 & -7.58 & 0.37 & 0.20480 & 0.51   & 5.9126 \\     
2454480.58502 & -3.22 & 0.42 & -4.69 & 0.27 & 0.20450 & 0.59   & 5.9116 \\     
2454483.59354 & 0.73 & 0.40 & -0.87 & 0.41  & 0.20461 & 0.53   & 5.9107 \\    
2454484.62713 & 2.01 & 0.43 & 1.11 & 0.53   & 0.20785 & 0.46   & 5.9128 \\   
2454486.58708 & -0.20 & 0.4 & -0.94 & 0.47  & 0.20710 & 0.60   & 5.9126 \\    
2454529.54157 & 1.59 & 0.43 & -0.50 & 0.48  & 0.20937 & 0.59   & 5.9166 \\    
2454555.49159 & -0.26 & 0.38 & -1.68 & 0.37 & 0.19768 & 0.52   & 5.9059 \\     
2454556.48326 & -3.20 & 0.40 & -5.69 & 0.44 & 0.19768 & 0.52   & 5.9029 \\     
2454557.48453 & -0.34 & 0.47 & -2.01 & 0.40 & 0.19579 & 0.67   & 5.9054 \\     
2454562.47919 & -0.9 & 0.35 & -1.38 & 0.46  & 0.19707 & 0.43   & 5.9017 \\    
2454566.47610 & -1.75 & 0.41 & -2.47 & 0.53 & 0.20228 & 0.56   & 5.9074 \\     
2454570.47483 & 5.08 & 0.45 & 3.16 & 0.44   & 0.20497 & 0.67   & 5.9156 \\   
2454736.87535 & 0.59 & 0.52 & -0.28 & 0.67  & 0.20753 & 0.79   & 5.9208 \\    
2454852.69375 & -0.19 & 0.46 & -2.03 & 0.53 & 0.22760 & 0.53   & 5.9260 \\     
2454854.61310 & -1.56 & 0.31 & -2.60 & 0.35 & 0.22883 & 0.54   & 5.9268 \\    
\end{longtable}
}

\end{document}